\documentclass[12pt]{iopart}
\usepackage{iopams}  
\usepackage[utf8]{inputenc}
\usepackage[english]{babel}
\usepackage[subpreambles=true, sort=true]{standalone}
\usepackage{csquotes}
\usepackage{amsmath, amsthm, amssymb, amsfonts}
\usepackage{dsfont}
\usepackage{mathrsfs}
\usepackage[numbers]{natbib}

\usepackage{hyperref}
\hypersetup{
    hidelinks,
    colorlinks=false,
    breaklinks=true,
    pdftitle={Internal structure of gauge-invariant projected entangled pair states}
}
\usepackage{eqparbox}
\usepackage{braket}
\usepackage{graphicx}
\usepackage{float}
\usepackage[shortlabels]{enumitem}
\setlist[enumerate,1]{label=(\roman*)}
\usepackage[page,toc]{appendix}
\usepackage{bbm}
\usepackage{tikz}
\usepackage[outline]{contour}
\usepackage{pgfkeys}

\theoremstyle{definition}
\newtheorem{definition}{Definition}[section]
\newtheorem{corollary}{Corollary}[section]
\newtheorem{lemma}{Lemma}[section]

\newtheorem{theorem}{Theorem}[section]
\newtheorem*{remark}{Remark}
\begin{document}
\title{Internal structure of gauge-invariant projected entangled pair states}
\date{\today}
\author{David Blanik}
\address{University of Vienna, Faculty of Physics, Boltzmanngasse 5, 1090 Vienna, Austria}
\author{Jos\'e Garre-Rubio}
\address{University of Vienna, Faculty of Mathematics, Oskar-Morgenstern-Platz 1, 1090 Vienna, Austria}
\address{Instituto de F\'isica Te\'orica, UAM/CSIC, C. Nicol\'as Cabrera 13-15, Cantoblanco, 28049 Madrid, Spain}
\author{Andr\'as Moln\'ar}
\address{University of Vienna, Faculty of Mathematics, Oskar-Morgenstern-Platz 1, 1090 Vienna, Austria}
\author{Erez Zohar}
\address{Racah Institute of Physics, The Hebrew University of Jerusalem, Givat Ram, Jerusalem 91904, Israel}

\begin{abstract}
  Projected entangled pair states (PEPS) are very useful in the description of strongly correlated systems, partly because they allow encoding symmetries, either global or local (gauge), naturally.
  In recent years, PEPS with local symmetries have increasingly been used in the study of non-perturbative regimes of lattice gauge theories, most prominently as a way to construct variational ansatz states depending only on a small number of parameters and yet capturing the relevant physical properties.
  For the case of one-dimensional PEPS (matrix product states --- MPS) a bidirectional connection was established between the internal structure of the tensor network, i.e.\ the mathematical properties of the constituent tensors, and the symmetry.
  In higher dimensions this has only been done for global symmetries, where in the local (gauge) case it is known only how to construct gauge-invariant states, but not what the symmetry implies on the internal structure of the PEPS.
  In the present work we complete this missing piece and study the internal structure of projected entangled pair states with a gauge symmetry. The PEPS we consider consist of matter and gauge field tensors placed on the vertices and edges, respectively, of arbitrary graphs.
\end{abstract}
\maketitle

\section{Introduction}

The role of gauge theories in modern physics cannot be underestimated. Having local symmetries at their core, they describe matter minimally coupled to gauge fields acting as interaction mediators. This quantum field theoretic description~\cite{peskin_introduction_1995} is central to the Standard Model of particle physics~\cite{langacker_standard_2011}, as well as several effective models of condensed matter~\cite{fradkin_field_2013}.

However, quantum gauge field theories, particularly non-abelian ones, are notoriously difficult to study in their non-perturbative regimes. One successful approach involves the latticization of the theory, as proposed by Wilson~\cite{wilson_confinement_1974}, which, when combined with Monte Carlo techniques~\cite{creutz_quarks_1983}, allows the computation of many important static quantities; for example, the hadronic spectrum of quantum chromodynamics (see, e.g., the  review~\cite{aoki_flag_2020}).
This approach, however, faces significant challenges when dealing with two specific, physically important scenarios. First, it does not allow the direct description of real-time evolution, since Monte Carlo computations are performed within a Euclidean, Wick-rotated spacetime. Second, at finite fermion density, the sign problem~\cite{troyer_computational_2005} may occur, blocking the way for Monte-Carlo studies of exotic phases, including many physically relevant regions of the QCD phase diagram~\cite{fukushima_phase_2011}.

One relatively recent way of tackling these issues, and in particular overcoming the sign problem, which is specific to path-integral Monte Carlo computations, is the use of tensor network states.
Tensor network states (TNS)~\cite{White92,Schollw11,PerezGarcia_MPSREP,ReviewTN} are constructed from local tensors placed at particle sites (usually vertices or edges of the lattice) mimicking the interactions between them. The size of these local tensors determines how well the TNS approximates a given state.
Crucially, TNS constructed from local tensors of only moderate size are already well suited to approximate states satisfying an entanglement area law, expected to hold for physically relevant states (e.g., ground states of local gapped Hamiltonians~\cite{Hastings07A,Hastings07B,Masanes09,Eisert10}), allowing for efficient numerical computations.
In the context of lattice gauge theories, two (and higher) dimensional tensor networks, and in particular projected entangled pair states (PEPS)~\cite{verstraete2004renormalization,ReviewTN}, have been applied to the Hamiltonian formulation, introduced by 
Kogut and Susskind~\cite{kogutHamiltonianFormulationWilson1975}. In one space dimension, the family of TNS known as matrix product states (MPS) have been successfully applied to several lattice gauge theory models (see, e.g., the earlier works~\cite{banuls_mass_2013,buyens_matrix_2014,rico_tensor_2014,kuhn_non-abelian_2015,banuls_thermal_2015,pichler_real-time_2016,buyens_hamiltonian_2016} as well as the review papers~\cite{dalmonte_lattice_2016,banulsSimulatingLatticeGauge2020,banuls_review_2020}).
In higher dimensions, either tree tensor networks or infinite PEPS are used~\cite{tagliacozzo_entanglement_2011,tagliacozzo_tensor_2014,crone_detecting_2020,robaina_simulating_2021,felser_two-dimensional_2020,magnifico_lattice_2021,montangero_loop-free_2022,cataldi202421d,magnifico2024tensornetworkslatticegauge} --- both are very powerful numerical techniques, though not necessarily tailored to the symmetries present in such models. A parallel, non-Hamiltonian approach is that of the Tensor Renormalization Group (TRG) method, as described in detail in the review~\cite{Meurice_2022}.

Other approaches are based on the gauging of globally symmetric PEPS~\cite{haegeman_gauging_2015,zohar_building_2016},
where one lifts the global symmetry to a local one, creating ansatz states useful for lattice gauge theory computations. In particular,~\cite{zohar_building_2016} suggested how this can be done in a way analogous to minimal coupling of free, globally invariant matter Hamiltonians or Lagrangians~\cite{peskin_introduction_1995}, where the lift is achieved by the introduction of gauge fields. It was shown that when the gauged state is a Gaussian fermionic PEPS~\cite{kraus_fermionic_2010}, giving rise to states known as gauged Gaussian PEPS ~\cite{zohar_fermionic_2015,zohar_projected_2016}, one can use variational Monte Carlo techniques for an efficient ground state search~\cite{zohar_combining_2018, emonts_variational_2020,emonts_finding_23,kelman2024gauged,kelman2024projected}.
Therefore such gauged PEPS are useful in physical contexts and could be utilized in the study of the non-perturbative regime of lattice gauge theories. While the above works do well in providing methods to construct PEPS with gauge symmetries, one would like to understand the reverse direction as well --- given a locally symmetric PEPS, how is the symmetry encoded in its local constituents?
The way global properties of TNS, e.g., symmetries, are encoded in the local tensors is of high relevance  to the study of such states (e.g., leading to the classification of 1D gapped phases~\cite{Chen11_complete, Schuch11_SPT}).

This question has previously been addressed for a single space dimension, that is, for MPS
\cite{kullClassificationMatrixProduct2017}, and in the present work we aim at extending their results to higher dimensions, arbitrary lattices, geometries or graphs. 
For several families of TNS, the local encoding of the symmetry can be extracted using a set of results known as Fundamental Theorems~\cite{molnarNormalProjectedEntangled2018},
however, fundamental obstructions~\cite{PhysRevLett.125.210504} prevent a fully general analysis.
Consequently, in this work we shall only consider PEPS with matter and gauge fields described by local tensors which become injective after blocking, a strong invertibility condition~\cite{PerezGarcia_MPSREP,molnarNormalProjectedEntangled2018}.
This family is important because an injective PEPS with only matter degrees of freedom is the unique ground state of an associated local, gapped Hamiltonian~\cite{perezgarcia07PHpeps},
known as the parent Hamiltonian.

The structure of the paper is as follows.
We begin in Section~\ref{sec:basics-tens-netw} by introducing the necessary background on tensor network states and their graphical notation.
Then in Section~\ref{sec:example-normal-peps} we analyze the very simplest non-trivial example of a PEPS on the square lattice.
This section only serves to illustrate the main ideas of our method in a simple setting, and it is followed by Section~\ref{sec:demo} where we present an illustration.
In Section~\ref{sec:main-technical-lemma} we state and prove our main technical lemma,
a slightly stronger variation of the Fundamental Theorem of injective MPS, established in~\cite{molnarNormalProjectedEntangled2018}.
Finally, in Section~\ref{sec:general-statement} we state and prove our results in the most general setting we have been able to analyze.

\section{Basics on tensor networks and graphical notation}
\label{sec:basics-tens-netw}

Tensor network states are constructed via the concatenation of local tensors, often placed on the vertices and/or edges of a regular lattice. We explain the key concepts of this construction on the very important example of matrix product states (MPS) in the setting of spin chains~\cite{PerezGarcia_MPSREP}.
A (translation invariant) MPS is defined by a $3$-index tensor $B$ where one index corresponds to the local Hilbert space $\mathbb{C}^d$ and the other two are associated with the so-called virtual or auxiliary level with bond dimension $D$. The tensor $B$ can be seen as a collection of $d$ matrices, each of size $D\times D$: $\{ B^i,i=1,\ldots, d\}$. Then the MPS on $n$ sites constructed with the tensor $B$ is given by:
\begin{equation}
  \label{eq:26}
  \ket{\Psi(B)}_n = \sum_{i_1,\ldots, i_n} \Tr{B^{i_1}\cdots B^{i_n} }\ket{i_1,\ldots, i_n}.
\end{equation}
From a slightly more abstract point of view, which we shall adopt below, one considers the MPS tensor \(B\)
to be an element of the tensor product of the physical and virtual Hilbert spaces, which will also be defined below.
The importance of this class of TNS stems from the ability of MPS to efficiently approximate ground states of local, gapped Hamiltonians, which follows from the fact that both satisfy an entanglement entropy area law~\cite{Hastings07A,Hastings07B}.

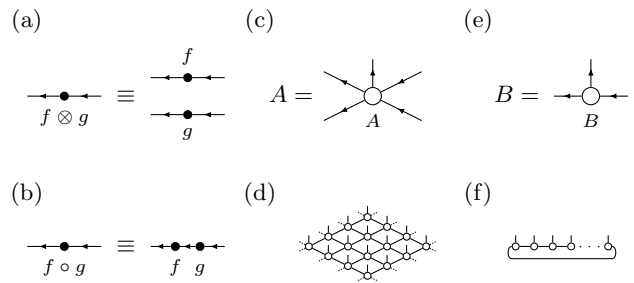
\begin{figure}[H]
  \centering
  \begin{tikzpicture}
  \begin{scope}[shift = {(0, 0)}]
    \node [empty] (L) at (-.5cm, 1cm) {(a)};
    \node[matrix, label = below:{\scriptsize\(f\otimes g\)}] (M1) at (0, 0) {};
    \draw [mid arrow] (1,0) -- (M1);
    \draw [mid arrow] (M1) -- (-1,0);
    \node [anchor=base, yshift=-\the\fontdimen22\textfont2] (E) at (5/3, 0) {\(\equiv\)};
    \node[matrix, label = above:{\scriptsize\(f\)}] (M2) at (3+1/3, 1/2) {};
    \draw [mid arrow] (4+1/3,1/2) -- (M2);
    \draw [mid arrow] (M2) -- (7/3,1/2);
    \node[matrix, label = below:{\scriptsize\(g\)}] (M3) at (3+1/3, -1/2) {};
    \draw [mid arrow] (4+1/3,-1/2) -- (M3);
    \draw [mid arrow] (M3) -- (7/3,-1/2);
  \end{scope}
  \begin{scope}[shift = {(0, -2cm)}]
    \node [empty] (L) at (-.5cm, .7cm) {(b)};
    \node[matrix, label = below:{\scriptsize\(f\circ g\)}] (M0) at (0, 0) {};
    \draw [mid arrow] (1,0) -- (M0);
    \draw [mid arrow] (M0) -- (-1,0);
    \node [anchor=base, yshift=-\the\fontdimen22\textfont2] (E) at (5/3, 0) {\(\equiv\)};
    \node[matrix, label = below:{\scriptsize\(f\)}] (M1) at (3, 0) {};
    \node[matrix, label = below:{\scriptsize\(g\vphantom{f}\)}] (M2) at (3+2/3, 0) {};
    \draw [mid arrow] (4+1/3,0) -- (M2);
    \draw [mid arrow] (M2) -- (M1);
    \draw [mid arrow] (M1) -- (7/3,0);
  \end{scope}
  \begin{scope}[shift = {(4.1cm,0)}, z={(0,1)}, y={(1.0em,0.5em)},x={(1.0em,-0.5em)}]
    \node [empty] (L) at (-1.5cm,1cm) {(c)};
    \node [anchor=base, yshift=-\the\fontdimen22\textfont2] (A) at (-5/3, -5/3) {\(A = \)};
    \node [tensor, label=below:{\scriptsize\(A\)}] (0) at (0, 0) {\(\)};
    \draw [mid arrow] (2, 0) -- (0);
    \draw [mid arrow] (0, 2) -- (0);
    \draw [mid arrow] (0) -- (0, -2);
    \draw [mid arrow] (0) -- (-2, 0);
    \draw [mid arrow] (0) --++ (0,0,1);
  \end{scope}
  \begin{scope}[shift = {(7cm,0)}]
    \node [empty] (L) at (-1.5cm ,1cm) {(e)};
    \node [anchor=base, yshift=-\the\fontdimen22\textfont2] (A) at (-2, 0) {\(B = \)};
    \node [tensor, label = below:{\scriptsize \(B\)}] (B) at (0, 0) {\(\)};
    \draw [mid arrow] (B) -- (0,1);
    \draw [mid arrow] (B) -- (-1,0);
    \draw [mid arrow] (1,0) -- (B);
  \end{scope}
  \begin{scope}[shift={(3.25cm,-2cm)} ,z={(0,1)}, y={(1.0em,0.5em)},x={(1.0em,-0.5em)}, scale = 0.2]
    \node [empty] (L) at (-3.25cm,3.5cm) {(d)};
    \foreach \x [remember=\x as \lx] in {0,...,3} 
      {
        \foreach \y [remember=\y as \ly] in {0,...,3}
        {
          \node [tensor, minimum width = .2em, text width = .2em, text height = .2em, minimum height = .2em] (\x\y) at (4*\x,4*\y) {};
          \draw (\x\y) --++ (0,0,1.5);
          \if0\x\else\draw (\x\y)--(\lx\y);\fi%
          \if0\y\else\draw (\x\y)--(\x\ly);\fi%
        }
      }
    \foreach \x in {0,...,3} 
      \draw [dash pattern=on .5pt off .5pt, line width = .3pt] (\x0) --++ (0,-2) (\x3) --++ (0,2);
    \foreach \y in {0,...,3} 
      \draw [dash pattern=on .5pt off .5pt, line width = .3pt] (0\y) --++ (-2,0) (3\y) --++ (2,0);
  \end{scope}
  \begin{scope}[shift = {(6cm,-2cm)}, scale = 0.5]
    \node [empty] (L) at (-1cm ,1.4cm) {(f)};
    \foreach \n in {0, 1, 2, 3, 5}
    {
      \node [tensor, minimum width = .2em, text width = .2em, text height = .2em, minimum height = .2em] (B\n) at (\n, 0) {};
      \draw (B\n) --++ (0,3/5);
    }
    \node [anchor=base, yshift=-\the\fontdimen22\textfont2] (D) at (4, 0) {\tiny\(\cdots\)};
    \draw (B3) -- (B2) -- (B1) -- (B0);
    \draw (B0.west) to[out=180,in=180] ([yshift=-1em]B0.west) to ([yshift=-1em]B5.east) to[out=0, in=0] (B5.east);
  \end{scope}
\end{tikzpicture}
  \caption{
    In graphical notation the tensor product (a) of operators is depicted
    by simply drawing the factors next to each other.
    Composition (b), or more generally contraction, is depicted as concatenation of diagrams,
    where the ordering is determined by the orientation of the legs.
    We depict a rank 5 tensor (c) and how it can be used to construct a PEPS on a square lattice (d),
    given the virtual Hilbert spaces are chosen appropriately.
    Similarly, a rank 3 tensor (e) can be used to construct a translation invariant MPS (f) with
    periodic boundary conditions.
  }\label{fig:intro}
\end{figure}

In order to generalize this construction to higher dimensions we will first introduce the graphical notation of tensors. A tensor is depicted by a shape, usually a circle, with as many legs as indices, where the indices should be thought of as corresponding to a choice of basis on the Hilbert space associated with the leg. There are two basic operations between the indices of (different) tensors. The first one is the contraction of two indices where the labels are identified and summed over. An example is matrix multiplication: $(f g)_{ij} = \sum_{kl} \delta_{kl} f_{ik} g_{lj}$. The graphical representation of the contraction is to glue or concatenate the legs of the involved indices together. This operation corresponds to the natural pairing between a Hilbert space and its dual and can in some cases be viewed as composition of functions, cf.\ Figure~\ref{fig:intro}(b). The second operation is blocking: here multiple indices, or Hilbert spaces, are considered as one, so that their total dimension is the product of the individual ones. An example is the tensor product of matrices or operators, $f \otimes g$, cf.\ Figure~\ref{fig:intro}(a). 
The higher-dimensional generalization of MPS are called projected entangled pair states (PEPS). For the two-dimensional square lattice the tensors of a PEPS are rank 5 (5 indices) and they are placed on the vertices, cf.\ Figure~\ref{fig:intro}(c,~d). Notice the simplicity of the graphical notation; the corresponding algebraic expression is too complicated to be written down here.

\begin{definition}[MPS]\label{def:mps}
  Formally, in TN diagrams a Hilbert space is assigned to each edge (or leg)
  and tensors, being nodes where multiple edges meet, are simply elements in the tensor product
  of the Hilbert spaces associated with its incident legs, e.g.,
  \begin{equation}
    \label{eq:24}
    B \equiv 
  \pgfkeys{
    label/.store in=\MPSTensorLabel,
    label=\(\), 
    label inner/.store in=\MPSTensorLabelInner, 
    label inner=\(\),
    rep top/.store in=\TopLegOperator,
    rep top = 1,
    vector top/.store in=\TopLegVector,
    vector top = 1,
    rep left/.store in=\LeftLegOperator,
    rep left = 1,
    rep right/.store in=\RightLegOperator,
    rep right = 1
  }
  \pgfkeys{label = \(B\)}%
  \begin{tikzpicture}[baseline={([yshift=-\the\fontdimen22\textfont2]A.center)}] 
\node[tensor, label=below:{\scriptsize\MPSTensorLabel}] (A) at (0, 0) {\MPSTensorLabelInner};
\if 1\RightLegOperator
  \node[empty] (Hv2) at (1, 0) {\scriptsize\MPSIndexRight};
  \draw [mid arrow] (Hv2) -- (A);
\else
  \node[matrix, label=above:{\scriptsize\RightLegOperator}, right = 1 em of A] (OpR) {};
  \node[empty] (Hv2) at (1.5, 0) {\scriptsize\MPSIndexRight};
  \draw [mid arrow] (OpR) -- (A);
  \draw (Hv2) -- (OpR);
\fi
\if 1\LeftLegOperator
  \node[empty] (Hv1) at (-1, 0) {\scriptsize\MPSIndexLeft};
  \draw [mid arrow] (A) -- (Hv1);
\else
  \node[matrix, label=above:{\scriptsize\LeftLegOperator}, left = 1 em of A] (OpL) {};
  \node[empty] (Hv1) at (-1.5, 0) {\scriptsize\MPSIndexLeft};
  \draw [mid arrow] (A) -- (OpL);
  \draw (OpL) -- (Hv1);
\fi
\if 1\TopLegOperator
  \if 1\TopLegVector
    \node[empty] (Hp) at (0, 1) {\scriptsize\MPSIndexTop};
    \draw [mid arrow] (A) -- (Hp);
  \else
    \node[matrix, label=right:{\scriptsize\TopLegVector}, above = 1 em of A] (VT) {};
    \draw [mid arrow] (A) -- (VT);
  \fi
\else
  \node[matrix, label=right:{\scriptsize\TopLegOperator}, above = 1 em of A] (OpT) {};
  \node[empty] (Hp) at (0, 1.5) {\scriptsize\MPSIndexTop};
  \draw [mid arrow] (A) -- (OpT);
  \draw (OpT) -- (Hp);
\fi
\end{tikzpicture}
 \equiv \begin{tikzpicture}[baseline={([yshift=-\the\fontdimen22\textfont2]B.center)}]
\node [tensor, label = below:{\scriptsize \(B\)}] (B) at (0, 0) {\(\)};
\node (p) at (0,3/2) {\scriptsize\(\mathcal{H}\)};
\node (v1) at (-3/2,0) {\scriptsize\(\mathcal{V}_{\mathrm{o}}\)};
\node (v2) at (3/2,0) {\scriptsize\(\mathcal{V}_{\mathrm{t}}\)};
\draw [mid arrow] (B) -- (0,1);
\draw [mid arrow] (B) -- (v1);
\draw [mid arrow] (v2) -- (B);
\end{tikzpicture}
    \in \mathcal{V}_{\mathrm{o}}\otimes \mathcal{H} \otimes \mathcal{V}_{\mathrm{t}}^{*},
  \end{equation}
  where the dual of a Hilbert space appears in the tensor product,
  if the edge it is associated with terminates at the tensor.
\end{definition}
Hilbert spaces denoted as \(\mathcal{H}\) we shall refer to as \emph{physical},
while Hilbert spaces denoted as \(\mathcal{V}\) we shall refer to as \emph{virtual}.
Equation~\eqref{eq:24} depicts the prototypical example of an MPS tensor, where
we denote the virtual Hilbert spaces \(\mathcal{V}_{\mathrm{o}}\) and \(\mathcal{V}_{\mathrm{t}}\),
because they are associated with edges that have the MPS tensor as their origin and terminus, respectively.
\begin{definition}[Injectivity]\label{def:lri}
  An MPS tensor \(B\), defined according to Equation~\eqref{eq:24}, is called
  \begin{enumerate}
  \item \emph{injective}, if it is injective as a map\footnote{\label{fn:tn-map}%
      When interpreting TN drawings as linear maps, we shall usually view them as going from bottom to top.}
      \begin{equation}\label{eq:15}
        
  \pgfkeys{
    label/.store in=\MPSTensorLabel,
    label=\(\), 
    label inner/.store in=\MPSTensorLabelInner, 
    label inner=\(\),
    rep top/.store in=\TopLegOperator,
    rep top = 1,
    vector top/.store in=\TopLegVector,
    vector top = 1,
    rep left/.store in=\LeftLegOperator,
    rep left = 1,
    rep right/.store in=\RightLegOperator,
    rep right = 1
  }
  \pgfkeys{label = \(B\)}%
  \begin{tikzpicture}[baseline={([yshift=-\the\fontdimen22\textfont2]A.center)}] 
\node[tensor, label=below:{\scriptsize\MPSTensorLabel}] (A) at (0, 0) {\MPSTensorLabelInner};
\if 1\RightLegOperator
  \node[empty] (Hv2) at (1, 0) {\scriptsize\MPSIndexRight};
  \draw [mid arrow] (Hv2) -- (A);
\else
  \node[matrix, label=above:{\scriptsize\RightLegOperator}, right = 1 em of A] (OpR) {};
  \node[empty] (Hv2) at (1.5, 0) {\scriptsize\MPSIndexRight};
  \draw [mid arrow] (OpR) -- (A);
  \draw (Hv2) -- (OpR);
\fi
\if 1\LeftLegOperator
  \node[empty] (Hv1) at (-1, 0) {\scriptsize\MPSIndexLeft};
  \draw [mid arrow] (A) -- (Hv1);
\else
  \node[matrix, label=above:{\scriptsize\LeftLegOperator}, left = 1 em of A] (OpL) {};
  \node[empty] (Hv1) at (-1.5, 0) {\scriptsize\MPSIndexLeft};
  \draw [mid arrow] (A) -- (OpL);
  \draw (OpL) -- (Hv1);
\fi
\if 1\TopLegOperator
  \if 1\TopLegVector
    \node[empty] (Hp) at (0, 1) {\scriptsize\MPSIndexTop};
    \draw [mid arrow] (A) -- (Hp);
  \else
    \node[matrix, label=right:{\scriptsize\TopLegVector}, above = 1 em of A] (VT) {};
    \draw [mid arrow] (A) -- (VT);
  \fi
\else
  \node[matrix, label=right:{\scriptsize\TopLegOperator}, above = 1 em of A] (OpT) {};
  \node[empty] (Hp) at (0, 1.5) {\scriptsize\MPSIndexTop};
  \draw [mid arrow] (A) -- (OpT);
  \draw (OpT) -- (Hp);
\fi
\end{tikzpicture}
:
        \mathcal{V}_{\mathrm{o}}^{*} \otimes \mathcal{V}_{\mathrm{t}} \longrightarrow \mathcal{H}.
      \end{equation}
    \item \emph{left-injective}, if it is injective as a map
      \begin{equation}\label{eq:16}
        \begin{tikzpicture}[baseline={([yshift=-\the\fontdimen22\textfont2]B.center)}]
\node [tensor, label = below:{\scriptsize \(B\)}] (B) at (0, 0) {\(\)};
\draw [mid arrow] (B) -- (0,1);
\draw [rounded corners=1em, mid arrow] (B) -- (-1,0) -- (-1,1);
\draw [mid arrow] (1,0) -- (B);
\end{tikzpicture}:
        \mathcal{V}_{\mathrm{t}} \longrightarrow \mathcal{V}_{\mathrm{o}}\otimes \mathcal{H}.
      \end{equation}
    \item \emph{right-injective}, if it is injective as a map
      \begin{equation}\label{eq:17}
        \begin{tikzpicture}[baseline={([yshift=-\the\fontdimen22\textfont2]B.center)}]
\node [tensor, label = below:{\scriptsize \(B\)}] (B) at (0, 0) {\(\)};
\draw [mid arrow] (B) -- (0,1);
\draw [mid arrow] (B) -- (-1,0);
\draw [rounded corners=1em, mid arrow] (1,1) -- (1,0) -- (B);
\end{tikzpicture}:
        \mathcal{V}_{\mathrm{o}}^{*} \longrightarrow \mathcal{H}\otimes \mathcal{V}_{\mathrm{t}}^{*}.
      \end{equation}
    \item \emph{unital}, if \(\mathcal{V}^{\mathrm{o}} = \mathcal{V}^{\mathrm{t}} =: \mathcal{V}\)
      and there exists \(x\in \mathcal{H}^{*}\) such that
      \begin{equation}
        \label{eq:9}
        \begin{tikzpicture}[baseline={([yshift=-\the\fontdimen22\textfont2]B.center)}]
\node [tensor, label = below:{\scriptsize \(B\)}] (B) at (0, 0) {\(\)};
\node [matrix, label=right:{\scriptsize \(x\)}] (x) at (0,1) {};
\draw [mid arrow] (B) -- (x);
\draw [mid arrow] (B) -- (-1,0);
\draw [mid arrow] (1,0) -- (B);
\end{tikzpicture} = \operatorname{id}_{\mathcal{V}}.
      \end{equation}
  \end{enumerate}
  If a tensor satisfies (ii) and (iii), we call it \emph{LRI}.
  In components, unitality just means \(\sum_i x_i B^i = \mathbbm{1}\).
\end{definition}
\begin{remark}
  Injective \(\Rightarrow\) unital \(\Rightarrow\) LRI.
\end{remark}
Part (i) of the previous definition is standard~\cite{PerezGarcia_MPSREP}
and Lemma \ref{lem:injective} justifies the introduction of notions (ii) and (iii).
Part (iv) is convenient to give examples.
Clearly, injectivity of the tensor \(B\) is equivalent to the existence of a tensor \(B^{-1}\),
often called the \emph{inverse} of \(B\), satisfying
\begin{equation}
  \label{eq:25}
  \begin{tikzpicture}[baseline={([yshift=-\the\fontdimen22\textfont2]Bc.center)}]
\node [tensor, label = below:{\scriptsize \(B\)}] (B) at (0, 0) {\(\)};
\node [tensor, label= above:{\scriptsize \(B^{-1}\)}] (Bi) at (0,3/2) {};
\node [empty] (Bc) at (0,3/4) {};
\draw [mid arrow] (B) -- (Bi);
\draw [mid arrow] (B) -- (-1,0);
\draw [mid arrow] (1,0) -- (B);
\draw [mid arrow] (Bi) -- (1,3/2);
\draw [mid arrow] (-1,3/2) -- (Bi);
\end{tikzpicture} = 
\begin{tikzpicture}[baseline={([yshift=-\the\fontdimen22\textfont2]Bc.center)}]
\node [empty] (Bc) at (0,3/4) {\(\)};
\draw [mid arrow] (-2/3,3/2) -- (-1/4, 3/2) -- (-1/4, 0) -- (-2/3,0);
\draw [mid arrow] (2/3,0) -- (1/4, 0) -- (1/4, 3/2) -- (2/3,3/2);
\end{tikzpicture}

  \equiv \operatorname{id}_{\mathcal{V}_{\mathrm{o}}^{*}}
  \otimes \operatorname{id}_{\mathcal{V}_{\mathrm{t}}},
\end{equation}
where in the last equality we view the TN diagram as a linear map going from bottom to top.
\begin{lemma}
  \label{lem:injective}
  Let \(T\) be an injective\footnote{
    A PEPS tensor is called \emph{injective} if it is injective as a map from virtual to physical Hilbert spaces, cf.\ Definition \ref{lem:blocking}.
  }
  tensor. For any MPS tensor \(B\) the following equivalences hold:
  \begin{align}
    \label{eq:6}
    \begin{tikzpicture}[baseline={([yshift=-\the\fontdimen22\textfont2]0.center)}, y={(1.0em,0.5em)},x={(1.0em,-0.5em)}, z={(0em,1em)}]
\node [tensor, label = below:{\scriptsize\(T\)}] (0) at (0, 0) {\(\)};
\node (L1) at (2, 0) {};
\node [stensor, label = below:{\scriptsize\(B\)}] (L2) at (0, 2) {};
\node (R1) at (0, -2) {};
\node (R2) at (-2, 0) {};
\draw (L2) --++ (0,.7,0);
\draw [mid arrow] (L1) -- (0);
\draw [mid arrow] (L2) -- (0);
\draw [mid arrow] (0) -- (R1);
\draw [mid arrow] (0) -- (R2);
\foreach \n in {0, L2} 
{
  \draw (\n) --++ (0,0,1);
}
\end{tikzpicture} &\text{ injective} & &\Leftrightarrow & B &\text{ is left-injective},\\
    \label{eq:22}
    \begin{tikzpicture}[baseline={([yshift=-\the\fontdimen22\textfont2]0.center)}, y={(1.0em,0.5em)},x={(1.0em,-0.5em)}, z={(0em,1em)}]
\node [tensor, label = below:{\scriptsize\(T\)}] (0) at (0, 0) {\(\)};
\node (L1) at (2, 0) {};
\node (L2) at (0, 2) {};
\node (R1) at (0, -2) {};
\node [stensor, label = below:{\scriptsize\(B\)}] (R2) at (-2, 0) {};
\draw (R2) --++ (-.7,0,0);
\draw [mid arrow] (L1) -- (0);
\draw [mid arrow] (L2) -- (0);
\draw [mid arrow] (0) -- (R1);
\draw [mid arrow] (0) -- (R2);
\foreach \n in {0, R2} 
{
  \draw (\n) --++ (0,0,1);
}
\end{tikzpicture}&\text{ injective} & &\Leftrightarrow & B &\text{ is right-injective}.
  \end{align}
\end{lemma}
\begin{proof}
  To prove (\ref{eq:6}, \(\Rightarrow\)), assume \(B\) is not left-injective,
  which implies the existence of \(x\in \mathcal{V}_{\mathrm{t}}\) such that
  \begin{equation}
    \label{eq:23}
    \begin{tikzpicture}[baseline={([yshift=-\the\fontdimen22\textfont2]B.center)}]
\node [tensor, label = below:{\scriptsize \(B\)}] (B) at (0, 0) {\(\)};
\draw [mid arrow] (B) -- (0,1);
\draw [mid arrow] (B) -- (-1,0);
\node [matrix, label = below:{\scriptsize \(x\)}] (x) at (1, 0) {\(\)};
\draw [mid arrow] (x) -- (B);
\end{tikzpicture} = 0,
  \end{equation}
  which is of course in contradiction to the injectivity of the composite tensor.
  The reverse implication (\ref{eq:6}, \(\Leftarrow\)) follows immediately after
  contracting the composite tensor with the inverse of \(T\).
  One proves \eqref{eq:22} in a completely analogous manner.
\end{proof}
Hence, blocking with \(B\) preserves injectivity precisely if it is left- and/or right-injective.

\section{Example: Normal PEPS on Square Lattice}
\label{sec:example-normal-peps}
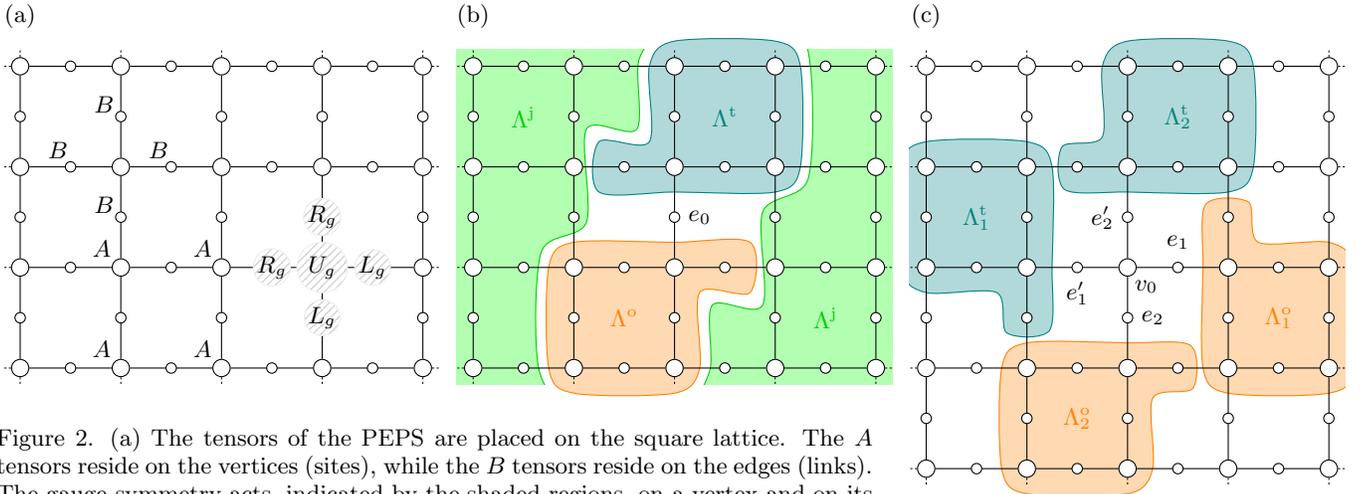
\begin{figure*}
 \scalebox{.75}{\begin{tikzpicture}[sop/.style ={circle, preaction={fill, white}, draw=black!7, pattern=north east lines, pattern color=black!20}, scale = 0.68]
\begin{scope}[]
  \node (l) at (0, 14) {(a)};
  \foreach \x [remember=\x as \lx] in {0,...,4} 
    {
      \foreach \y [remember=\y as \ly] in {0,...,3}
      {
        \node [tensor] (\x\y) at (4*\x,4*\y) {};
        \if0\x\else\draw [mid bullet] (\x\y)--(\lx\y);\fi%
        \if0\y\else\draw [mid bullet] (\x\y)--(\x\ly);\fi%
      }
    }
  \foreach \x in {0,...,4} 
    \draw [dash pattern=on .82pt off .90pt, line width = .3pt] (\x0) --++ (0,-1em) (\x3) --++ (0,1em);
  \foreach \y in {0,...,3} 
    \draw [dash pattern=on .82pt off .90pt, line width = .3pt] (0\y) --++ (-1em,0) (4\y) --++ (1em,0);
  
  \foreach \a in {10, 20, 11, 21} 
    \node (l\a) at ([shift={(-3/4,3/4)}]\a) {\(A\)};
  \foreach \b in {(3/2,2/3), (-2/3,5/2), (-5/2, 2/3), (-2/3,-3/2)} 
    \node (\b) at ([shift={\b}]12) {\(B\)};
  \node[sop, inner sep = 2pt] (U)  at (12, 4) {\(U_g\)};
  \node[sop, inner sep = 0pt] (R1) at (10, 4) {\(R_g\)};
  \node[sop, inner sep = 0pt] (R2) at (12, 2) {\(L_g\)};
  \node[sop, inner sep = 0pt] (L1) at (14, 4) {\(L_g\)};
  \node[sop, inner sep = 0pt] (L2) at (12, 6) {\(R_g\)};
\end{scope}
\begin{scope}[shift = {(18,0)}]
  \node (l) at (0, 14) {(b)};
  \begin{scope}
    \clip (-1em,-1em) rectangle ($(16,12)+(1em,1em)$);
    \draw [white, fill = green!30!white] (-1,-1) rectangle (17,13);
    \draw [green!80!black, fill=white] plot [smooth cycle, tension=.5] coordinates {(13.3,7.5) (13.4,12) (12.5, 13.5) (8,13.5) (6.6,12.3) (6.5,9.5) (4.5,9.5) (4.5,5.8) (2.7,4.6) (2.6, 0) (4, -1.5) (8.2,-1.5) (9.4,-0.1) (9.5, 2.5) (11.5,2.5) (11.5, 6.2)};
  \end{scope}
  \draw [teal, fill=teal!30!white] plot [smooth cycle, tension=0.5] coordinates {(8,7) (12,7) (13,8) (13,12) (12, 13) (8,13) (7,12) (7,9) (5,9) (5,7)};
  \begin{scope}[rotate=180, shift={(-16,-12)}]
    \draw [orange, fill=orange!30!white] plot [smooth cycle, tension=0.5] coordinates {(8,7) (12,7) (13,8) (13,12) (12, 13) (8,13) (7,12) (7,9) (5,9) (5,7)};
  \end{scope}
  \foreach \x [remember=\x as \lx] in {0,...,4} 
      \foreach \y [remember=\y as \ly] in {0,...,3}
      {
        \node [tensor, fill=white] (\x\y) at (4*\x,4*\y) {};
        \if0\x\else\draw [mid bullet] (\x\y)--(\lx\y);\fi%
        \if0\y\else\draw [mid bullet] (\x\y)--(\x\ly);\fi%
      }
  \foreach \x in {0,...,4} 
    \draw [dash pattern=on .82pt off .90pt, line width = .3pt] (\x0) --++ (0,-1em) (\x3) --++ (0,1em);
  \foreach \y in {0,...,3} 
    \draw [dash pattern=on .82pt off .90pt, line width = .3pt] (0\y) --++ (-1em,0) (4\y) --++ (1em,0);
  
  \node (e) at ([shift={(1,2)}]21) {\(e_0\)};
  \node (o) at (10,10) {\(\color{teal}\Lambda^{\mathrm{t}}\)};
  \node (o) at (6,2) {\(\color{orange}\Lambda^{\mathrm{o}}\)};
  \node (o) at (14,2) {\(\color{green!80!black}\Lambda^{\mathrm{j}}\)};
  \node (o) at (2, 10) {\(\color{green!80!black}\Lambda^{\mathrm{j}}\)};
\end{scope}
\begin{scope}[shift = {(36,-4)}]
  \node (l) at (0, 18) {(c)};
  \clip (-1em,-1.2) rectangle ($(16,16)+(1em,1.2)$);
  \begin{scope}[shift={(0,4)}]
    \draw [teal, fill=teal!30!white] plot [smooth cycle, tension=0.5] coordinates {(8,7) (12,7) (13,8) (13,12) (12, 13) (8,13) (7,12) (7,9.2) (5.4,8.8) (5.5,7.1)};
  \end{scope}
  \begin{scope}[rotate=180, shift={(-16,-12)}]
    \draw [orange, fill=orange!30!white] plot [smooth cycle, tension=0.5] coordinates {(8,7) (12,7) (13,8) (13,12) (12, 13) (8,13) (7,12) (7,9.2) (5.4,8.8) (5.5,7.1)};
  \end{scope}
  \begin{scope}[rotate=270, shift={(-16,4)}]
    \draw [orange, fill=orange!30!white] plot [smooth cycle, tension=0.5] coordinates {(8,7) (12,7) (13,8) (13,12) (12, 13) (8,13) (7,12) (7,9.2) (5.4,8.8) (5.5,7.1)};
  \end{scope}
  \begin{scope}[rotate=90, shift={(0,-12)}]
    \draw [teal, fill=teal!30!white] plot [smooth cycle, tension=0.5] coordinates {(8,7) (12,7) (13,8) (13,12) (12, 13) (8,13) (7,12) (7,9.2) (5.4,8.8) (5.5,7.1)};
  \end{scope}
  \foreach \x [remember=\x as \lx] in {0,...,4} 
      \foreach \y [remember=\y as \ly] in {0,...,4}
      {
        \node [tensor, fill=white] (\x\y) at (4*\x,4*\y) {};
        \if0\x\else\draw [mid bullet] (\x\y)--(\lx\y);\fi%
        \if0\y\else\draw [mid bullet] (\x\y)--(\x\ly);\fi%
      }
  \foreach \x in {0,...,4} 
    \draw [dash pattern=on .82pt off .90pt, line width = .3pt] (\x0) --++ (0,-1em) (\x4) --++ (0,1em);
  \foreach \y in {0,...,4} 
    \draw [dash pattern=on .82pt off .90pt, line width = .3pt] (0\y) --++ (-1em,0) (4\y) --++ (1em,0);
  
  \node (v) at ([shift={(3/4,-3/4)}]22) {\(v_0\)};
  \node (e1p) at ([shift={(1,2)}]21) {\(e_2\)};
  \node (e2p) at ([shift={(2,-1)}]12) {\(e_1^{\prime}\)};
  \node (e1) at ([shift={(2,1)}]22) {\(e_1\)};
  \node (e2) at ([shift={(-1,2)}]22) {\(e_2^{\prime}\)};
  \node (o) at (10,14) {\(\color{teal}\Lambda^{\mathrm{t}}_2\)};
  \node (o) at (14,6) {\(\color{orange}\Lambda^{\mathrm{o}}_1\)};
  \node (o) at (6,2) {\(\color{orange}\Lambda^{\mathrm{o}}_2\)};
  \node (o) at (2,10) {\(\color{teal}\Lambda^{\mathrm{t}}_1\)};
\end{scope}
\end{tikzpicture}}
  \vspace{-1.8cm}
  \caption{\label{fig:square-lattice-combined}
    \parshape=4
    0pt 0.65\textwidth
    0pt 0.65\textwidth
    0pt 0.65\textwidth
    0pt \textwidth
    (a) The tensors of the PEPS are placed on the square lattice.
    The \(A\) tensors reside on the vertices (sites), while the \(B\) tensors reside on the edges (links).
    The gauge symmetry acts, indicated by the shaded regions, on a vertex and on its surrounding edges.
    (b) Isolating a single gauge tensor \(B\) at edge \(e_0\) from the rest of the tensor network
    is achieved by selecting adjacent regions \(\Lambda^{\mathrm{o}}\) and \(\Lambda^{\mathrm{t}}\),
    as indicated in the figure,
    in such a way that blocking the tensors contained within results in injective tensors.
    The region \(\Lambda^{\mathrm{j}}\) contains the remainder of the lattice and can either be connected
    or disconnected. It is used only for bookkeeping purposes.
    The superscripts are meant to reference \emph{origin}, \emph{terminus} and \emph{junk}, respectively,
    and we use the colors orange, teal and jade as a visual aid.
    When we don't explicitly indicate the orientation of the edges in drawings, they are usually
    implied to be oriented left to right and bottom to top.
    (c) To separate a single matter tensor \(A\) at vertex \(v_0\)
    together with its surrounding gauge tensors, from the remaining tensor network
    we can employ the same regions we used above when isolating the edge tensors incident to \(v_0\),
    since blocking the tensors contained within
    \(\Lambda^{\mathrm{t}}_1\cup\Lambda^{\mathrm{t}}_2\cup\Lambda^{\mathrm{o}}_1\cup\Lambda^{\mathrm{o}}_2\),
    results in an injective tensor.
  }
\end{figure*}
As a simple first example,
to illustrate the ideas appearing in the proof of the general statement
in Section~\ref{sec:general-statement},
we consider a gauge-invariant PEPS on a square lattice with periodic boundary conditions,
i.e.\ defined on the underlying geometry of a torus.

Such a state is constructed by placing rank 5 tensors \(A_v\),
which we think of as describing the matter degrees of freedom,
at each vertex \(v\) and rank 3 tensors \(B_e\),
which we think of as describing the gauge degrees of freedom,
at each edge \(e\) of the lattice.
The physical local Hilbert spaces,
always associated with the first leg of each tensor,
are denoted \(\mathcal{H}_v\) and \(\mathcal{H}_e\), respectively.

Contracting the remaining legs, corresponding to virtual degrees of freedom,
according to the lattice geometry depicted in Figure~\ref{fig:square-lattice-combined}(a),
results in the state:
\begin{equation}
  \ket{\Psi(A,B)} \in \bigotimes_v \mathcal{H}_v \otimes \bigotimes_e \mathcal{H}_e.
\end{equation}
Let \(G\) be a finite, or compact Lie group and consider representations\footnote{
  Homomorphisms from \(G\) to the general linear group on the respective Hilbert space.
}
\(U_v : G \rightarrow \operatorname{GL}(\mathcal{H}_v)\)
and
\(L_e,R_e:G \rightarrow \operatorname{GL}(\mathcal{H}_e)\),
for all vertices (sites) \(v\) and edges (links) \(e\),
where we require \(L\) and \(R\) to be commuting,
i.e.\ $[R_e(h),L_e(g)]=0$ for all edges \(e\) and \(g,h\in G\).
Saying the state \(\ket{\Psi(A,B)}\) has a gauge symmetry simply means that 
(disregarding the case of static charges)
it satisfies
\begin{equation}\label{eq:local-sym}
  \widehat{U}_v(g)
  \ket{\Psi(A,B)} = \ket{\Psi(A,B)}
\end{equation}
for all vertices \(v\), with
\begin{equation}
  \widehat{U}_v :=
  R_{v_{\mathrm{l}}} \otimes R_{v_{\mathrm{t}}}
  \otimes U_v \otimes
  L_{v_{\mathrm{r}}} \otimes L_{v_{\mathrm{b}}},
\end{equation}
where the subscripts denote the edges to the left, top, right and below of the vertex \(v\),
cf.\ Figure \ref{fig:square-lattice-combined}~(a).

To simplify notation, we shall assume that all vertex tensors
and edge tensors are equal,
to be precise \(A_v \equiv A\) and \(B_e \equiv B\) for all \(v\) and \(e\).
This restriction is by no means necessary,
but there is simply no insight to be gained from considering the general case here.
Graphically, we can depict property \eqref{eq:local-sym} for a single vertex as
\begin{equation}
  \label{eq:11}
  \begin{tikzpicture}[baseline={([yshift=-\the\fontdimen22\textfont2]0.center)}, y={(1.0em,0.5em)},x={(1.0em,-0.5em)}, z={(0em,1em)}]
\node [tensor] (0) at (0, 0) {\(\)};
\node [stensor] (L1) at (2, 0) {};
\node [stensor] (L2) at (0, 2) {};
\node [stensor] (R1) at (0, -2) {};
\node [stensor] (R2) at (-2, 0) {};
\draw [dotted] (L1) --++ (1.4,0,0);
\draw (L1) --++ (.7,0,0);
\draw [dotted] (L2) --++ (0,1.4,0);
\draw (L2) --++ (0,.7,0);
\draw [dotted] (R1) --++ (0,-1.4,0);
\draw (R1) --++ (0,-.7,0);
\draw [dotted] (R2) --++ (-1.4,0,0);
\draw (R2) --++ (-.7,0,0);
\draw [mid arrow] (L1) -- (0);
\draw [mid arrow] (L2) -- (0);
\draw [mid arrow] (0) -- (R1);
\draw [mid arrow] (0) -- (R2);
\foreach \x/\n in {R/1, R/2, L/1, L/2} 
{
  \node [matrix, shift={(0,0,.8)}, label = {[label distance = -3.8pt]left:{\scriptsize\(\x_g\)}}] (u\x\n) at (\x\n) {};
  \draw (\x\n) -- (u\x\n) --++ (0,0,.6);
}
\node [matrix, shift={(0,0,1)}, label = {[label distance = -3pt]left:{\scriptsize\(U_g\)}}] (u0) at (0) {};
\draw (0) -- (u0) --++ (0,0,.7);
\end{tikzpicture} = \begin{tikzpicture}[baseline={([yshift=-\the\fontdimen22\textfont2]0.center)}, y={(1.0em,0.5em)},x={(1.0em,-0.5em)}, z={(0em,1em)}]
\node [tensor] (0) at (0, 0) {\(\)};
\node [stensor] (L1) at (2, 0) {};
\node [stensor] (L2) at (0, 2) {};
\node [stensor] (R1) at (0, -2) {};
\node [stensor] (R2) at (-2, 0) {};
\draw [dotted] (L1) --++ (1.4,0,0);
\draw (L1) --++ (.7,0,0);
\draw [dotted] (L2) --++ (0,1.4,0);
\draw (L2) --++ (0,.7,0);
\draw [dotted] (R1) --++ (0,-1.4,0);
\draw (R1) --++ (0,-.7,0);
\draw [dotted] (R2) --++ (-1.4,0,0);
\draw (R2) --++ (-.7,0,0);
\draw [mid arrow] (L1) -- (0);
\draw [mid arrow] (L2) -- (0);
\draw [mid arrow] (0) -- (R1);
\draw [mid arrow] (0) -- (R2);
\foreach \n in {0, L1, L2, R1, R2} 
{
  \draw (\n) --++ (0,0,1);
}
\end{tikzpicture},
\end{equation}
where we use dotted lines to indicate that the equality involves the whole tensor network
and not just the indicated tensors.
We want to understand how the gauge symmetry of the PEPS
is realized at the virtual level of the constituent tensors \(A\) and \(B\),
similar to what has already been established for MPS
\cite{kullClassificationMatrixProduct2017}.
Given the well known fundamental obstructions~\cite{PhysRevLett.125.210504},
we do not expect to be able to solve this problem in full generality,
but some extra assumptions on the tensors are needed to make progress.
For this example we shall impose the following (unnecessarily strong) conditions:
\begin{enumerate}
\item The vertex tensors become injective after blocking rectangular (say \(2\times2\)) regions,
  i.e.\ \(A\) is normal.
\item The edge tensor \(B\) is unital, cf.\ Definition~\ref{def:lri}(iv).
\end{enumerate} 
These conditions imply in particular that blocked tensors of the shape
\begin{equation}\label{eq:10}
  \begin{tikzpicture}[baseline={([yshift=-\the\fontdimen22\textfont2]11.center)}, y={(1em,0.5em)},x={(1em,-0.5em)}, z={(0em,1em)}]
\node [tensor] (00) at (0, 0) {\(\)};
\node [tensor] (01) at (0, 4) {};
\node [tensor] (10) at (4, 0) {};
\node [tensor] (11) at (4, 4) {};
\node [stensor] (a) at (0, 2) {};
\node [stensor] (d) at (2, 0) {};
\node [stensor] (b) at (2, 4) {};
\node [stensor] (c) at (4, 2) {};
\node [stensor] (e) at (4, 6) {};
\draw (00) -- (a) -- (01) -- (b) -- (11) -- (c) -- (10) -- (d) -- (00);
\draw (00) --++ (0, -1, 0) (01) --++ (0, 1, 0);
\draw (10) --++ (0, -1, 0) (11) --++ (0, 1, 0);
\draw (01) --++ (-1, 0, 0) (11) --++ (1, 0, 0);
\draw (00) --++ (-1, 0, 0) (10) --++ (1, 0, 0);
\draw (11) -- (e) --++ (0, 1, 0);
\foreach \n in {00, 01, 10, 11, a, b, c, d, e} 
  \draw (\n) --++ (0,0,1);
\end{tikzpicture}
\end{equation}
are injective.
When we prove the general statement we will see that we can weaken these conditions substantially.
Continuing now with the above conditions on the PEPS tensors
it is straightforward to derive their local transformation properties
under the gauge symmetry,
by focusing first on the behavior of a single edge tensor.
Fix an edge \(e_0\) and consider regions \(\Lambda^{\mathrm{o}}\), \(\Lambda^{\mathrm{t}}\) and \(\Lambda^{\mathrm{j}}\),
as indicated in Figure \ref{fig:square-lattice-combined}~(b), and let
\(\mathbb{T}^{\mathrm{o}}\),
\(\mathbb{T}^{\mathrm{t}}\) and
\(\mathbb{T}^{\mathrm{j}}\),
respectively, denote the tensors one gets after blocking the tensors inside
these regions.
Clearly the PEPS can now be written as an MPS
\begin{align*}
  \label{eq:3}
  \ket{\Psi(A,B)}
  = \begin{tikzpicture}[baseline={([yshift=-\the\fontdimen22\textfont2]e.center)}] 
\begin{scope}[scale=.42]
\clip (1,-1.2) rectangle (15,13.2);
\begin{scope}
  \clip (-1em,-1em) rectangle ($(16,12)+(1em,1em)$);
  \draw [white, fill = green!30!white] (-1,-1) rectangle (17,13);
  \draw [green!80!black, fill=white] plot [smooth cycle, tension=.5] coordinates {(13.3,7.5) (13.4,12) (12.5, 13.5) (8,13.5) (6.6,12.3) (6.5,9.5) (4.5,9.5) (4.5,5.8) (2.7,4.6) (2.6, 0) (4, -1.5) (8.2,-1.5) (9.4,-0.1) (9.5, 2.5) (11.5,2.5) (11.5, 6.2)};
\end{scope}
\begin{scope}
  \clip plot [smooth cycle, tension=.5] coordinates {(13.3,7.5) (13.4,12) (12.5, 13.5) (8,13.5) (6.6,12.3) (6.5,9.5) (4.5,9.5) (4.5,5.8) (2.7,4.6) (2.6, 0) (4, -1.5) (8.2,-1.5) (9.4,-0.1) (9.5, 2.5) (11.5,2.5) (11.5, 6.2)};
  \foreach \x [remember=\x as \lx] in {0,...,4} 
      \foreach \y [remember=\y as \ly] in {0,...,3}
      {
        \node [stensor, fill=white] (\x\y) at (4*\x,4*\y) {};
        \if0\x\else\draw [mid bullet] (\x\y)--(\lx\y);\fi%
        \if0\y\else\draw [mid bullet] (\x\y)--(\x\ly);\fi%
      }
  \foreach \x in {0,...,4} 
    \draw [dash pattern=on .82pt off .90pt, line width = .3pt] (\x0) --++ (0,-1em) (\x3) --++ (0,1em);
  \foreach \y in {0,...,3} 
    \draw [dash pattern=on .82pt off .90pt, line width = .3pt] (0\y) --++ (-1em,0) (4\y) --++ (1em,0);
\end{scope}
\draw [teal, fill=teal!30!white] plot [smooth cycle, tension=0.5] coordinates {(8,7) (12,7) (13,8) (13,12) (12, 13) (8,13) (7,12) (7,9) (5,9) (5,7)};
\begin{scope}[rotate=180, shift={(-16,-12)}]
  \draw [orange, fill=orange!30!white] plot [smooth cycle, tension=0.5] coordinates {(8,7) (12,7) (13,8) (13,12) (12, 13) (8,13) (7,12) (7,9) (5,9) (5,7)};
\end{scope}
\node (e) at ([shift={(7/4,2)}]21) {\scriptsize\(B_{e_0}\)};
\node (o) at (10,10) {\scriptsize\(\color{teal}\mathbb{T}^{\mathrm{t}}\)};
\node (o) at (6,2) {\scriptsize\(\color{orange}\mathbb{T}^{\mathrm{o}}\)};
\node (o) at (13,2) {\scriptsize\(\color{green!80!black}\mathbb{T}^{\mathrm{j}}\)};
\node (o) at (3, 10) {\scriptsize\(\color{green!80!black}\mathbb{T}^{\mathrm{j}}\)};
\end{scope}
\end{tikzpicture}
  = 
  \pgfkeys{
    label left/.store in=\MPSTensorLabelL,
    label left=\(\), 
    label center/.store in=\MPSTensorLabelC,
    label center=\(\), 
    label right/.store in=\MPSTensorLabelR,
    label right=\(\), 
    rep left/.store in=\RepL,
    rep left = 1,
    op left/.store in=\LegOperatorL,
    op left = 1,
    op center/.store in=\LegOperatorC,
    op center = 1,
    op right/.store in=\LegOperatorR,
    op right = 1,
    op bottom/.store in=\LegOperatorB,
    op bottom = 1
  }
  \pgfkeys{label left= \(\mathbb{T}^{\mathrm{t}}\),
            label center=\(B_{e_0}\),
            label right = \(\mathbb{T}^{\mathrm{o}}\),
            op bottom = \(\mathbb{T}^{\mathrm{j}}\)}%
  \begin{tikzpicture}[baseline={([yshift=-\the\fontdimen22\textfont2]A0.center)}] 
\node[tensor, label=below:{\scriptsize\MPSTensorLabelL}] (A0) at (0 cm, 0) {};
\node[tensor, label=below:{\scriptsize\MPSTensorLabelC}] (A1) at (1 cm, 0) {};
\node[tensor, label=below:{\scriptsize\MPSTensorLabelR}] (A2) at (2 cm, 0) {};
\foreach \i in {0, ..., 2}
{
  \node[empty] (H\i) at (\i cm, 1.4) {};
  \draw [mid arrow] (A\i) -- (H\i);
}
\if1\LegOperatorL
\else
  \node[matrix, label={[xshift=2pt]left:{\scriptsize\LegOperatorL}}] (O0) at (0 cm, .8) {};
\fi
\if1\LegOperatorC
\else
  \node[matrix, label={[xshift=2pt]left:{\scriptsize\LegOperatorC}}] (O1) at (1 cm, .8) {};
\fi
\if1\LegOperatorR
\else
  \node[matrix, label={[xshift=2pt]left:{\scriptsize\LegOperatorR}}] (O2) at (2 cm, .8) {};
\fi

\draw [mid arrow] (A2.west) to (A1.east);
\draw [mid arrow] (A1.west) to (A0.east);

\if1\LegOperatorB
  \draw [mid arrow] (A0.west) to[out=180,in=180] (-.3em,-2.2em) to (2cm + .3em, -2.2em) to[out=0,in=0] (A2.east);
\else
  \node[matrix, label=below:{\scriptsize\LegOperatorB}] (OB) at (1 cm, -2.2em) {};
  \draw (A0.west) to[out=180,in=180] (-.3em,-2.2em);
  \draw [mid arrow] (-.3em,-2.2em) to (OB.east);
  \draw [mid arrow] (OB.west) to (2cm + .3em, -2.2em);
  \draw (2cm + .3em, -2.2em) to[out=0,in=0] (A2.east);
\fi
\if1\RepL
\else
  \node[matrix, label=below:{\scriptsize\RepL}] (OB) at (0.5 cm, 0) {};
\fi
\end{tikzpicture}
,
\end{align*}
where we do not draw the (many) physical legs of \(\mathbb{T}^{\mathrm{j}}\).
Notice that the blocked tensors
\(\mathbb{T}^{\mathrm{o}}\) and \(\mathbb{T}^{\mathrm{t}}\)
are of the form \eqref{eq:10}, hence injective,
and that the gauge transformations involving \(e_0\)
act either on \(\mathbb{T}^{\mathrm{o}}\) or \(\mathbb{T}^{\mathrm{t}}\)
but never on both and never on \(\mathbb{T}^{\mathrm{j}}\),
hence, defining
\(\widetilde{U}_{\mathrm{t}}(g) \otimes L(g) := \widehat{U}_{\operatorname{t}(e_0)}(g)\)
and \(R(g) \otimes \widetilde{U}_{\mathrm{o}}(g) := \widehat{U}_{\operatorname{o}(e_0)}(g)\)
allows us to write
\begin{equation}
  \label{eq:4}
  
  \pgfkeys{
    label left/.store in=\MPSTensorLabelL,
    label left=\(\), 
    label center/.store in=\MPSTensorLabelC,
    label center=\(\), 
    label right/.store in=\MPSTensorLabelR,
    label right=\(\), 
    rep left/.store in=\RepL,
    rep left = 1,
    op left/.store in=\LegOperatorL,
    op left = 1,
    op center/.store in=\LegOperatorC,
    op center = 1,
    op right/.store in=\LegOperatorR,
    op right = 1,
    op bottom/.store in=\LegOperatorB,
    op bottom = 1
  }
  \pgfkeys{label left= \(\mathbb{T}^{\mathrm{t}}\),
        label center=\(B\),
        label right = \(\mathbb{T}^{\mathrm{o}}\),
        op bottom = \(\mathbb{T}^{\mathrm{j}}\),
        op left = \(\widetilde{U}_{\mathrm{t}}(g)\),
        op center = \(L(g)\),
        }%
  \begin{tikzpicture}[baseline={([yshift=-\the\fontdimen22\textfont2]A0.center)}] 
\node[tensor, label=below:{\scriptsize\MPSTensorLabelL}] (A0) at (0 cm, 0) {};
\node[tensor, label=below:{\scriptsize\MPSTensorLabelC}] (A1) at (1 cm, 0) {};
\node[tensor, label=below:{\scriptsize\MPSTensorLabelR}] (A2) at (2 cm, 0) {};
\foreach \i in {0, ..., 2}
{
  \node[empty] (H\i) at (\i cm, 1.4) {};
  \draw [mid arrow] (A\i) -- (H\i);
}
\if1\LegOperatorL
\else
  \node[matrix, label={[xshift=2pt]left:{\scriptsize\LegOperatorL}}] (O0) at (0 cm, .8) {};
\fi
\if1\LegOperatorC
\else
  \node[matrix, label={[xshift=2pt]left:{\scriptsize\LegOperatorC}}] (O1) at (1 cm, .8) {};
\fi
\if1\LegOperatorR
\else
  \node[matrix, label={[xshift=2pt]left:{\scriptsize\LegOperatorR}}] (O2) at (2 cm, .8) {};
\fi

\draw [mid arrow] (A2.west) to (A1.east);
\draw [mid arrow] (A1.west) to (A0.east);

\if1\LegOperatorB
  \draw [mid arrow] (A0.west) to[out=180,in=180] (-.3em,-2.2em) to (2cm + .3em, -2.2em) to[out=0,in=0] (A2.east);
\else
  \node[matrix, label=below:{\scriptsize\LegOperatorB}] (OB) at (1 cm, -2.2em) {};
  \draw (A0.west) to[out=180,in=180] (-.3em,-2.2em);
  \draw [mid arrow] (-.3em,-2.2em) to (OB.east);
  \draw [mid arrow] (OB.west) to (2cm + .3em, -2.2em);
  \draw (2cm + .3em, -2.2em) to[out=0,in=0] (A2.east);
\fi
\if1\RepL
\else
  \node[matrix, label=below:{\scriptsize\RepL}] (OB) at (0.5 cm, 0) {};
\fi
\end{tikzpicture}
 =
  
  \pgfkeys{
    label left/.store in=\MPSTensorLabelL,
    label left=\(\), 
    label center/.store in=\MPSTensorLabelC,
    label center=\(\), 
    label right/.store in=\MPSTensorLabelR,
    label right=\(\), 
    rep left/.store in=\RepL,
    rep left = 1,
    op left/.store in=\LegOperatorL,
    op left = 1,
    op center/.store in=\LegOperatorC,
    op center = 1,
    op right/.store in=\LegOperatorR,
    op right = 1,
    op bottom/.store in=\LegOperatorB,
    op bottom = 1
  }
  \pgfkeys{label left= \(\mathbb{T}^{\mathrm{t}}\),
        label center=\(B\),
        label right = \(\mathbb{T}^{\mathrm{o}}\),
        op bottom = \(\mathbb{T}^{\mathrm{j}}\)}%
  \begin{tikzpicture}[baseline={([yshift=-\the\fontdimen22\textfont2]A0.center)}] 
\node[tensor, label=below:{\scriptsize\MPSTensorLabelL}] (A0) at (0 cm, 0) {};
\node[tensor, label=below:{\scriptsize\MPSTensorLabelC}] (A1) at (1 cm, 0) {};
\node[tensor, label=below:{\scriptsize\MPSTensorLabelR}] (A2) at (2 cm, 0) {};
\foreach \i in {0, ..., 2}
{
  \node[empty] (H\i) at (\i cm, 1.4) {};
  \draw [mid arrow] (A\i) -- (H\i);
}
\if1\LegOperatorL
\else
  \node[matrix, label={[xshift=2pt]left:{\scriptsize\LegOperatorL}}] (O0) at (0 cm, .8) {};
\fi
\if1\LegOperatorC
\else
  \node[matrix, label={[xshift=2pt]left:{\scriptsize\LegOperatorC}}] (O1) at (1 cm, .8) {};
\fi
\if1\LegOperatorR
\else
  \node[matrix, label={[xshift=2pt]left:{\scriptsize\LegOperatorR}}] (O2) at (2 cm, .8) {};
\fi

\draw [mid arrow] (A2.west) to (A1.east);
\draw [mid arrow] (A1.west) to (A0.east);

\if1\LegOperatorB
  \draw [mid arrow] (A0.west) to[out=180,in=180] (-.3em,-2.2em) to (2cm + .3em, -2.2em) to[out=0,in=0] (A2.east);
\else
  \node[matrix, label=below:{\scriptsize\LegOperatorB}] (OB) at (1 cm, -2.2em) {};
  \draw (A0.west) to[out=180,in=180] (-.3em,-2.2em);
  \draw [mid arrow] (-.3em,-2.2em) to (OB.east);
  \draw [mid arrow] (OB.west) to (2cm + .3em, -2.2em);
  \draw (2cm + .3em, -2.2em) to[out=0,in=0] (A2.east);
\fi
\if1\RepL
\else
  \node[matrix, label=below:{\scriptsize\RepL}] (OB) at (0.5 cm, 0) {};
\fi
\end{tikzpicture}

\end{equation}
 and
\begin{equation}
  \label{eq:5}
  
  \pgfkeys{
    label left/.store in=\MPSTensorLabelL,
    label left=\(\), 
    label center/.store in=\MPSTensorLabelC,
    label center=\(\), 
    label right/.store in=\MPSTensorLabelR,
    label right=\(\), 
    rep left/.store in=\RepL,
    rep left = 1,
    op left/.store in=\LegOperatorL,
    op left = 1,
    op center/.store in=\LegOperatorC,
    op center = 1,
    op right/.store in=\LegOperatorR,
    op right = 1,
    op bottom/.store in=\LegOperatorB,
    op bottom = 1
  }
  \pgfkeys{label left= \(\mathbb{T}^{\mathrm{t}}\),
        label center=\(B\),
        label right = \(\mathbb{T}^{\mathrm{o}}\),
        op bottom = \(\mathbb{T}^{\mathrm{j}}\),
        op right = \(\widetilde{U}_{\mathrm{o}}(g)\),
        op center = \(R(g)\),
        }%
  \begin{tikzpicture}[baseline={([yshift=-\the\fontdimen22\textfont2]A0.center)}] 
\node[tensor, label=below:{\scriptsize\MPSTensorLabelL}] (A0) at (0 cm, 0) {};
\node[tensor, label=below:{\scriptsize\MPSTensorLabelC}] (A1) at (1 cm, 0) {};
\node[tensor, label=below:{\scriptsize\MPSTensorLabelR}] (A2) at (2 cm, 0) {};
\foreach \i in {0, ..., 2}
{
  \node[empty] (H\i) at (\i cm, 1.4) {};
  \draw [mid arrow] (A\i) -- (H\i);
}
\if1\LegOperatorL
\else
  \node[matrix, label={[xshift=2pt]left:{\scriptsize\LegOperatorL}}] (O0) at (0 cm, .8) {};
\fi
\if1\LegOperatorC
\else
  \node[matrix, label={[xshift=2pt]left:{\scriptsize\LegOperatorC}}] (O1) at (1 cm, .8) {};
\fi
\if1\LegOperatorR
\else
  \node[matrix, label={[xshift=2pt]left:{\scriptsize\LegOperatorR}}] (O2) at (2 cm, .8) {};
\fi

\draw [mid arrow] (A2.west) to (A1.east);
\draw [mid arrow] (A1.west) to (A0.east);

\if1\LegOperatorB
  \draw [mid arrow] (A0.west) to[out=180,in=180] (-.3em,-2.2em) to (2cm + .3em, -2.2em) to[out=0,in=0] (A2.east);
\else
  \node[matrix, label=below:{\scriptsize\LegOperatorB}] (OB) at (1 cm, -2.2em) {};
  \draw (A0.west) to[out=180,in=180] (-.3em,-2.2em);
  \draw [mid arrow] (-.3em,-2.2em) to (OB.east);
  \draw [mid arrow] (OB.west) to (2cm + .3em, -2.2em);
  \draw (2cm + .3em, -2.2em) to[out=0,in=0] (A2.east);
\fi
\if1\RepL
\else
  \node[matrix, label=below:{\scriptsize\RepL}] (OB) at (0.5 cm, 0) {};
\fi
\end{tikzpicture}
 =
  
  \pgfkeys{
    label left/.store in=\MPSTensorLabelL,
    label left=\(\), 
    label center/.store in=\MPSTensorLabelC,
    label center=\(\), 
    label right/.store in=\MPSTensorLabelR,
    label right=\(\), 
    rep left/.store in=\RepL,
    rep left = 1,
    op left/.store in=\LegOperatorL,
    op left = 1,
    op center/.store in=\LegOperatorC,
    op center = 1,
    op right/.store in=\LegOperatorR,
    op right = 1,
    op bottom/.store in=\LegOperatorB,
    op bottom = 1
  }
  \pgfkeys{label left= \(\mathbb{T}^{\mathrm{t}}\),
        label center=\(B\),
        label right = \(\mathbb{T}^{\mathrm{o}}\),
        op bottom = \(\mathbb{T}^{\mathrm{j}}\)}%
  \begin{tikzpicture}[baseline={([yshift=-\the\fontdimen22\textfont2]A0.center)}] 
\node[tensor, label=below:{\scriptsize\MPSTensorLabelL}] (A0) at (0 cm, 0) {};
\node[tensor, label=below:{\scriptsize\MPSTensorLabelC}] (A1) at (1 cm, 0) {};
\node[tensor, label=below:{\scriptsize\MPSTensorLabelR}] (A2) at (2 cm, 0) {};
\foreach \i in {0, ..., 2}
{
  \node[empty] (H\i) at (\i cm, 1.4) {};
  \draw [mid arrow] (A\i) -- (H\i);
}
\if1\LegOperatorL
\else
  \node[matrix, label={[xshift=2pt]left:{\scriptsize\LegOperatorL}}] (O0) at (0 cm, .8) {};
\fi
\if1\LegOperatorC
\else
  \node[matrix, label={[xshift=2pt]left:{\scriptsize\LegOperatorC}}] (O1) at (1 cm, .8) {};
\fi
\if1\LegOperatorR
\else
  \node[matrix, label={[xshift=2pt]left:{\scriptsize\LegOperatorR}}] (O2) at (2 cm, .8) {};
\fi

\draw [mid arrow] (A2.west) to (A1.east);
\draw [mid arrow] (A1.west) to (A0.east);

\if1\LegOperatorB
  \draw [mid arrow] (A0.west) to[out=180,in=180] (-.3em,-2.2em) to (2cm + .3em, -2.2em) to[out=0,in=0] (A2.east);
\else
  \node[matrix, label=below:{\scriptsize\LegOperatorB}] (OB) at (1 cm, -2.2em) {};
  \draw (A0.west) to[out=180,in=180] (-.3em,-2.2em);
  \draw [mid arrow] (-.3em,-2.2em) to (OB.east);
  \draw [mid arrow] (OB.west) to (2cm + .3em, -2.2em);
  \draw (2cm + .3em, -2.2em) to[out=0,in=0] (A2.east);
\fi
\if1\RepL
\else
  \node[matrix, label=below:{\scriptsize\RepL}] (OB) at (0.5 cm, 0) {};
\fi
\end{tikzpicture}
,
\end{equation}
for all \(g \in G\).
This means that the conditions of Lemma \ref{lem:fundamental},
to be stated and proved in the next section, are satisfied and
we conclude the existence of unique, invertible matrices \(V_{\mathrm{L}}(g)\) and \(V_{\mathrm{R}}(g)\)
such that
\begin{align}
  
  \pgfkeys{
    label/.store in=\MPSTensorLabel,
    label=\(\), 
    label inner/.store in=\MPSTensorLabelInner, 
    label inner=\(\),
    rep top/.store in=\TopLegOperator,
    rep top = 1,
    vector top/.store in=\TopLegVector,
    vector top = 1,
    rep left/.store in=\LeftLegOperator,
    rep left = 1,
    rep right/.store in=\RightLegOperator,
    rep right = 1
  }
  \pgfkeys{label = \(B\), rep top = \(L(g)\)}%
  \begin{tikzpicture}[baseline={([yshift=-\the\fontdimen22\textfont2]A.center)}] 
\node[tensor, label=below:{\scriptsize\MPSTensorLabel}] (A) at (0, 0) {\MPSTensorLabelInner};
\if 1\RightLegOperator
  \node[empty] (Hv2) at (1, 0) {\scriptsize\MPSIndexRight};
  \draw [mid arrow] (Hv2) -- (A);
\else
  \node[matrix, label=above:{\scriptsize\RightLegOperator}, right = 1 em of A] (OpR) {};
  \node[empty] (Hv2) at (1.5, 0) {\scriptsize\MPSIndexRight};
  \draw [mid arrow] (OpR) -- (A);
  \draw (Hv2) -- (OpR);
\fi
\if 1\LeftLegOperator
  \node[empty] (Hv1) at (-1, 0) {\scriptsize\MPSIndexLeft};
  \draw [mid arrow] (A) -- (Hv1);
\else
  \node[matrix, label=above:{\scriptsize\LeftLegOperator}, left = 1 em of A] (OpL) {};
  \node[empty] (Hv1) at (-1.5, 0) {\scriptsize\MPSIndexLeft};
  \draw [mid arrow] (A) -- (OpL);
  \draw (OpL) -- (Hv1);
\fi
\if 1\TopLegOperator
  \if 1\TopLegVector
    \node[empty] (Hp) at (0, 1) {\scriptsize\MPSIndexTop};
    \draw [mid arrow] (A) -- (Hp);
  \else
    \node[matrix, label=right:{\scriptsize\TopLegVector}, above = 1 em of A] (VT) {};
    \draw [mid arrow] (A) -- (VT);
  \fi
\else
  \node[matrix, label=right:{\scriptsize\TopLegOperator}, above = 1 em of A] (OpT) {};
  \node[empty] (Hp) at (0, 1.5) {\scriptsize\MPSIndexTop};
  \draw [mid arrow] (A) -- (OpT);
  \draw (OpT) -- (Hp);
\fi
\end{tikzpicture}

  = 
  \pgfkeys{
    label/.store in=\MPSTensorLabel,
    label=\(\), 
    label inner/.store in=\MPSTensorLabelInner, 
    label inner=\(\),
    rep top/.store in=\TopLegOperator,
    rep top = 1,
    vector top/.store in=\TopLegVector,
    vector top = 1,
    rep left/.store in=\LeftLegOperator,
    rep left = 1,
    rep right/.store in=\RightLegOperator,
    rep right = 1
  }
  \pgfkeys{label = \(B\), rep left = \(V_{\mathrm{L}}(g)^{-1}\)}%
  \begin{tikzpicture}[baseline={([yshift=-\the\fontdimen22\textfont2]A.center)}] 
\node[tensor, label=below:{\scriptsize\MPSTensorLabel}] (A) at (0, 0) {\MPSTensorLabelInner};
\if 1\RightLegOperator
  \node[empty] (Hv2) at (1, 0) {\scriptsize\MPSIndexRight};
  \draw [mid arrow] (Hv2) -- (A);
\else
  \node[matrix, label=above:{\scriptsize\RightLegOperator}, right = 1 em of A] (OpR) {};
  \node[empty] (Hv2) at (1.5, 0) {\scriptsize\MPSIndexRight};
  \draw [mid arrow] (OpR) -- (A);
  \draw (Hv2) -- (OpR);
\fi
\if 1\LeftLegOperator
  \node[empty] (Hv1) at (-1, 0) {\scriptsize\MPSIndexLeft};
  \draw [mid arrow] (A) -- (Hv1);
\else
  \node[matrix, label=above:{\scriptsize\LeftLegOperator}, left = 1 em of A] (OpL) {};
  \node[empty] (Hv1) at (-1.5, 0) {\scriptsize\MPSIndexLeft};
  \draw [mid arrow] (A) -- (OpL);
  \draw (OpL) -- (Hv1);
\fi
\if 1\TopLegOperator
  \if 1\TopLegVector
    \node[empty] (Hp) at (0, 1) {\scriptsize\MPSIndexTop};
    \draw [mid arrow] (A) -- (Hp);
  \else
    \node[matrix, label=right:{\scriptsize\TopLegVector}, above = 1 em of A] (VT) {};
    \draw [mid arrow] (A) -- (VT);
  \fi
\else
  \node[matrix, label=right:{\scriptsize\TopLegOperator}, above = 1 em of A] (OpT) {};
  \node[empty] (Hp) at (0, 1.5) {\scriptsize\MPSIndexTop};
  \draw [mid arrow] (A) -- (OpT);
  \draw (OpT) -- (Hp);
\fi
\end{tikzpicture}
,&&
  
  \pgfkeys{
    label/.store in=\MPSTensorLabel,
    label=\(\), 
    label inner/.store in=\MPSTensorLabelInner, 
    label inner=\(\),
    rep top/.store in=\TopLegOperator,
    rep top = 1,
    vector top/.store in=\TopLegVector,
    vector top = 1,
    rep left/.store in=\LeftLegOperator,
    rep left = 1,
    rep right/.store in=\RightLegOperator,
    rep right = 1
  }
  \pgfkeys{label = \(B\), rep top = \(R(g)\)}%
  \begin{tikzpicture}[baseline={([yshift=-\the\fontdimen22\textfont2]A.center)}] 
\node[tensor, label=below:{\scriptsize\MPSTensorLabel}] (A) at (0, 0) {\MPSTensorLabelInner};
\if 1\RightLegOperator
  \node[empty] (Hv2) at (1, 0) {\scriptsize\MPSIndexRight};
  \draw [mid arrow] (Hv2) -- (A);
\else
  \node[matrix, label=above:{\scriptsize\RightLegOperator}, right = 1 em of A] (OpR) {};
  \node[empty] (Hv2) at (1.5, 0) {\scriptsize\MPSIndexRight};
  \draw [mid arrow] (OpR) -- (A);
  \draw (Hv2) -- (OpR);
\fi
\if 1\LeftLegOperator
  \node[empty] (Hv1) at (-1, 0) {\scriptsize\MPSIndexLeft};
  \draw [mid arrow] (A) -- (Hv1);
\else
  \node[matrix, label=above:{\scriptsize\LeftLegOperator}, left = 1 em of A] (OpL) {};
  \node[empty] (Hv1) at (-1.5, 0) {\scriptsize\MPSIndexLeft};
  \draw [mid arrow] (A) -- (OpL);
  \draw (OpL) -- (Hv1);
\fi
\if 1\TopLegOperator
  \if 1\TopLegVector
    \node[empty] (Hp) at (0, 1) {\scriptsize\MPSIndexTop};
    \draw [mid arrow] (A) -- (Hp);
  \else
    \node[matrix, label=right:{\scriptsize\TopLegVector}, above = 1 em of A] (VT) {};
    \draw [mid arrow] (A) -- (VT);
  \fi
\else
  \node[matrix, label=right:{\scriptsize\TopLegOperator}, above = 1 em of A] (OpT) {};
  \node[empty] (Hp) at (0, 1.5) {\scriptsize\MPSIndexTop};
  \draw [mid arrow] (A) -- (OpT);
  \draw (OpT) -- (Hp);
\fi
\end{tikzpicture}

  = 
  \pgfkeys{
    label/.store in=\MPSTensorLabel,
    label=\(\), 
    label inner/.store in=\MPSTensorLabelInner, 
    label inner=\(\),
    rep top/.store in=\TopLegOperator,
    rep top = 1,
    vector top/.store in=\TopLegVector,
    vector top = 1,
    rep left/.store in=\LeftLegOperator,
    rep left = 1,
    rep right/.store in=\RightLegOperator,
    rep right = 1
  }
  \pgfkeys{label = \(B\), rep right = \(V_{\mathrm{R}}(g)\)}%
  \begin{tikzpicture}[baseline={([yshift=-\the\fontdimen22\textfont2]A.center)}] 
\node[tensor, label=below:{\scriptsize\MPSTensorLabel}] (A) at (0, 0) {\MPSTensorLabelInner};
\if 1\RightLegOperator
  \node[empty] (Hv2) at (1, 0) {\scriptsize\MPSIndexRight};
  \draw [mid arrow] (Hv2) -- (A);
\else
  \node[matrix, label=above:{\scriptsize\RightLegOperator}, right = 1 em of A] (OpR) {};
  \node[empty] (Hv2) at (1.5, 0) {\scriptsize\MPSIndexRight};
  \draw [mid arrow] (OpR) -- (A);
  \draw (Hv2) -- (OpR);
\fi
\if 1\LeftLegOperator
  \node[empty] (Hv1) at (-1, 0) {\scriptsize\MPSIndexLeft};
  \draw [mid arrow] (A) -- (Hv1);
\else
  \node[matrix, label=above:{\scriptsize\LeftLegOperator}, left = 1 em of A] (OpL) {};
  \node[empty] (Hv1) at (-1.5, 0) {\scriptsize\MPSIndexLeft};
  \draw [mid arrow] (A) -- (OpL);
  \draw (OpL) -- (Hv1);
\fi
\if 1\TopLegOperator
  \if 1\TopLegVector
    \node[empty] (Hp) at (0, 1) {\scriptsize\MPSIndexTop};
    \draw [mid arrow] (A) -- (Hp);
  \else
    \node[matrix, label=right:{\scriptsize\TopLegVector}, above = 1 em of A] (VT) {};
    \draw [mid arrow] (A) -- (VT);
  \fi
\else
  \node[matrix, label=right:{\scriptsize\TopLegOperator}, above = 1 em of A] (OpT) {};
  \node[empty] (Hp) at (0, 1.5) {\scriptsize\MPSIndexTop};
  \draw [mid arrow] (A) -- (OpT);
  \draw (OpT) -- (Hp);
\fi
\end{tikzpicture}
,
  \label{eqB}
\end{align}
for all \(g \in G\).
Furthermore,
from the uniqueness of \(V_{\mathrm{L}}(g)\) and \(V_{\mathrm{R}}(g)\)
we conclude that \(V_{\mathrm{L}}\) and \(V_{\mathrm{R}}\)
constitute representations of \(G\).
Having established how the tensor \(B\) transforms under \(L\) and \(R\),
only leaves the transformation property of the tensor \(A\),
which is remarkably simple to derive now.

We shall again focus on a fixed vertex \(v_0\).
Notice that blocking the tensors contained in the regions
\(\Lambda^{\mathrm{o}}_1\),
\(\Lambda^{\mathrm{o}}_2\),
\(\Lambda^{\mathrm{t}}_1\) and
\(\Lambda^{\mathrm{t}}_2\),
as indicated in Figure \ref{fig:square-lattice-combined}~(c),
yields four injective tensors, which we can invert to 
decouple the vertex tensor \(A\) at \(v_0\) and its surrounding \(B\) tensors
from the rest of the lattice.
Knowing how the tensors \(B\) transform under the symmetry operations we can write Equation~\eqref{eq:11} as
\begin{equation}\label{eq:12}
  \begin{tikzpicture}[baseline={([yshift=-\the\fontdimen22\textfont2]0.center)}, y={(1.0em,0.5em)},x={(1.0em,-0.5em)}, z={(0em,1em)}]
\begin{scope}[scale=1.4]
  \draw [white, shading=rectangle, left color=teal!50!white, right color=orange!50!white] plot [smooth cycle, tension=1.2] coordinates {(-3,0) (-1.4,1.4) (0,3) (1.4,1.4) (3,0) (1.4,-1.4) (0,-3) (-1.4,-1.4)};
\end{scope}
\draw [white, fill=white] plot [smooth cycle, tension=.8] coordinates {(-3,0) (-1.4,1.4) (0,3) (1.4,1.4) (3,0) (1.4,-1.4) (0,-3) (-1.4,-1.4)};
\node [tensor] (0) at (0, 0) {\(\)};
\node [stensor] (L1) at (2, 0) {};
\node [stensor] (L2) at (0, 2) {};
\node [stensor] (R1) at (0, -2) {};
\node [stensor] (R2) at (-2, 0) {};
\draw [dotted] (L1) --++ (1.4,0,0);
\draw (L1) --++ (.7,0,0);
\draw [dotted] (L2) --++ (0,1.4,0);
\draw (L2) --++ (0,.7,0);
\draw [dotted] (R1) --++ (0,-1.4,0);
\draw (R1) --++ (0,-.7,0);
\draw [dotted] (R2) --++ (-1.4,0,0);
\draw (R2) --++ (-.7,0,0);
\draw [mid arrow] (L1) -- (0);
\draw [mid arrow] (L2) -- (0);
\draw [mid arrow] (0) -- (R1);
\draw [mid arrow] (0) -- (R2);
\foreach \x/\n in {R/1, R/2, L/1, L/2} 
{
  \draw (\x\n) --++ (0,0,.7);
}
\node [matrix, shift={(0,0,1)}, label = {[label distance = -3pt]left:{\scriptsize\(U\)}}] (u0) at (0) {};
\draw (0) -- (u0) --++ (0,0,.7);
\end{tikzpicture} \hspace{-1em}=\hspace{-1em}
  \begin{tikzpicture}[baseline={([yshift=-\the\fontdimen22\textfont2]0.center)}, y={(1.0em,0.5em)},x={(1.0em,-0.5em)}, z={(0em,1em)}]
\begin{scope}[scale=1.4]
  \draw [white, shading=rectangle, left color=teal!50!white, right color=orange!50!white] plot [smooth cycle, tension=1.2] coordinates {(-3,0) (-1.4,1.4) (0,3) (1.4,1.4) (3,0) (1.4,-1.4) (0,-3) (-1.4,-1.4)};
\end{scope}
\draw [white, fill=white] plot [smooth cycle, tension=.8] coordinates {(-3,0) (-1.4,1.4) (0,3) (1.4,1.4) (3,0) (1.4,-1.4) (0,-3) (-1.4,-1.4)};
\node [tensor] (0) at (0, 0) {\(\)};
\node [stensor] (L1) at (2, 0) {};
\node [stensor] (L2) at (0, 2) {};
\node [stensor] (R1) at (0, -2) {};
\node [stensor] (R2) at (-2, 0) {};
\draw [dotted] (L1) --++ (1.4,0,0);
\draw (L1) --++ (.7,0,0);
\draw [dotted] (L2) --++ (0,1.4,0);
\draw (L2) --++ (0,.7,0);
\draw [dotted] (R1) --++ (0,-1.4,0);
\draw (R1) --++ (0,-.7,0);
\draw [dotted] (R2) --++ (-1.4,0,0);
\draw (R2) --++ (-.7,0,0);
\foreach \x/\n in {R/1, R/2, L/1, L/2} 
{
  \draw (\x\n) --++ (0,0,.7);
  \node [matrix, label = {[label distance = -2pt]\if1\n below\else above\fi:{\contour{white}{\scriptsize\(V_{\mathrm{\x}}^{\if R\x -1\hspace{-5pt}\else\vphantom{1}\fi}\)}}}] (V\x\n) at ($(0)!0.5!(\x\n)$) {};
  \draw (0) -- (\x\n);
}
\draw (0) --++ (0,0,0.9);
\end{tikzpicture}\hspace{-1em},
\end{equation}
where we have suppressed the explicit \(g\) dependence.
Now we invert the surrounding injective region as well as the surrounding edge tensors and,
after discarding the rest of the lattice, are left with
\begin{equation}\label{eq:13}
  \begin{tikzpicture}[baseline={([yshift=-\the\fontdimen22\textfont2]0.center)}, y={(1.0em,0.5em)},x={(1.0em,-0.5em)}, z={(0em,1em)}]
\node [tensor] (0) at (0, 0) {\(\)};
\draw [mid arrow] (2, 0) -- (0);
\draw [mid arrow] (0, 2) -- (0);
\draw [mid arrow] (0) -- (0, -2);
\draw [mid arrow] (0) -- (-2, 0);
\node [matrix, shift={(0,0,1.4)}, label = {[label distance = -3pt]left:{\scriptsize\(U(g)\)}}] (u0) at (0) {};
\draw (0) -- (u0) --++ (0,0,.9);
\end{tikzpicture}= \begin{tikzpicture}[baseline={([yshift=-\the\fontdimen22\textfont2]0.center)}, y={(1.0em,0.5em)},x={(1.0em,-0.5em)}, z={(0em,1em)}]
\node [tensor] (0) at (0, 0) {\(\)};
\node [matrix, label=below:{\scriptsize\(V_{\mathrm{L}}(g)^{\vphantom{1}}\)}] (L1) at (1.4, 0) {};
\node [matrix, label=above:{\scriptsize\(V_{\mathrm{L}}(g)^{\vphantom{1}}\)}] (L2) at (0, 1.4) {};
\node [matrix, label=below:{\scriptsize\(V_{\mathrm{R}}(g)^{-1}\)}] (R1) at (0, -1.4) {};
\node [matrix, label=above:{\scriptsize\(V_{\mathrm{R}}(g)^{-1}\)}] (R2) at (-1.4, 0) {};
\draw [mid arrow] (0) --++ (0,0,1.4);
\draw  (2, 0)  -- (L1) -- (0);
\draw  (0, 2)  -- (L2) -- (0);
\draw  (0, -2) -- (R1) -- (0);
\draw  (-2, 0) -- (R2) -- (0);
\end{tikzpicture},
\end{equation}
for all \(g \in G\).
Equations \eqref{eqB} and \eqref{eq:13} can be recognised as the symmetry conditions of gauge-invariant PEPS, cf.~\cite{zohar_building_2016}. There, it was shown that these conditions guarantee a construction of a gauge-invariant PEPS; here we prove the opposite direction.

\section{Demonstration}
\label{sec:demo}

Let us first recall the original construction of injective PEPS in the square lattice. Every edge is associated with a maximally entangled pair state between two qudits, $\sum_{i=0}^{d-1} \ket{ii}$, such that each vertex is associated with four qudits. A generic injective PEPS is then constructed by applying an invertible operator $Q$, that acts on $(\mathbb{C}^{d})^{\otimes 4 }$ on each vertex:
$$ \ket{\Psi_{Q}}= 
    \begin{tikzpicture}[baseline=6pt]
    \foreach \x in {0,2}{ 
    \foreach \y in {0,2}{
    \filldraw[thick] (\x-0.35,\y) circle (2pt)--++ (-0.65,0);
    \filldraw[thick] (\x+0.35,\y) circle (2pt)--++ (0.65,0);
    \filldraw[thick] (\x,\y+0.35) circle (2pt)--++ (0,0.65);
    \filldraw[thick] (\x,\y-0.35) circle (2pt)--++ (0,-0.65);
    \draw[thick, red] (\x,\y) circle (0.6);
    }}
    \end{tikzpicture} \ , \quad
        \begin{tikzpicture}
    \filldraw[thick] (0,0) circle (2pt)--++ (1,0) circle (2pt);
    \end{tikzpicture}
    = \sum_{i=0}^{d-1} \ket{ii} \ ,\
    \begin{tikzpicture}[baseline=-3pt] \draw[thick, red] (0,0) circle (0.6);
    \end{tikzpicture} 
    = Q
    $$

As such, the PEPS tensor is $Q$ where the physical Hilbert space is  $(\mathbb{C}^{d})^{\otimes 4 }$ and each virtual leg is $\mathbb{C}^{d}$.

The following family of states corresponds to injective PEPS with a global symmetry of the finite group $G$ (not necessarily abelian). Let as first associate $\mathbb{C}^{d}$ with the group algebra $\mathbb{C}[G]$ spanned by $\{ \ket{g}, g\in G\}$ so that the maximally entangled pair reads $\sum_g \ket{gg}$. This space is endowed with a left and right regular representation given by $L_g\ket{h}=\ket{gh}$ and $R_g\ket{h}=\ket{hg^{-1}}$. Let's us further assume that $Q$ is a unitary operator so that we can define the following unitary representation of $G$ acting on $(\mathbb{C}^{d})^{\otimes 4 }$:

$$U(g) = Q(L_g\otimes L_g\otimes R^\dagger_g \otimes R^\dagger_g ) Q^\dagger$$

Then, Eq.~\eqref{eq:13} is satisfied with $V_R(g) = R_g$ and $V_L(g) = L_g$. A PEPS with a local symmetry can be constructed by inserting appropiate gauge field tensors. With the following choice:
\begin{equation}
        
  \pgfkeys{
    label/.store in=\MPSTensorLabel,
    label=\(\), 
    label inner/.store in=\MPSTensorLabelInner, 
    label inner=\(\),
    rep top/.store in=\TopLegOperator,
    rep top = 1,
    vector top/.store in=\TopLegVector,
    vector top = 1,
    rep left/.store in=\LeftLegOperator,
    rep left = 1,
    rep right/.store in=\RightLegOperator,
    rep right = 1
  }
  \pgfkeys{label = \(B\)}%
  \begin{tikzpicture}[baseline={([yshift=-\the\fontdimen22\textfont2]A.center)}] 
\node[tensor, label=below:{\scriptsize\MPSTensorLabel}] (A) at (0, 0) {\MPSTensorLabelInner};
\if 1\RightLegOperator
  \node[empty] (Hv2) at (1, 0) {\scriptsize\MPSIndexRight};
  \draw [mid arrow] (Hv2) -- (A);
\else
  \node[matrix, label=above:{\scriptsize\RightLegOperator}, right = 1 em of A] (OpR) {};
  \node[empty] (Hv2) at (1.5, 0) {\scriptsize\MPSIndexRight};
  \draw [mid arrow] (OpR) -- (A);
  \draw (Hv2) -- (OpR);
\fi
\if 1\LeftLegOperator
  \node[empty] (Hv1) at (-1, 0) {\scriptsize\MPSIndexLeft};
  \draw [mid arrow] (A) -- (Hv1);
\else
  \node[matrix, label=above:{\scriptsize\LeftLegOperator}, left = 1 em of A] (OpL) {};
  \node[empty] (Hv1) at (-1.5, 0) {\scriptsize\MPSIndexLeft};
  \draw [mid arrow] (A) -- (OpL);
  \draw (OpL) -- (Hv1);
\fi
\if 1\TopLegOperator
  \if 1\TopLegVector
    \node[empty] (Hp) at (0, 1) {\scriptsize\MPSIndexTop};
    \draw [mid arrow] (A) -- (Hp);
  \else
    \node[matrix, label=right:{\scriptsize\TopLegVector}, above = 1 em of A] (VT) {};
    \draw [mid arrow] (A) -- (VT);
  \fi
\else
  \node[matrix, label=right:{\scriptsize\TopLegOperator}, above = 1 em of A] (OpT) {};
  \node[empty] (Hp) at (0, 1.5) {\scriptsize\MPSIndexTop};
  \draw [mid arrow] (A) -- (OpT);
  \draw (OpT) -- (Hp);
\fi
\end{tikzpicture}
 = \sum_{g,h}\ket{gh}\otimes |g)(h| \ 
\end{equation}
(where the regular ket is physical and the circular ket and bra - virtual), Eq.~\eqref{eqB} is satisfied and then Eq.~\eqref{eq:11} is too.

This agrees with the construction of \cite{zohar_building_2016}, formulated in a rather more physical way, applied for actual LGT calculations, both analytically and numerically, in \cite{zohar_fermionic_2015,zohar_projected_2016,zohar_combining_2018,emonts_variational_2020,emonts_finding_23,kelman2024gauged,kelman2024projected}.

\section{Main Technical Lemma}
\label{sec:main-technical-lemma}
In this section we prove a slightly stronger variation of the Fundamental Theorem of injective MPS~\cite{molnarNormalProjectedEntangled2018}.
This lemma will be used to prove our general statement in Section \ref{sec:general-statement}.
\begin{lemma}[Fundamental Lemma]\label{lem:fundamental}
  Let \(A_{\mathrm{L}}\) and \(A_{\mathrm{R}}\) be injective MPS tensors
  and \(X\) an arbitrary non-zero tensor,
  which may include additional non-contracted (usually physical) legs.
  If \(B\) is a right-injective MPS tensor and the symmetry property
  \begin{equation}\label{eq:19}
    
  \pgfkeys{
    label left/.store in=\MPSTensorLabelL,
    label left=\(\), 
    label center/.store in=\MPSTensorLabelC,
    label center=\(\), 
    label right/.store in=\MPSTensorLabelR,
    label right=\(\), 
    rep left/.store in=\RepL,
    rep left = 1,
    op left/.store in=\LegOperatorL,
    op left = 1,
    op center/.store in=\LegOperatorC,
    op center = 1,
    op right/.store in=\LegOperatorR,
    op right = 1,
    op bottom/.store in=\LegOperatorB,
    op bottom = 1
  }
  \pgfkeys{label left= \(A_{\mathrm{L}}\),
          label center=\(B\),
          label right = \(A_{\mathrm{R}}\),
          op left = \(U_{\mathrm{L}}\),
          op center = \(L\),
          op bottom = \(X\)}%
  \begin{tikzpicture}[baseline={([yshift=-\the\fontdimen22\textfont2]A0.center)}] 
\node[tensor, label=below:{\scriptsize\MPSTensorLabelL}] (A0) at (0 cm, 0) {};
\node[tensor, label=below:{\scriptsize\MPSTensorLabelC}] (A1) at (1 cm, 0) {};
\node[tensor, label=below:{\scriptsize\MPSTensorLabelR}] (A2) at (2 cm, 0) {};
\foreach \i in {0, ..., 2}
{
  \node[empty] (H\i) at (\i cm, 1.4) {};
  \draw [mid arrow] (A\i) -- (H\i);
}
\if1\LegOperatorL
\else
  \node[matrix, label={[xshift=2pt]left:{\scriptsize\LegOperatorL}}] (O0) at (0 cm, .8) {};
\fi
\if1\LegOperatorC
\else
  \node[matrix, label={[xshift=2pt]left:{\scriptsize\LegOperatorC}}] (O1) at (1 cm, .8) {};
\fi
\if1\LegOperatorR
\else
  \node[matrix, label={[xshift=2pt]left:{\scriptsize\LegOperatorR}}] (O2) at (2 cm, .8) {};
\fi

\draw [mid arrow] (A2.west) to (A1.east);
\draw [mid arrow] (A1.west) to (A0.east);

\if1\LegOperatorB
  \draw [mid arrow] (A0.west) to[out=180,in=180] (-.3em,-2.2em) to (2cm + .3em, -2.2em) to[out=0,in=0] (A2.east);
\else
  \node[matrix, label=below:{\scriptsize\LegOperatorB}] (OB) at (1 cm, -2.2em) {};
  \draw (A0.west) to[out=180,in=180] (-.3em,-2.2em);
  \draw [mid arrow] (-.3em,-2.2em) to (OB.east);
  \draw [mid arrow] (OB.west) to (2cm + .3em, -2.2em);
  \draw (2cm + .3em, -2.2em) to[out=0,in=0] (A2.east);
\fi
\if1\RepL
\else
  \node[matrix, label=below:{\scriptsize\RepL}] (OB) at (0.5 cm, 0) {};
\fi
\end{tikzpicture}
 =
    
  \pgfkeys{
    label left/.store in=\MPSTensorLabelL,
    label left=\(\), 
    label center/.store in=\MPSTensorLabelC,
    label center=\(\), 
    label right/.store in=\MPSTensorLabelR,
    label right=\(\), 
    rep left/.store in=\RepL,
    rep left = 1,
    op left/.store in=\LegOperatorL,
    op left = 1,
    op center/.store in=\LegOperatorC,
    op center = 1,
    op right/.store in=\LegOperatorR,
    op right = 1,
    op bottom/.store in=\LegOperatorB,
    op bottom = 1
  }
  \pgfkeys{label left= \(A_{\mathrm{L}}\),
          label center=\(B\),
          label right = \(A_{\mathrm{R}}\),
          op bottom = \(X\)}%
  \begin{tikzpicture}[baseline={([yshift=-\the\fontdimen22\textfont2]A0.center)}] 
\node[tensor, label=below:{\scriptsize\MPSTensorLabelL}] (A0) at (0 cm, 0) {};
\node[tensor, label=below:{\scriptsize\MPSTensorLabelC}] (A1) at (1 cm, 0) {};
\node[tensor, label=below:{\scriptsize\MPSTensorLabelR}] (A2) at (2 cm, 0) {};
\foreach \i in {0, ..., 2}
{
  \node[empty] (H\i) at (\i cm, 1.4) {};
  \draw [mid arrow] (A\i) -- (H\i);
}
\if1\LegOperatorL
\else
  \node[matrix, label={[xshift=2pt]left:{\scriptsize\LegOperatorL}}] (O0) at (0 cm, .8) {};
\fi
\if1\LegOperatorC
\else
  \node[matrix, label={[xshift=2pt]left:{\scriptsize\LegOperatorC}}] (O1) at (1 cm, .8) {};
\fi
\if1\LegOperatorR
\else
  \node[matrix, label={[xshift=2pt]left:{\scriptsize\LegOperatorR}}] (O2) at (2 cm, .8) {};
\fi

\draw [mid arrow] (A2.west) to (A1.east);
\draw [mid arrow] (A1.west) to (A0.east);

\if1\LegOperatorB
  \draw [mid arrow] (A0.west) to[out=180,in=180] (-.3em,-2.2em) to (2cm + .3em, -2.2em) to[out=0,in=0] (A2.east);
\else
  \node[matrix, label=below:{\scriptsize\LegOperatorB}] (OB) at (1 cm, -2.2em) {};
  \draw (A0.west) to[out=180,in=180] (-.3em,-2.2em);
  \draw [mid arrow] (-.3em,-2.2em) to (OB.east);
  \draw [mid arrow] (OB.west) to (2cm + .3em, -2.2em);
  \draw (2cm + .3em, -2.2em) to[out=0,in=0] (A2.east);
\fi
\if1\RepL
\else
  \node[matrix, label=below:{\scriptsize\RepL}] (OB) at (0.5 cm, 0) {};
\fi
\end{tikzpicture}

  \end{equation}
  holds, where \(L\) and \(U_{\mathrm{L}}\) are invertible operators on the physical Hilbert spaces,
  then there exists a unique invertible matrix \(V_{\mathrm{L}}\) such that
  \begin{equation}
    
  \pgfkeys{
    label/.store in=\MPSTensorLabel,
    label=\(\), 
    label inner/.store in=\MPSTensorLabelInner, 
    label inner=\(\),
    rep top/.store in=\TopLegOperator,
    rep top = 1,
    vector top/.store in=\TopLegVector,
    vector top = 1,
    rep left/.store in=\LeftLegOperator,
    rep left = 1,
    rep right/.store in=\RightLegOperator,
    rep right = 1
  }
  \pgfkeys{label = \(B\), rep top = \(L\)}%
  \begin{tikzpicture}[baseline={([yshift=-\the\fontdimen22\textfont2]A.center)}] 
\node[tensor, label=below:{\scriptsize\MPSTensorLabel}] (A) at (0, 0) {\MPSTensorLabelInner};
\if 1\RightLegOperator
  \node[empty] (Hv2) at (1, 0) {\scriptsize\MPSIndexRight};
  \draw [mid arrow] (Hv2) -- (A);
\else
  \node[matrix, label=above:{\scriptsize\RightLegOperator}, right = 1 em of A] (OpR) {};
  \node[empty] (Hv2) at (1.5, 0) {\scriptsize\MPSIndexRight};
  \draw [mid arrow] (OpR) -- (A);
  \draw (Hv2) -- (OpR);
\fi
\if 1\LeftLegOperator
  \node[empty] (Hv1) at (-1, 0) {\scriptsize\MPSIndexLeft};
  \draw [mid arrow] (A) -- (Hv1);
\else
  \node[matrix, label=above:{\scriptsize\LeftLegOperator}, left = 1 em of A] (OpL) {};
  \node[empty] (Hv1) at (-1.5, 0) {\scriptsize\MPSIndexLeft};
  \draw [mid arrow] (A) -- (OpL);
  \draw (OpL) -- (Hv1);
\fi
\if 1\TopLegOperator
  \if 1\TopLegVector
    \node[empty] (Hp) at (0, 1) {\scriptsize\MPSIndexTop};
    \draw [mid arrow] (A) -- (Hp);
  \else
    \node[matrix, label=right:{\scriptsize\TopLegVector}, above = 1 em of A] (VT) {};
    \draw [mid arrow] (A) -- (VT);
  \fi
\else
  \node[matrix, label=right:{\scriptsize\TopLegOperator}, above = 1 em of A] (OpT) {};
  \node[empty] (Hp) at (0, 1.5) {\scriptsize\MPSIndexTop};
  \draw [mid arrow] (A) -- (OpT);
  \draw (OpT) -- (Hp);
\fi
\end{tikzpicture}
 =
    
  \pgfkeys{
    label/.store in=\MPSTensorLabel,
    label=\(\), 
    label inner/.store in=\MPSTensorLabelInner, 
    label inner=\(\),
    rep top/.store in=\TopLegOperator,
    rep top = 1,
    vector top/.store in=\TopLegVector,
    vector top = 1,
    rep left/.store in=\LeftLegOperator,
    rep left = 1,
    rep right/.store in=\RightLegOperator,
    rep right = 1
  }
  \pgfkeys{label = \(B\), rep left = \(V_{\mathrm{L}}^{-1}\)}%
  \begin{tikzpicture}[baseline={([yshift=-\the\fontdimen22\textfont2]A.center)}] 
\node[tensor, label=below:{\scriptsize\MPSTensorLabel}] (A) at (0, 0) {\MPSTensorLabelInner};
\if 1\RightLegOperator
  \node[empty] (Hv2) at (1, 0) {\scriptsize\MPSIndexRight};
  \draw [mid arrow] (Hv2) -- (A);
\else
  \node[matrix, label=above:{\scriptsize\RightLegOperator}, right = 1 em of A] (OpR) {};
  \node[empty] (Hv2) at (1.5, 0) {\scriptsize\MPSIndexRight};
  \draw [mid arrow] (OpR) -- (A);
  \draw (Hv2) -- (OpR);
\fi
\if 1\LeftLegOperator
  \node[empty] (Hv1) at (-1, 0) {\scriptsize\MPSIndexLeft};
  \draw [mid arrow] (A) -- (Hv1);
\else
  \node[matrix, label=above:{\scriptsize\LeftLegOperator}, left = 1 em of A] (OpL) {};
  \node[empty] (Hv1) at (-1.5, 0) {\scriptsize\MPSIndexLeft};
  \draw [mid arrow] (A) -- (OpL);
  \draw (OpL) -- (Hv1);
\fi
\if 1\TopLegOperator
  \if 1\TopLegVector
    \node[empty] (Hp) at (0, 1) {\scriptsize\MPSIndexTop};
    \draw [mid arrow] (A) -- (Hp);
  \else
    \node[matrix, label=right:{\scriptsize\TopLegVector}, above = 1 em of A] (VT) {};
    \draw [mid arrow] (A) -- (VT);
  \fi
\else
  \node[matrix, label=right:{\scriptsize\TopLegOperator}, above = 1 em of A] (OpT) {};
  \node[empty] (Hp) at (0, 1.5) {\scriptsize\MPSIndexTop};
  \draw [mid arrow] (A) -- (OpT);
  \draw (OpT) -- (Hp);
\fi
\end{tikzpicture}
.
  \end{equation}
  Analogously, if \(B\) is a left-injective MPS tensor and the symmetry property
  \begin{equation}\label{eq:20}
    
  \pgfkeys{
    label left/.store in=\MPSTensorLabelL,
    label left=\(\), 
    label center/.store in=\MPSTensorLabelC,
    label center=\(\), 
    label right/.store in=\MPSTensorLabelR,
    label right=\(\), 
    rep left/.store in=\RepL,
    rep left = 1,
    op left/.store in=\LegOperatorL,
    op left = 1,
    op center/.store in=\LegOperatorC,
    op center = 1,
    op right/.store in=\LegOperatorR,
    op right = 1,
    op bottom/.store in=\LegOperatorB,
    op bottom = 1
  }
  \pgfkeys{label left= \(A_{\mathrm{L}}\),
          label center=\(B\),
          label right = \(A_{\mathrm{R}}\),
          op center = \(R\),
          op right = \(U_{\mathrm{R}}\),
          op bottom = \(X\)}%
  \begin{tikzpicture}[baseline={([yshift=-\the\fontdimen22\textfont2]A0.center)}] 
\node[tensor, label=below:{\scriptsize\MPSTensorLabelL}] (A0) at (0 cm, 0) {};
\node[tensor, label=below:{\scriptsize\MPSTensorLabelC}] (A1) at (1 cm, 0) {};
\node[tensor, label=below:{\scriptsize\MPSTensorLabelR}] (A2) at (2 cm, 0) {};
\foreach \i in {0, ..., 2}
{
  \node[empty] (H\i) at (\i cm, 1.4) {};
  \draw [mid arrow] (A\i) -- (H\i);
}
\if1\LegOperatorL
\else
  \node[matrix, label={[xshift=2pt]left:{\scriptsize\LegOperatorL}}] (O0) at (0 cm, .8) {};
\fi
\if1\LegOperatorC
\else
  \node[matrix, label={[xshift=2pt]left:{\scriptsize\LegOperatorC}}] (O1) at (1 cm, .8) {};
\fi
\if1\LegOperatorR
\else
  \node[matrix, label={[xshift=2pt]left:{\scriptsize\LegOperatorR}}] (O2) at (2 cm, .8) {};
\fi

\draw [mid arrow] (A2.west) to (A1.east);
\draw [mid arrow] (A1.west) to (A0.east);

\if1\LegOperatorB
  \draw [mid arrow] (A0.west) to[out=180,in=180] (-.3em,-2.2em) to (2cm + .3em, -2.2em) to[out=0,in=0] (A2.east);
\else
  \node[matrix, label=below:{\scriptsize\LegOperatorB}] (OB) at (1 cm, -2.2em) {};
  \draw (A0.west) to[out=180,in=180] (-.3em,-2.2em);
  \draw [mid arrow] (-.3em,-2.2em) to (OB.east);
  \draw [mid arrow] (OB.west) to (2cm + .3em, -2.2em);
  \draw (2cm + .3em, -2.2em) to[out=0,in=0] (A2.east);
\fi
\if1\RepL
\else
  \node[matrix, label=below:{\scriptsize\RepL}] (OB) at (0.5 cm, 0) {};
\fi
\end{tikzpicture}
 = 
    
  \pgfkeys{
    label left/.store in=\MPSTensorLabelL,
    label left=\(\), 
    label center/.store in=\MPSTensorLabelC,
    label center=\(\), 
    label right/.store in=\MPSTensorLabelR,
    label right=\(\), 
    rep left/.store in=\RepL,
    rep left = 1,
    op left/.store in=\LegOperatorL,
    op left = 1,
    op center/.store in=\LegOperatorC,
    op center = 1,
    op right/.store in=\LegOperatorR,
    op right = 1,
    op bottom/.store in=\LegOperatorB,
    op bottom = 1
  }
  \pgfkeys{label left= \(A_{\mathrm{L}}\),
          label center=\(B\),
          label right = \(A_{\mathrm{R}}\),
          op bottom = \(X\)}%
  \begin{tikzpicture}[baseline={([yshift=-\the\fontdimen22\textfont2]A0.center)}] 
\node[tensor, label=below:{\scriptsize\MPSTensorLabelL}] (A0) at (0 cm, 0) {};
\node[tensor, label=below:{\scriptsize\MPSTensorLabelC}] (A1) at (1 cm, 0) {};
\node[tensor, label=below:{\scriptsize\MPSTensorLabelR}] (A2) at (2 cm, 0) {};
\foreach \i in {0, ..., 2}
{
  \node[empty] (H\i) at (\i cm, 1.4) {};
  \draw [mid arrow] (A\i) -- (H\i);
}
\if1\LegOperatorL
\else
  \node[matrix, label={[xshift=2pt]left:{\scriptsize\LegOperatorL}}] (O0) at (0 cm, .8) {};
\fi
\if1\LegOperatorC
\else
  \node[matrix, label={[xshift=2pt]left:{\scriptsize\LegOperatorC}}] (O1) at (1 cm, .8) {};
\fi
\if1\LegOperatorR
\else
  \node[matrix, label={[xshift=2pt]left:{\scriptsize\LegOperatorR}}] (O2) at (2 cm, .8) {};
\fi

\draw [mid arrow] (A2.west) to (A1.east);
\draw [mid arrow] (A1.west) to (A0.east);

\if1\LegOperatorB
  \draw [mid arrow] (A0.west) to[out=180,in=180] (-.3em,-2.2em) to (2cm + .3em, -2.2em) to[out=0,in=0] (A2.east);
\else
  \node[matrix, label=below:{\scriptsize\LegOperatorB}] (OB) at (1 cm, -2.2em) {};
  \draw (A0.west) to[out=180,in=180] (-.3em,-2.2em);
  \draw [mid arrow] (-.3em,-2.2em) to (OB.east);
  \draw [mid arrow] (OB.west) to (2cm + .3em, -2.2em);
  \draw (2cm + .3em, -2.2em) to[out=0,in=0] (A2.east);
\fi
\if1\RepL
\else
  \node[matrix, label=below:{\scriptsize\RepL}] (OB) at (0.5 cm, 0) {};
\fi
\end{tikzpicture}

  \end{equation}
  holds, where \(R\) and \(U_{\mathrm{R}}\) are invertible operators on the physical Hilbert spaces,
  then there exists a unique invertible matrix \(V_{\mathrm{R}}\) such that
  \begin{equation}
    
  \pgfkeys{
    label/.store in=\MPSTensorLabel,
    label=\(\), 
    label inner/.store in=\MPSTensorLabelInner, 
    label inner=\(\),
    rep top/.store in=\TopLegOperator,
    rep top = 1,
    vector top/.store in=\TopLegVector,
    vector top = 1,
    rep left/.store in=\LeftLegOperator,
    rep left = 1,
    rep right/.store in=\RightLegOperator,
    rep right = 1
  }
  \pgfkeys{label = \(B\), rep top = \(R\)}%
  \begin{tikzpicture}[baseline={([yshift=-\the\fontdimen22\textfont2]A.center)}] 
\node[tensor, label=below:{\scriptsize\MPSTensorLabel}] (A) at (0, 0) {\MPSTensorLabelInner};
\if 1\RightLegOperator
  \node[empty] (Hv2) at (1, 0) {\scriptsize\MPSIndexRight};
  \draw [mid arrow] (Hv2) -- (A);
\else
  \node[matrix, label=above:{\scriptsize\RightLegOperator}, right = 1 em of A] (OpR) {};
  \node[empty] (Hv2) at (1.5, 0) {\scriptsize\MPSIndexRight};
  \draw [mid arrow] (OpR) -- (A);
  \draw (Hv2) -- (OpR);
\fi
\if 1\LeftLegOperator
  \node[empty] (Hv1) at (-1, 0) {\scriptsize\MPSIndexLeft};
  \draw [mid arrow] (A) -- (Hv1);
\else
  \node[matrix, label=above:{\scriptsize\LeftLegOperator}, left = 1 em of A] (OpL) {};
  \node[empty] (Hv1) at (-1.5, 0) {\scriptsize\MPSIndexLeft};
  \draw [mid arrow] (A) -- (OpL);
  \draw (OpL) -- (Hv1);
\fi
\if 1\TopLegOperator
  \if 1\TopLegVector
    \node[empty] (Hp) at (0, 1) {\scriptsize\MPSIndexTop};
    \draw [mid arrow] (A) -- (Hp);
  \else
    \node[matrix, label=right:{\scriptsize\TopLegVector}, above = 1 em of A] (VT) {};
    \draw [mid arrow] (A) -- (VT);
  \fi
\else
  \node[matrix, label=right:{\scriptsize\TopLegOperator}, above = 1 em of A] (OpT) {};
  \node[empty] (Hp) at (0, 1.5) {\scriptsize\MPSIndexTop};
  \draw [mid arrow] (A) -- (OpT);
  \draw (OpT) -- (Hp);
\fi
\end{tikzpicture}
 =
    
  \pgfkeys{
    label/.store in=\MPSTensorLabel,
    label=\(\), 
    label inner/.store in=\MPSTensorLabelInner, 
    label inner=\(\),
    rep top/.store in=\TopLegOperator,
    rep top = 1,
    vector top/.store in=\TopLegVector,
    vector top = 1,
    rep left/.store in=\LeftLegOperator,
    rep left = 1,
    rep right/.store in=\RightLegOperator,
    rep right = 1
  }
  \pgfkeys{label = \(B\), rep right = \(V_{\mathrm{R}}\)}%
  \begin{tikzpicture}[baseline={([yshift=-\the\fontdimen22\textfont2]A.center)}] 
\node[tensor, label=below:{\scriptsize\MPSTensorLabel}] (A) at (0, 0) {\MPSTensorLabelInner};
\if 1\RightLegOperator
  \node[empty] (Hv2) at (1, 0) {\scriptsize\MPSIndexRight};
  \draw [mid arrow] (Hv2) -- (A);
\else
  \node[matrix, label=above:{\scriptsize\RightLegOperator}, right = 1 em of A] (OpR) {};
  \node[empty] (Hv2) at (1.5, 0) {\scriptsize\MPSIndexRight};
  \draw [mid arrow] (OpR) -- (A);
  \draw (Hv2) -- (OpR);
\fi
\if 1\LeftLegOperator
  \node[empty] (Hv1) at (-1, 0) {\scriptsize\MPSIndexLeft};
  \draw [mid arrow] (A) -- (Hv1);
\else
  \node[matrix, label=above:{\scriptsize\LeftLegOperator}, left = 1 em of A] (OpL) {};
  \node[empty] (Hv1) at (-1.5, 0) {\scriptsize\MPSIndexLeft};
  \draw [mid arrow] (A) -- (OpL);
  \draw (OpL) -- (Hv1);
\fi
\if 1\TopLegOperator
  \if 1\TopLegVector
    \node[empty] (Hp) at (0, 1) {\scriptsize\MPSIndexTop};
    \draw [mid arrow] (A) -- (Hp);
  \else
    \node[matrix, label=right:{\scriptsize\TopLegVector}, above = 1 em of A] (VT) {};
    \draw [mid arrow] (A) -- (VT);
  \fi
\else
  \node[matrix, label=right:{\scriptsize\TopLegOperator}, above = 1 em of A] (OpT) {};
  \node[empty] (Hp) at (0, 1.5) {\scriptsize\MPSIndexTop};
  \draw [mid arrow] (A) -- (OpT);
  \draw (OpT) -- (Hp);
\fi
\end{tikzpicture}
.
  \end{equation}
\end{lemma}
\begin{proof}
  We prove the statement for \(L\) and \(V_{\mathrm{L}}\).
  The proof for \(R\) and \(V_{\mathrm{R}}\) is completely analogous.
  From Equation~\eqref{eq:19} and the fact that \(A_{\mathrm{R}}\) is injective, we conclude
  \begin{equation}
    \label{eq:18}
    \begin{tikzpicture}[baseline={([yshift=-\the\fontdimen22\textfont2]A0.center)}] 
\node[tensor, label=below:{\scriptsize\(A_{\mathrm{L}}\)}] (A0) at (0 cm, 0) {};
\node[tensor, label=below:{\scriptsize\(B\)}] (A1) at (1 cm, 0) {};
\node[tensor] (C1) at (1 cm, 3/2) {};
\node[empty] (H0) at (0, 1.4) {};
\draw[mid arrow] (A0) -- (H0);
\draw (A1) -- (C1);
\node[matrix, label={[xshift=2pt]left:{\scriptsize\(L^{-1}\)}}] (O1) at ($ (A1)!0.5!(C1) $) {};

\draw [mid arrow] (A1) to (A0);
\node[matrix, label=below:{\scriptsize\(X\)}] (OB) at (1 cm, -2.2em) {};
\draw (A0.west) to[out=180,in=180] (-.3em,-2.2em);
\draw [mid arrow] (-.3em,-2.2em) to (OB.east);
\draw [mid arrow, rounded corners=1em] (OB) -- (2cm, -2.2em) -- (2cm, 0);

\draw [rounded corners=.5em, mid arrow] (.7cm,5/2) -- (.7cm,3/2) -- (C1);
\draw[mid arrow] (C1) to[out=0,in=90] ({1cm + 1.5em},3/4) to[out=-90, in = 0] (A1);
\end{tikzpicture} \;= \begin{tikzpicture}[baseline={([yshift=-\the\fontdimen22\textfont2]A0.center)}] 
\node[tensor, label=below:{\scriptsize\(A_{\mathrm{L}}\)}] (A0) at (0 cm, 0) {};
\node[tensor, label=below:{\scriptsize\(B\)}] (A1) at (1 cm, 0) {};
\node[tensor] (C1) at (1 cm, 3/2) {};
\node[tensor, label=below:{\scriptsize\(A_{\mathrm{R}}\)}] (A2) at (2 cm, 0) {};
\node[tensor, label=above:{\scriptsize\(A_{\mathrm{R}}^{-1}\)}] (C2) at (2 cm, 3/2) {};
\node[empty] (H0) at (0, 1.4) {};
\draw (A0) -- (H0);
\draw [mid arrow] (A1) -- (C1);
\draw [mid arrow] (A2) -- (C2);
\draw [mid arrow] (C1) -- (C2);
\node[matrix, label={[xshift=2pt]left:{\scriptsize\(U_{\mathrm{L}}\)}}] (O1) at ($ (A0.north)!0.5!(H0.south) $) {};

\draw [mid arrow] (A2) to (A1);
\draw [mid arrow] (A1) to (A0);
\node[matrix, label=below:{\scriptsize\(X\)}] (OB) at (1 cm, -2.2em) {};
\draw (A0.west) to[out=180,in=180] (-.3em,-2.2em);
\draw [mid arrow] (-.3em,-2.2em) to (OB.east);
\draw [mid arrow] (OB.west) to (2cm + .3em, -2.2em);
\draw (2cm + .3em, -2.2em) to[out=0,in=0] (A2.east);

\draw [rounded corners=.5em, mid arrow] (.7cm,5/2) -- (.7cm,3/2) -- (C1);
\draw [rounded corners=.5em, mid arrow] (C2) -- (2.3cm,3/2) -- (2.3cm,5/2);
\end{tikzpicture},
  \end{equation}
  where the unlabeled tensor can be arbitrary.
  Since we required \(B\) to be left-injective,
  the inverse of \begin{tikzpicture}[baseline={([yshift=-\the\fontdimen22\textfont2]B.center)}]
\node [tensor, label = below:{\scriptsize \(B\)}] (B) at (0, 0) {\(\)};
\node [tensor, label = below:{\scriptsize \(A_{\mathrm{R}}\)}] (A) at (1, 0) {\(\)};
\draw [] (A) -- (B);
\draw [mid arrow] (B) -- (0,1);
\draw [mid arrow] (B) -- (-1,0);
\draw [mid arrow] (2,0) -- (A);
\draw [mid arrow] (A) -- (1,1);
\end{tikzpicture} is of the form \begin{tikzpicture}[baseline={([yshift=-\the\fontdimen22\textfont2]B.center)}]
\node [tensor, label = above:{\scriptsize \(C_{\vphantom{\mathrm{R}}}^{\vphantom{-1}}\)}] (B) at (0, 0) {\(\)};
\node [tensor, label = above:{\scriptsize \(A_{\mathrm{R}}^{-1}\)}] (A) at (1, 0) {\(\)};
\draw [] (A) -- (B);
\draw [mid arrow] (0,-1) -- (B);
\draw [mid arrow] (-1,0) -- (B);
\draw [mid arrow] (A) -- (2,0);
\draw [mid arrow] (1,-1) -- (A);
\end{tikzpicture},
  and Equation~\eqref{eq:18} implies
  \begin{align}\label{eq:21}
    
  \pgfkeys{
    label/.store in=\MPSTensorLabel,
    label=\(\), 
    label inner/.store in=\MPSTensorLabelInner, 
    label inner=\(\),
    rep top/.store in=\TopLegOperator,
    rep top = 1,
    vector top/.store in=\TopLegVector,
    vector top = 1,
    rep left/.store in=\LeftLegOperator,
    rep left = 1,
    rep right/.store in=\RightLegOperator,
    rep right = 1
  }
  \pgfkeys{label= \(A_{\mathrm{L}}\), rep left = \(X^{\intercal}\), rep top = \(U_{\mathrm{L}}\)}%
  \begin{tikzpicture}[baseline={([yshift=-\the\fontdimen22\textfont2]A.center)}] 
\node[tensor, label=below:{\scriptsize\MPSTensorLabel}] (A) at (0, 0) {\MPSTensorLabelInner};
\if 1\RightLegOperator
  \node[empty] (Hv2) at (1, 0) {\scriptsize\MPSIndexRight};
  \draw [mid arrow] (Hv2) -- (A);
\else
  \node[matrix, label=above:{\scriptsize\RightLegOperator}, right = 1 em of A] (OpR) {};
  \node[empty] (Hv2) at (1.5, 0) {\scriptsize\MPSIndexRight};
  \draw [mid arrow] (OpR) -- (A);
  \draw (Hv2) -- (OpR);
\fi
\if 1\LeftLegOperator
  \node[empty] (Hv1) at (-1, 0) {\scriptsize\MPSIndexLeft};
  \draw [mid arrow] (A) -- (Hv1);
\else
  \node[matrix, label=above:{\scriptsize\LeftLegOperator}, left = 1 em of A] (OpL) {};
  \node[empty] (Hv1) at (-1.5, 0) {\scriptsize\MPSIndexLeft};
  \draw [mid arrow] (A) -- (OpL);
  \draw (OpL) -- (Hv1);
\fi
\if 1\TopLegOperator
  \if 1\TopLegVector
    \node[empty] (Hp) at (0, 1) {\scriptsize\MPSIndexTop};
    \draw [mid arrow] (A) -- (Hp);
  \else
    \node[matrix, label=right:{\scriptsize\TopLegVector}, above = 1 em of A] (VT) {};
    \draw [mid arrow] (A) -- (VT);
  \fi
\else
  \node[matrix, label=right:{\scriptsize\TopLegOperator}, above = 1 em of A] (OpT) {};
  \node[empty] (Hp) at (0, 1.5) {\scriptsize\MPSIndexTop};
  \draw [mid arrow] (A) -- (OpT);
  \draw (OpT) -- (Hp);
\fi
\end{tikzpicture}
 &= 
    
  \pgfkeys{
    label/.store in=\MPSTensorLabel,
    label=\(\), 
    label inner/.store in=\MPSTensorLabelInner, 
    label inner=\(\),
    rep top/.store in=\TopLegOperator,
    rep top = 1,
    vector top/.store in=\TopLegVector,
    vector top = 1,
    rep left/.store in=\LeftLegOperator,
    rep left = 1,
    rep right/.store in=\RightLegOperator,
    rep right = 1
  }
  \pgfkeys{label= \(A_{\mathrm{L}}\), rep left = \(X^{\intercal}\), rep right = \(V_{\mathrm{L}}\)}%
  \begin{tikzpicture}[baseline={([yshift=-\the\fontdimen22\textfont2]A.center)}] 
\node[tensor, label=below:{\scriptsize\MPSTensorLabel}] (A) at (0, 0) {\MPSTensorLabelInner};
\if 1\RightLegOperator
  \node[empty] (Hv2) at (1, 0) {\scriptsize\MPSIndexRight};
  \draw [mid arrow] (Hv2) -- (A);
\else
  \node[matrix, label=above:{\scriptsize\RightLegOperator}, right = 1 em of A] (OpR) {};
  \node[empty] (Hv2) at (1.5, 0) {\scriptsize\MPSIndexRight};
  \draw [mid arrow] (OpR) -- (A);
  \draw (Hv2) -- (OpR);
\fi
\if 1\LeftLegOperator
  \node[empty] (Hv1) at (-1, 0) {\scriptsize\MPSIndexLeft};
  \draw [mid arrow] (A) -- (Hv1);
\else
  \node[matrix, label=above:{\scriptsize\LeftLegOperator}, left = 1 em of A] (OpL) {};
  \node[empty] (Hv1) at (-1.5, 0) {\scriptsize\MPSIndexLeft};
  \draw [mid arrow] (A) -- (OpL);
  \draw (OpL) -- (Hv1);
\fi
\if 1\TopLegOperator
  \if 1\TopLegVector
    \node[empty] (Hp) at (0, 1) {\scriptsize\MPSIndexTop};
    \draw [mid arrow] (A) -- (Hp);
  \else
    \node[matrix, label=right:{\scriptsize\TopLegVector}, above = 1 em of A] (VT) {};
    \draw [mid arrow] (A) -- (VT);
  \fi
\else
  \node[matrix, label=right:{\scriptsize\TopLegOperator}, above = 1 em of A] (OpT) {};
  \node[empty] (Hp) at (0, 1.5) {\scriptsize\MPSIndexTop};
  \draw [mid arrow] (A) -- (OpT);
  \draw (OpT) -- (Hp);
\fi
\end{tikzpicture}
,
    &
  \pgfkeys{
    label left/.store in=\MPSLabelLeft,
    label left=\(\), 
    label right/.store in=\MPSLabelRight,
    label right=\(\), 
    width/.store in=\widthTN,
    width=1cm, 
    map 1/.store in=\mapL,
    map 1=1,
    map 2/.store in=\mapR,
    map 2=1
  }
  \pgfkeys{map 1 = \(V_{\mathrm{L}}\)}%
  \begin{tikzpicture}[baseline={([yshift=-\the\fontdimen22\textfont2]Hr.center)}] 
\node[empty] (Hl) at (-\widthTN/2, 0) {\MPSLabelLeft};
\node[empty] (Hr) at ( \widthTN/2, 0) {\MPSLabelRight};
\if1\mapL
  \draw [mid arrow] (Hr) -- (Hl);
\else
  \if1\mapR
    \node[matrix, label = below:{\scriptsize\mapL}] (M1) at (0, 0) {};
    \draw [mid arrow] (Hr) -- (M1);
    \draw [mid arrow] (M1) -- (Hl);
  \else
    \node[matrix, label = below:{\scriptsize\mapL}] (M1) at (-\widthTN/6, 0) {};
    \node[matrix, label = below:{\scriptsize\mapR}] (M2) at ( \widthTN/6, 0) {};
    \draw [mid arrow] (Hr) -- (M2);
    \draw [mid arrow] (M2) -- (M1);
    \draw [mid arrow] (M1) -- (Hl);
  \fi
\fi
\end{tikzpicture}
 &:= \begin{tikzpicture}[baseline={([yshift=-\the\fontdimen22\textfont2]Bt.center)}] 
\node[tensor, label=above:{\scriptsize \(C\)}] (Bt) at (0, 0) {};
\node[tensor, label=below:{\scriptsize \(B\)}] (B) at (0, -.7cm) {};
\node[matrix, label=left:{\scriptsize \(L^{-1}\)}] (Li) at (0, -.35cm) {};
\draw (B.north) to (Li) to (Bt.south);
\draw[mid arrow] (Bt.east) to[out=0,in=90] (.5cm,-.35cm) to[out=-90, in = 0] (B.east);
\draw[mid arrow] (.5cm, .7cm) to ([yshift=.7cm]Bt.west) to[out=180,in=90] (-.5cm,.35cm) to[out=-90, in = 180] (Bt.west);
\draw[mid arrow] (B.west) to (-.5cm,-.7cm);
\end{tikzpicture}.
  \end{align}
  Inserting Equation~\eqref{eq:21} back into \eqref{eq:19} gives
  \begin{equation}
    
  \pgfkeys{
    label left/.store in=\MPSTensorLabelL,
    label left=\(\), 
    label center/.store in=\MPSTensorLabelC,
    label center=\(\), 
    label right/.store in=\MPSTensorLabelR,
    label right=\(\), 
    rep left/.store in=\RepL,
    rep left = 1,
    op left/.store in=\LegOperatorL,
    op left = 1,
    op center/.store in=\LegOperatorC,
    op center = 1,
    op right/.store in=\LegOperatorR,
    op right = 1,
    op bottom/.store in=\LegOperatorB,
    op bottom = 1
  }
  \pgfkeys{label left= \(A_{\mathrm{L}}\),
          label center=\(B\),
          label right = \(A_{\mathrm{R}}\),
          rep left = \(V_{\mathrm{L}}\),
          op center = \(L\),
          op bottom = \(X\)}%
  \begin{tikzpicture}[baseline={([yshift=-\the\fontdimen22\textfont2]A0.center)}] 
\node[tensor, label=below:{\scriptsize\MPSTensorLabelL}] (A0) at (0 cm, 0) {};
\node[tensor, label=below:{\scriptsize\MPSTensorLabelC}] (A1) at (1 cm, 0) {};
\node[tensor, label=below:{\scriptsize\MPSTensorLabelR}] (A2) at (2 cm, 0) {};
\foreach \i in {0, ..., 2}
{
  \node[empty] (H\i) at (\i cm, 1.4) {};
  \draw [mid arrow] (A\i) -- (H\i);
}
\if1\LegOperatorL
\else
  \node[matrix, label={[xshift=2pt]left:{\scriptsize\LegOperatorL}}] (O0) at (0 cm, .8) {};
\fi
\if1\LegOperatorC
\else
  \node[matrix, label={[xshift=2pt]left:{\scriptsize\LegOperatorC}}] (O1) at (1 cm, .8) {};
\fi
\if1\LegOperatorR
\else
  \node[matrix, label={[xshift=2pt]left:{\scriptsize\LegOperatorR}}] (O2) at (2 cm, .8) {};
\fi

\draw [mid arrow] (A2.west) to (A1.east);
\draw [mid arrow] (A1.west) to (A0.east);

\if1\LegOperatorB
  \draw [mid arrow] (A0.west) to[out=180,in=180] (-.3em,-2.2em) to (2cm + .3em, -2.2em) to[out=0,in=0] (A2.east);
\else
  \node[matrix, label=below:{\scriptsize\LegOperatorB}] (OB) at (1 cm, -2.2em) {};
  \draw (A0.west) to[out=180,in=180] (-.3em,-2.2em);
  \draw [mid arrow] (-.3em,-2.2em) to (OB.east);
  \draw [mid arrow] (OB.west) to (2cm + .3em, -2.2em);
  \draw (2cm + .3em, -2.2em) to[out=0,in=0] (A2.east);
\fi
\if1\RepL
\else
  \node[matrix, label=below:{\scriptsize\RepL}] (OB) at (0.5 cm, 0) {};
\fi
\end{tikzpicture}
=
    
  \pgfkeys{
    label left/.store in=\MPSTensorLabelL,
    label left=\(\), 
    label center/.store in=\MPSTensorLabelC,
    label center=\(\), 
    label right/.store in=\MPSTensorLabelR,
    label right=\(\), 
    rep left/.store in=\RepL,
    rep left = 1,
    op left/.store in=\LegOperatorL,
    op left = 1,
    op center/.store in=\LegOperatorC,
    op center = 1,
    op right/.store in=\LegOperatorR,
    op right = 1,
    op bottom/.store in=\LegOperatorB,
    op bottom = 1
  }
  \pgfkeys{label left= \(A_{\mathrm{L}}\),
          label center=\(B\),
          label right = \(A_{\mathrm{R}}\),
          op bottom = \(X\)}%
  \begin{tikzpicture}[baseline={([yshift=-\the\fontdimen22\textfont2]A0.center)}] 
\node[tensor, label=below:{\scriptsize\MPSTensorLabelL}] (A0) at (0 cm, 0) {};
\node[tensor, label=below:{\scriptsize\MPSTensorLabelC}] (A1) at (1 cm, 0) {};
\node[tensor, label=below:{\scriptsize\MPSTensorLabelR}] (A2) at (2 cm, 0) {};
\foreach \i in {0, ..., 2}
{
  \node[empty] (H\i) at (\i cm, 1.4) {};
  \draw [mid arrow] (A\i) -- (H\i);
}
\if1\LegOperatorL
\else
  \node[matrix, label={[xshift=2pt]left:{\scriptsize\LegOperatorL}}] (O0) at (0 cm, .8) {};
\fi
\if1\LegOperatorC
\else
  \node[matrix, label={[xshift=2pt]left:{\scriptsize\LegOperatorC}}] (O1) at (1 cm, .8) {};
\fi
\if1\LegOperatorR
\else
  \node[matrix, label={[xshift=2pt]left:{\scriptsize\LegOperatorR}}] (O2) at (2 cm, .8) {};
\fi

\draw [mid arrow] (A2.west) to (A1.east);
\draw [mid arrow] (A1.west) to (A0.east);

\if1\LegOperatorB
  \draw [mid arrow] (A0.west) to[out=180,in=180] (-.3em,-2.2em) to (2cm + .3em, -2.2em) to[out=0,in=0] (A2.east);
\else
  \node[matrix, label=below:{\scriptsize\LegOperatorB}] (OB) at (1 cm, -2.2em) {};
  \draw (A0.west) to[out=180,in=180] (-.3em,-2.2em);
  \draw [mid arrow] (-.3em,-2.2em) to (OB.east);
  \draw [mid arrow] (OB.west) to (2cm + .3em, -2.2em);
  \draw (2cm + .3em, -2.2em) to[out=0,in=0] (A2.east);
\fi
\if1\RepL
\else
  \node[matrix, label=below:{\scriptsize\RepL}] (OB) at (0.5 cm, 0) {};
\fi
\end{tikzpicture}

  \end{equation}
  and after inverting the injective tensors \(A_{\mathrm{L}}\) and \(A_{\mathrm{R}}\) we are left with
  (recall \(X\neq 0\))
  \begin{align}
    \label{eq:1}
    
  \pgfkeys{
    label/.store in=\MPSTensorLabel,
    label=\(\), 
    label inner/.store in=\MPSTensorLabelInner, 
    label inner=\(\),
    rep top/.store in=\TopLegOperator,
    rep top = 1,
    vector top/.store in=\TopLegVector,
    vector top = 1,
    rep left/.store in=\LeftLegOperator,
    rep left = 1,
    rep right/.store in=\RightLegOperator,
    rep right = 1
  }
  \pgfkeys{label = \(B\), rep left = \(V_{\mathrm{L}}\), rep top = \(L\)}%
  \begin{tikzpicture}[baseline={([yshift=-\the\fontdimen22\textfont2]A.center)}] 
\node[tensor, label=below:{\scriptsize\MPSTensorLabel}] (A) at (0, 0) {\MPSTensorLabelInner};
\if 1\RightLegOperator
  \node[empty] (Hv2) at (1, 0) {\scriptsize\MPSIndexRight};
  \draw [mid arrow] (Hv2) -- (A);
\else
  \node[matrix, label=above:{\scriptsize\RightLegOperator}, right = 1 em of A] (OpR) {};
  \node[empty] (Hv2) at (1.5, 0) {\scriptsize\MPSIndexRight};
  \draw [mid arrow] (OpR) -- (A);
  \draw (Hv2) -- (OpR);
\fi
\if 1\LeftLegOperator
  \node[empty] (Hv1) at (-1, 0) {\scriptsize\MPSIndexLeft};
  \draw [mid arrow] (A) -- (Hv1);
\else
  \node[matrix, label=above:{\scriptsize\LeftLegOperator}, left = 1 em of A] (OpL) {};
  \node[empty] (Hv1) at (-1.5, 0) {\scriptsize\MPSIndexLeft};
  \draw [mid arrow] (A) -- (OpL);
  \draw (OpL) -- (Hv1);
\fi
\if 1\TopLegOperator
  \if 1\TopLegVector
    \node[empty] (Hp) at (0, 1) {\scriptsize\MPSIndexTop};
    \draw [mid arrow] (A) -- (Hp);
  \else
    \node[matrix, label=right:{\scriptsize\TopLegVector}, above = 1 em of A] (VT) {};
    \draw [mid arrow] (A) -- (VT);
  \fi
\else
  \node[matrix, label=right:{\scriptsize\TopLegOperator}, above = 1 em of A] (OpT) {};
  \node[empty] (Hp) at (0, 1.5) {\scriptsize\MPSIndexTop};
  \draw [mid arrow] (A) -- (OpT);
  \draw (OpT) -- (Hp);
\fi
\end{tikzpicture}
=
    
  \pgfkeys{
    label/.store in=\MPSTensorLabel,
    label=\(\), 
    label inner/.store in=\MPSTensorLabelInner, 
    label inner=\(\),
    rep top/.store in=\TopLegOperator,
    rep top = 1,
    vector top/.store in=\TopLegVector,
    vector top = 1,
    rep left/.store in=\LeftLegOperator,
    rep left = 1,
    rep right/.store in=\RightLegOperator,
    rep right = 1
  }
  \pgfkeys{label = \(B\)}%
  \begin{tikzpicture}[baseline={([yshift=-\the\fontdimen22\textfont2]A.center)}] 
\node[tensor, label=below:{\scriptsize\MPSTensorLabel}] (A) at (0, 0) {\MPSTensorLabelInner};
\if 1\RightLegOperator
  \node[empty] (Hv2) at (1, 0) {\scriptsize\MPSIndexRight};
  \draw [mid arrow] (Hv2) -- (A);
\else
  \node[matrix, label=above:{\scriptsize\RightLegOperator}, right = 1 em of A] (OpR) {};
  \node[empty] (Hv2) at (1.5, 0) {\scriptsize\MPSIndexRight};
  \draw [mid arrow] (OpR) -- (A);
  \draw (Hv2) -- (OpR);
\fi
\if 1\LeftLegOperator
  \node[empty] (Hv1) at (-1, 0) {\scriptsize\MPSIndexLeft};
  \draw [mid arrow] (A) -- (Hv1);
\else
  \node[matrix, label=above:{\scriptsize\LeftLegOperator}, left = 1 em of A] (OpL) {};
  \node[empty] (Hv1) at (-1.5, 0) {\scriptsize\MPSIndexLeft};
  \draw [mid arrow] (A) -- (OpL);
  \draw (OpL) -- (Hv1);
\fi
\if 1\TopLegOperator
  \if 1\TopLegVector
    \node[empty] (Hp) at (0, 1) {\scriptsize\MPSIndexTop};
    \draw [mid arrow] (A) -- (Hp);
  \else
    \node[matrix, label=right:{\scriptsize\TopLegVector}, above = 1 em of A] (VT) {};
    \draw [mid arrow] (A) -- (VT);
  \fi
\else
  \node[matrix, label=right:{\scriptsize\TopLegOperator}, above = 1 em of A] (OpT) {};
  \node[empty] (Hp) at (0, 1.5) {\scriptsize\MPSIndexTop};
  \draw [mid arrow] (A) -- (OpT);
  \draw (OpT) -- (Hp);
\fi
\end{tikzpicture}
.
  \end{align}
  Notice that by the right-injectivity of \(B\),
  the matrix \(V_{\mathrm{L}}\) satisfying Equation~\eqref{eq:1} has to be unique.
  What remains is to show that \(V_{\mathrm{L}}\) is invertible.
  By the same reasoning that led to Equation~\eqref{eq:1},
  but inverting \(U_{\mathrm{L}}\) instead of \(L\) in Equation~\eqref{eq:19},
  we get
  \begin{align}
    \label{eq:2}
    
  \pgfkeys{
    label/.store in=\MPSTensorLabel,
    label=\(\), 
    label inner/.store in=\MPSTensorLabelInner, 
    label inner=\(\),
    rep top/.store in=\TopLegOperator,
    rep top = 1,
    vector top/.store in=\TopLegVector,
    vector top = 1,
    rep left/.store in=\LeftLegOperator,
    rep left = 1,
    rep right/.store in=\RightLegOperator,
    rep right = 1
  }
  \pgfkeys{label = \(B\), rep left = \(V^{\prime}\), rep top = \(L^{-1}\)}%
  \begin{tikzpicture}[baseline={([yshift=-\the\fontdimen22\textfont2]A.center)}] 
\node[tensor, label=below:{\scriptsize\MPSTensorLabel}] (A) at (0, 0) {\MPSTensorLabelInner};
\if 1\RightLegOperator
  \node[empty] (Hv2) at (1, 0) {\scriptsize\MPSIndexRight};
  \draw [mid arrow] (Hv2) -- (A);
\else
  \node[matrix, label=above:{\scriptsize\RightLegOperator}, right = 1 em of A] (OpR) {};
  \node[empty] (Hv2) at (1.5, 0) {\scriptsize\MPSIndexRight};
  \draw [mid arrow] (OpR) -- (A);
  \draw (Hv2) -- (OpR);
\fi
\if 1\LeftLegOperator
  \node[empty] (Hv1) at (-1, 0) {\scriptsize\MPSIndexLeft};
  \draw [mid arrow] (A) -- (Hv1);
\else
  \node[matrix, label=above:{\scriptsize\LeftLegOperator}, left = 1 em of A] (OpL) {};
  \node[empty] (Hv1) at (-1.5, 0) {\scriptsize\MPSIndexLeft};
  \draw [mid arrow] (A) -- (OpL);
  \draw (OpL) -- (Hv1);
\fi
\if 1\TopLegOperator
  \if 1\TopLegVector
    \node[empty] (Hp) at (0, 1) {\scriptsize\MPSIndexTop};
    \draw [mid arrow] (A) -- (Hp);
  \else
    \node[matrix, label=right:{\scriptsize\TopLegVector}, above = 1 em of A] (VT) {};
    \draw [mid arrow] (A) -- (VT);
  \fi
\else
  \node[matrix, label=right:{\scriptsize\TopLegOperator}, above = 1 em of A] (OpT) {};
  \node[empty] (Hp) at (0, 1.5) {\scriptsize\MPSIndexTop};
  \draw [mid arrow] (A) -- (OpT);
  \draw (OpT) -- (Hp);
\fi
\end{tikzpicture}
=
    
  \pgfkeys{
    label/.store in=\MPSTensorLabel,
    label=\(\), 
    label inner/.store in=\MPSTensorLabelInner, 
    label inner=\(\),
    rep top/.store in=\TopLegOperator,
    rep top = 1,
    vector top/.store in=\TopLegVector,
    vector top = 1,
    rep left/.store in=\LeftLegOperator,
    rep left = 1,
    rep right/.store in=\RightLegOperator,
    rep right = 1
  }
  \pgfkeys{label = \(B\)}%
  \begin{tikzpicture}[baseline={([yshift=-\the\fontdimen22\textfont2]A.center)}] 
\node[tensor, label=below:{\scriptsize\MPSTensorLabel}] (A) at (0, 0) {\MPSTensorLabelInner};
\if 1\RightLegOperator
  \node[empty] (Hv2) at (1, 0) {\scriptsize\MPSIndexRight};
  \draw [mid arrow] (Hv2) -- (A);
\else
  \node[matrix, label=above:{\scriptsize\RightLegOperator}, right = 1 em of A] (OpR) {};
  \node[empty] (Hv2) at (1.5, 0) {\scriptsize\MPSIndexRight};
  \draw [mid arrow] (OpR) -- (A);
  \draw (Hv2) -- (OpR);
\fi
\if 1\LeftLegOperator
  \node[empty] (Hv1) at (-1, 0) {\scriptsize\MPSIndexLeft};
  \draw [mid arrow] (A) -- (Hv1);
\else
  \node[matrix, label=above:{\scriptsize\LeftLegOperator}, left = 1 em of A] (OpL) {};
  \node[empty] (Hv1) at (-1.5, 0) {\scriptsize\MPSIndexLeft};
  \draw [mid arrow] (A) -- (OpL);
  \draw (OpL) -- (Hv1);
\fi
\if 1\TopLegOperator
  \if 1\TopLegVector
    \node[empty] (Hp) at (0, 1) {\scriptsize\MPSIndexTop};
    \draw [mid arrow] (A) -- (Hp);
  \else
    \node[matrix, label=right:{\scriptsize\TopLegVector}, above = 1 em of A] (VT) {};
    \draw [mid arrow] (A) -- (VT);
  \fi
\else
  \node[matrix, label=right:{\scriptsize\TopLegOperator}, above = 1 em of A] (OpT) {};
  \node[empty] (Hp) at (0, 1.5) {\scriptsize\MPSIndexTop};
  \draw [mid arrow] (A) -- (OpT);
  \draw (OpT) -- (Hp);
\fi
\end{tikzpicture}
,
  \end{align}
  for some matrix \(V^{\prime}\).
  Combining Equations (\ref{eq:1}) and (\ref{eq:2}) with
  the right-injectivity of \(B\) yields
  \begin{align}
    
  \pgfkeys{
    label left/.store in=\MPSLabelLeft,
    label left=\(\), 
    label right/.store in=\MPSLabelRight,
    label right=\(\), 
    width/.store in=\widthTN,
    width=1cm, 
    map 1/.store in=\mapL,
    map 1=1,
    map 2/.store in=\mapR,
    map 2=1
  }
  \pgfkeys{map 1 = \(V_{\mathrm{L}}^{\vphantom{\prime}}\), map 2 = \(V^{\prime}\), width = 1.5cm}%
  \begin{tikzpicture}[baseline={([yshift=-\the\fontdimen22\textfont2]Hr.center)}] 
\node[empty] (Hl) at (-\widthTN/2, 0) {\MPSLabelLeft};
\node[empty] (Hr) at ( \widthTN/2, 0) {\MPSLabelRight};
\if1\mapL
  \draw [mid arrow] (Hr) -- (Hl);
\else
  \if1\mapR
    \node[matrix, label = below:{\scriptsize\mapL}] (M1) at (0, 0) {};
    \draw [mid arrow] (Hr) -- (M1);
    \draw [mid arrow] (M1) -- (Hl);
  \else
    \node[matrix, label = below:{\scriptsize\mapL}] (M1) at (-\widthTN/6, 0) {};
    \node[matrix, label = below:{\scriptsize\mapR}] (M2) at ( \widthTN/6, 0) {};
    \draw [mid arrow] (Hr) -- (M2);
    \draw [mid arrow] (M2) -- (M1);
    \draw [mid arrow] (M1) -- (Hl);
  \fi
\fi
\end{tikzpicture}

    = 
  \pgfkeys{
    label left/.store in=\MPSLabelLeft,
    label left=\(\), 
    label right/.store in=\MPSLabelRight,
    label right=\(\), 
    width/.store in=\widthTN,
    width=1cm, 
    map 1/.store in=\mapL,
    map 1=1,
    map 2/.store in=\mapR,
    map 2=1
  }
  \pgfkeys{}%
  \begin{tikzpicture}[baseline={([yshift=-\the\fontdimen22\textfont2]Hr.center)}] 
\node[empty] (Hl) at (-\widthTN/2, 0) {\MPSLabelLeft};
\node[empty] (Hr) at ( \widthTN/2, 0) {\MPSLabelRight};
\if1\mapL
  \draw [mid arrow] (Hr) -- (Hl);
\else
  \if1\mapR
    \node[matrix, label = below:{\scriptsize\mapL}] (M1) at (0, 0) {};
    \draw [mid arrow] (Hr) -- (M1);
    \draw [mid arrow] (M1) -- (Hl);
  \else
    \node[matrix, label = below:{\scriptsize\mapL}] (M1) at (-\widthTN/6, 0) {};
    \node[matrix, label = below:{\scriptsize\mapR}] (M2) at ( \widthTN/6, 0) {};
    \draw [mid arrow] (Hr) -- (M2);
    \draw [mid arrow] (M2) -- (M1);
    \draw [mid arrow] (M1) -- (Hl);
  \fi
\fi
\end{tikzpicture}
 = 
    
  \pgfkeys{
    label left/.store in=\MPSLabelLeft,
    label left=\(\), 
    label right/.store in=\MPSLabelRight,
    label right=\(\), 
    width/.store in=\widthTN,
    width=1cm, 
    map 1/.store in=\mapL,
    map 1=1,
    map 2/.store in=\mapR,
    map 2=1
  }
  \pgfkeys{map 2 = \(V_{\mathrm{L}}^{\vphantom{\prime}}\), map 1 = \(V^{\prime}\), width = 1.5cm}%
  \begin{tikzpicture}[baseline={([yshift=-\the\fontdimen22\textfont2]Hr.center)}] 
\node[empty] (Hl) at (-\widthTN/2, 0) {\MPSLabelLeft};
\node[empty] (Hr) at ( \widthTN/2, 0) {\MPSLabelRight};
\if1\mapL
  \draw [mid arrow] (Hr) -- (Hl);
\else
  \if1\mapR
    \node[matrix, label = below:{\scriptsize\mapL}] (M1) at (0, 0) {};
    \draw [mid arrow] (Hr) -- (M1);
    \draw [mid arrow] (M1) -- (Hl);
  \else
    \node[matrix, label = below:{\scriptsize\mapL}] (M1) at (-\widthTN/6, 0) {};
    \node[matrix, label = below:{\scriptsize\mapR}] (M2) at ( \widthTN/6, 0) {};
    \draw [mid arrow] (Hr) -- (M2);
    \draw [mid arrow] (M2) -- (M1);
    \draw [mid arrow] (M1) -- (Hl);
  \fi
\fi
\end{tikzpicture}
,
  \end{align}
  i.e.\ \(V^{\prime}\) is the inverse of \(V_L\).
\end{proof}
\section{General statement}
\label{sec:general-statement}
\subsection{PEPS construction}
In this section we give a general PEPS construction,
where matter and gauge degrees of freedom are associated with the vertices and edges, respectively,
of an arbitrary oriented graph.
Throughout we use standard graph theory definitions and notation, cf.~\cite{serreTrees1980},
which we summarize in Appendix \ref{sec:graphs}.
\label{sec:peps-construction}
\begin{definition}[PEPS]
  Let \(\Gamma = (V, E)\) be an oriented\footnote{
    A graph is said to be oriented if it has no double edges or loops and each edge is given an orientation.
  }
  graph with Hilbert spaces \(\mathcal{H}_v\) and \(\mathcal{V}_e\) assigned to every
  vertex \(v\in V\) and edge \(e \in E\), respectively.
  We define a collection of \emph{PEPS tensors} on \(\Gamma\) to be tensors
  \begin{equation}
    A_v \in
    \bigotimes_{o\in E_v^{\mathrm{o}}} \mathcal{V}_o
    \otimes \mathcal{H}_v \otimes
    \bigotimes_{t \in E_v^{\mathrm{t}}} \mathcal{V}_t^{*}
  \end{equation}
  associated with each \(v \in V\).
  The corresponding state, simply referred to as \emph{PEPS},
  is defined to be
  \begin{equation}
    \ket{\Psi_{\Gamma}(A)}
    := \operatorname{tr}_E \bigotimes_{v\in V} A_v \in \bigotimes_{v\in V} \mathcal{H}_v,
  \end{equation}
  where we use \(\operatorname{tr}_E\) to denote tensor contraction over all Hilbert spaces
  labeled by edges.
\end{definition}
The Hilbert spaces denoted by \(\mathcal{H}\)
are referred to as \emph{physical}, while
the Hilbert spaces denoted by \(\mathcal{V}\),
introduced only for the purpose of constructing the PEPS,
are referred to as \emph{virtual}.
As indicated by the notation,
when we speak of a PEPS \(\ket{\Psi_{\Gamma}(A)}\),
we have implicitly fixed a collection of PEPS tensors giving rise to said state.
\begin{definition}[Global symmetry]
  We say a PEPS \(\ket{\Psi_{\Gamma}(A)}\) has a \emph{global symmetry},
  or is \emph{globally symmetric},
  with respect to given representations
  \((\mathcal{H}_v, U_v)\in \operatorname{Rep}G\),
  for all \(v\in \operatorname{vert}\Gamma\),
  if it satisfies
  \begin{align}
    \bigotimes_{v\in \Gamma}  U_v(g) 
    \ket{\Psi_{\Gamma}(A)} = \ket{\Psi_{\Gamma}(A)},
  \end{align}
  for all \(g\in G\).
\end{definition}
\begin{definition}[Local symmetry]
  We say a PEPS \(\ket{\Psi_{\Gamma}(A)}\) has a \emph{local symmetry},
  or is \emph{locally symmetric},
  with respect to given representations
  \((\mathcal{H}_v, U_v)\in \operatorname{Rep}G\),
  for all \(v\in \operatorname{vert}\Gamma\),
  if it satisfies
  \begin{align}
    U_v(g) \ket{\Psi_{\Gamma}(A)} = \ket{\Psi_{\Gamma}(A)},
  \end{align}
  for all \(g\in G\) and \(v\in \operatorname{vert}\Gamma\).
\end{definition}
Clearly, the above definitions make sense for any state in \(\bigotimes_v \mathcal{H}_v\), not just PEPS.
We introduce these terms to explicitly distinguish them from gauge-invariant PEPS,
which are introduced later.
\begin{definition}[Blocking]\label{lem:blocking}
  Let \(\ket{\Psi_{\Gamma}(A)}\) be a PEPS and
  \(\Gamma^{\prime} = (V^{\prime}, E^{\prime}) \subseteq \Gamma\)
  a subgraph. The tensor
  \begin{equation}
    \mathbb{T}_{\Gamma^{\prime}}(A) := \operatorname{tr}_{E^{\prime}} \bigotimes_{v\in V^{\prime}} A_v,
  \end{equation}
  where \(\operatorname{tr}_{E^{\prime}}\) denotes tensor contraction over all Hilbert spaces
  labeled by edges in \(E^{\prime}\),
  is called \emph{blocked PEPS tensor along} \(\Gamma^{\prime}\).
  The tensor \(\mathbb{T}_{\Gamma^{\prime}}(A)\) is said to be \emph{injective} if it is injective as a map
  \begin{equation}
    \bigotimes_{e \in\partial_{\mathrm{o}}\Gamma^{\prime}} \mathcal{V}_e^{*}\otimes
    \bigotimes_{f \in\partial_{\mathrm{i}}\Gamma^{\prime}} \mathcal{V}_f^{*}\otimes\mathcal{V}_f
    \otimes \bigotimes_{d \in\partial_{\mathrm{t}}\Gamma^{\prime}} \mathcal{V}_d 
    \longrightarrow \bigotimes_{v\in V^{\prime}} \mathcal{H}_v,
  \end{equation}
  where
  \(\partial_{\mathrm{t}}\Gamma^{\prime}\),
  \(\partial_{\mathrm{o}}\Gamma^{\prime}\) and
  \(\partial_{\mathrm{i}}\Gamma^{\prime}\)
  denote the terminal, original and internal boundaries\footnote{
    The terminal (original) boundary is comprised of all incoming (outgoing) edges,
    cf.\ Definition \ref{def:bdry-subgraph}.
  }
  of \(\Gamma^{\prime}\), respectively.
  In this context we will often say that \(\Gamma^{\prime}\) defines an \emph{injective region}.
\end{definition}
Commonly one considers blocking along a full\footnote{
  A full subgraph contains all edges in \(E\) connecting vertices in \(V^{\prime}\),
  cf.\ Definition \ref{def:full-subgraph}.
}
subgraph, which simplifies the injectivity condition on the blocked tensor,
since full subgraphs have empty internal boundary.
Notice that for the special case where we choose a single vertex as the subgraph,
the above notion of injectivity reduces to the usual one.

Rephrasing~\cite[Lemma 4]{molnarNormalProjectedEntangled2018} in the language of PEPS
defined on oriented graphs gives the following useful lemma.
\begin{lemma}[Union of injective regions is injective]\label{lem:union-injective}
  Let \(\ket{\Psi_{\Gamma}(A)}\) be a PEPS and \(\Gamma_1, \Gamma_2 \subseteq \Gamma\) injective regions,
  i.e.\ the tensors \(\mathbb{T}_{\Gamma_1}(A)\) and \(\mathbb{T}_{\Gamma_2}(A)\) are injective.
  Then the union of the two subgraphs also defines an injective region,
  i.e.\ the blocked tensor \(\mathbb{T}_{\Gamma_1\cup \Gamma_2}(A)\) is injective.
\end{lemma}
Recall that in the example of Section \ref{sec:example-normal-peps},
we considered a square lattice, where we assigned
physical Hilbert spaces and rank 3 tensors also to the edges of the lattice,
which is the accepted way to define gauge theories on a lattice.
In our general setting of PEPS defined on graphs,
we only want to associate virtual Hilbert spaces with edges,
hence it is most natural to introduce an additional vertex at each edge,
resulting in the bipartite graph \(\Gamma_{\mathrm{b}} = (V, E, \mathcal{E})\),
and carrying out the PEPS construction for \(\Gamma_{\mathrm{b}}\).
\begin{definition}[gPEPS]
  The special case of a PEPS defined on
  \(\Gamma_{\mathrm{b}} = (V,E,\mathcal{E})\),
  the barycentric\footnote{
    The barycentric subdivision is obtained by inserting an additional vertex in the middle of each edge,
    resulting in the new set of edges \(\mathcal{E}\),
    cf.\ Definition \ref{def:barycentric}.
  }
  subdivision of the oriented graph \(\Gamma = (V,E)\),
  we call \emph{gPEPS} on \(\Gamma\).
  To reflect the bipartition of \(\Gamma_{\mathrm{b}}\)
  we adopt slightly modified notation for the PEPS tensors
  \begin{equation}
    B_e \in
    \mathcal{V}_{e^{\mathrm{t}}}
    \otimes \mathcal{H}_e
    \otimes \mathcal{V}_{e^{\mathrm{o}}}^{*}
  \end{equation}
  that are labeled by vertices in \(E\),
  where \(e^{\mathrm{o}}\) and \(e^{\mathrm{t}}\),
  defined in Equations~\eqref{eq:7},
  can be thought of as ``half edges'' of the original graph,
  and for the corresponding state
  \begin{equation}
    \ket{\Psi_{\Gamma}(A, B)}\in
    \bigotimes_{v\in V} \mathcal{H}_v
    \otimes \bigotimes_{e\in E} \mathcal{H}_e.
  \end{equation}
\end{definition}
To wit, given a graph \(\Gamma = (V,E)\) we have associated
to each vertex \(v\in V\) a Hilbert space \(\mathcal{H}_v\),
which we think of as describing matter degrees of freedom,
and to each edge \(e\in E\) a Hilbert space \(\mathcal{H}_e\),
describing gauge degrees of freedom.
We use the notation
\begin{align}
  \mathcal{H}^{\mathrm{m}} := \bigotimes_{v\in V} \mathcal{H}_v,&&
  \mathcal{H}^{\mathrm{g}} := \bigotimes_{e\in E} \mathcal{H}_e,
\end{align}
for collectively referring to all matter (resp.\ gauge) degrees of freedom.
The general PEPS construction, applied to \(\Gamma_{\mathrm{b}}\), then yields
\begin{align}
  A_v &\in  \bigotimes_{o\in \mathcal{E}_v^{\mathrm{o}}} \mathcal{V}_o \otimes \mathcal{H}_v \otimes 
  \bigotimes_{t \in \mathcal{E}_v^{\mathrm{t}}} \mathcal{V}_t^{*},\\
  \label{eq:27}
  B_e &\in
  \mathcal{V}_{e^{\mathrm{t}}}
  \otimes \mathcal{H}_e
  \otimes \mathcal{V}_{e^{\mathrm{o}}}^{*},
\end{align}
for each \(v\in V\) and \(e\in E\).
When comparing \eqref{eq:27} with Equation~\eqref{eq:24},
keep in mind that \(e^{\mathrm{t}}\) originates at \(e\)
while \(e^{\mathrm{o}}\) terminates at \(e\).
We will sometimes refer to the tensors \(A\) and \(B\) as
\emph{vertex tensors} and \emph{edge tensors}, respectively.
\begin{definition}[Gauge-invariant PEPS]
  A gPEPS \(\ket{\Psi_{\Gamma}(A,B)}\) is said to be \emph{gauge-invariant},
  referred to as \emph{gauge-invariant PEPS} below,
  with respect to given representations
  \((\mathcal{H}_v, U_v),
    (\mathcal{H}_e, L_e),
    (\mathcal{H}_e, R_e) \in \operatorname{Rep} G\),
  for all \(v \in V\) and \(e \in E\), if
  \begin{equation}
    \widehat{U}_v (g) \ket{\Psi_{\Gamma}(A,B)} = \ket{\Psi_{\Gamma}(A,B)},
  \end{equation}
  for all \(g \in G\) and \(v\in V\), where
  \begin{equation}
    \widehat{U}_v :=
    \bigotimes_{e\in E^{\mathrm{o}}_v} R_e
    \otimes U_v \otimes 
    \bigotimes_{e\in E^{\mathrm{t}}_v} L_e,
  \end{equation}
  is called \emph{gauge operator centered at} \(v\).
\end{definition}
When we speak of a gauge-invariant PEPS \(\ket{\Psi_{\Gamma}(A, B)}\),
we will often leave the defining group representations implicit.
\subsection{Main theorem}
We prove that under mild assumptions the local tensors defining a gauge-invariant PEPS 
have to satisfy intertwiner relations analogous to the ones encountered in the square lattice example.
\begin{definition}[Invertible neighborhoods]\label{def:inj-nbhd}
  Let \(\ket{\Psi_{\Gamma}(A,B)}\) be a gauge-invariant PEPS.
  We say that \(e\in \operatorname{edge} \Gamma\)
  has \emph{invertible neighborhoods},\footnote{Not to be confused with the neighbourhood of a vertex,
    which has a different meaning in graph theory.}
  if there are two disjoint subgraphs 
  \(\Lambda^{\mathrm{o}}, \Lambda^{\mathrm{t}}\subseteq \Gamma_{\mathrm{b}}\setminus \left\{ e \right\}\),
  with
  \(\operatorname{o}(e)\in \operatorname{vert}\Lambda^{\mathrm{o}}\) 
  and
  \(\operatorname{t}(e)\in \operatorname{vert}\Lambda^{\mathrm{t}}\),
  such that the blocked tensors
  \(\mathbb{T}_{\Lambda^{\mathrm{o}}} (A,B)\)
  and
  \(\mathbb{T}_{\Lambda^{\mathrm{t}}} (A,B)\)
  are injective.
\end{definition}
\begin{theorem}\label{thm:1}
  Let \(\ket{\Psi_{\Gamma}(A,B)}\) be a gauge-invariant PEPS.
  If \(e \in \operatorname{edge}\Gamma\) has invertible neighborhoods, 
  then there exist linear representations
  \(
    (\mathcal{V}_{e^{\mathrm{o}}}, V_{e^{\mathrm{o}}}),
    (\mathcal{V}_{e^{\mathrm{t}}}, V_{e^{\mathrm{t}}})
    \in \operatorname{Rep}G
  \),
  such that the corresponding edge tensor \(B_e\) satisfies the intertwiner relations
  \begin{align}
    
  \pgfkeys{
    label/.store in=\MPSTensorLabel,
    label=\(\), 
    label inner/.store in=\MPSTensorLabelInner, 
    label inner=\(\),
    rep top/.store in=\TopLegOperator,
    rep top = 1,
    vector top/.store in=\TopLegVector,
    vector top = 1,
    rep left/.store in=\LeftLegOperator,
    rep left = 1,
    rep right/.store in=\RightLegOperator,
    rep right = 1
  }
  \pgfkeys{label = \(B_e\), rep top = \(L_e(g)\)}%
  \begin{tikzpicture}[baseline={([yshift=-\the\fontdimen22\textfont2]A.center)}] 
\node[tensor, label=below:{\scriptsize\MPSTensorLabel}] (A) at (0, 0) {\MPSTensorLabelInner};
\if 1\RightLegOperator
  \node[empty] (Hv2) at (1, 0) {\scriptsize\MPSIndexRight};
  \draw [mid arrow] (Hv2) -- (A);
\else
  \node[matrix, label=above:{\scriptsize\RightLegOperator}, right = 1 em of A] (OpR) {};
  \node[empty] (Hv2) at (1.5, 0) {\scriptsize\MPSIndexRight};
  \draw [mid arrow] (OpR) -- (A);
  \draw (Hv2) -- (OpR);
\fi
\if 1\LeftLegOperator
  \node[empty] (Hv1) at (-1, 0) {\scriptsize\MPSIndexLeft};
  \draw [mid arrow] (A) -- (Hv1);
\else
  \node[matrix, label=above:{\scriptsize\LeftLegOperator}, left = 1 em of A] (OpL) {};
  \node[empty] (Hv1) at (-1.5, 0) {\scriptsize\MPSIndexLeft};
  \draw [mid arrow] (A) -- (OpL);
  \draw (OpL) -- (Hv1);
\fi
\if 1\TopLegOperator
  \if 1\TopLegVector
    \node[empty] (Hp) at (0, 1) {\scriptsize\MPSIndexTop};
    \draw [mid arrow] (A) -- (Hp);
  \else
    \node[matrix, label=right:{\scriptsize\TopLegVector}, above = 1 em of A] (VT) {};
    \draw [mid arrow] (A) -- (VT);
  \fi
\else
  \node[matrix, label=right:{\scriptsize\TopLegOperator}, above = 1 em of A] (OpT) {};
  \node[empty] (Hp) at (0, 1.5) {\scriptsize\MPSIndexTop};
  \draw [mid arrow] (A) -- (OpT);
  \draw (OpT) -- (Hp);
\fi
\end{tikzpicture}
 =
    
  \pgfkeys{
    label/.store in=\MPSTensorLabel,
    label=\(\), 
    label inner/.store in=\MPSTensorLabelInner, 
    label inner=\(\),
    rep top/.store in=\TopLegOperator,
    rep top = 1,
    vector top/.store in=\TopLegVector,
    vector top = 1,
    rep left/.store in=\LeftLegOperator,
    rep left = 1,
    rep right/.store in=\RightLegOperator,
    rep right = 1
  }
  \pgfkeys{label = \(B_e\), rep left = \(V_{e^{\mathrm{t}}}^{-1}(g)\)}%
  \begin{tikzpicture}[baseline={([yshift=-\the\fontdimen22\textfont2]A.center)}] 
\node[tensor, label=below:{\scriptsize\MPSTensorLabel}] (A) at (0, 0) {\MPSTensorLabelInner};
\if 1\RightLegOperator
  \node[empty] (Hv2) at (1, 0) {\scriptsize\MPSIndexRight};
  \draw [mid arrow] (Hv2) -- (A);
\else
  \node[matrix, label=above:{\scriptsize\RightLegOperator}, right = 1 em of A] (OpR) {};
  \node[empty] (Hv2) at (1.5, 0) {\scriptsize\MPSIndexRight};
  \draw [mid arrow] (OpR) -- (A);
  \draw (Hv2) -- (OpR);
\fi
\if 1\LeftLegOperator
  \node[empty] (Hv1) at (-1, 0) {\scriptsize\MPSIndexLeft};
  \draw [mid arrow] (A) -- (Hv1);
\else
  \node[matrix, label=above:{\scriptsize\LeftLegOperator}, left = 1 em of A] (OpL) {};
  \node[empty] (Hv1) at (-1.5, 0) {\scriptsize\MPSIndexLeft};
  \draw [mid arrow] (A) -- (OpL);
  \draw (OpL) -- (Hv1);
\fi
\if 1\TopLegOperator
  \if 1\TopLegVector
    \node[empty] (Hp) at (0, 1) {\scriptsize\MPSIndexTop};
    \draw [mid arrow] (A) -- (Hp);
  \else
    \node[matrix, label=right:{\scriptsize\TopLegVector}, above = 1 em of A] (VT) {};
    \draw [mid arrow] (A) -- (VT);
  \fi
\else
  \node[matrix, label=right:{\scriptsize\TopLegOperator}, above = 1 em of A] (OpT) {};
  \node[empty] (Hp) at (0, 1.5) {\scriptsize\MPSIndexTop};
  \draw [mid arrow] (A) -- (OpT);
  \draw (OpT) -- (Hp);
\fi
\end{tikzpicture}
,&&
    
  \pgfkeys{
    label/.store in=\MPSTensorLabel,
    label=\(\), 
    label inner/.store in=\MPSTensorLabelInner, 
    label inner=\(\),
    rep top/.store in=\TopLegOperator,
    rep top = 1,
    vector top/.store in=\TopLegVector,
    vector top = 1,
    rep left/.store in=\LeftLegOperator,
    rep left = 1,
    rep right/.store in=\RightLegOperator,
    rep right = 1
  }
  \pgfkeys{label = \(B_e\), rep top = \(R_e(g)\)}%
  \begin{tikzpicture}[baseline={([yshift=-\the\fontdimen22\textfont2]A.center)}] 
\node[tensor, label=below:{\scriptsize\MPSTensorLabel}] (A) at (0, 0) {\MPSTensorLabelInner};
\if 1\RightLegOperator
  \node[empty] (Hv2) at (1, 0) {\scriptsize\MPSIndexRight};
  \draw [mid arrow] (Hv2) -- (A);
\else
  \node[matrix, label=above:{\scriptsize\RightLegOperator}, right = 1 em of A] (OpR) {};
  \node[empty] (Hv2) at (1.5, 0) {\scriptsize\MPSIndexRight};
  \draw [mid arrow] (OpR) -- (A);
  \draw (Hv2) -- (OpR);
\fi
\if 1\LeftLegOperator
  \node[empty] (Hv1) at (-1, 0) {\scriptsize\MPSIndexLeft};
  \draw [mid arrow] (A) -- (Hv1);
\else
  \node[matrix, label=above:{\scriptsize\LeftLegOperator}, left = 1 em of A] (OpL) {};
  \node[empty] (Hv1) at (-1.5, 0) {\scriptsize\MPSIndexLeft};
  \draw [mid arrow] (A) -- (OpL);
  \draw (OpL) -- (Hv1);
\fi
\if 1\TopLegOperator
  \if 1\TopLegVector
    \node[empty] (Hp) at (0, 1) {\scriptsize\MPSIndexTop};
    \draw [mid arrow] (A) -- (Hp);
  \else
    \node[matrix, label=right:{\scriptsize\TopLegVector}, above = 1 em of A] (VT) {};
    \draw [mid arrow] (A) -- (VT);
  \fi
\else
  \node[matrix, label=right:{\scriptsize\TopLegOperator}, above = 1 em of A] (OpT) {};
  \node[empty] (Hp) at (0, 1.5) {\scriptsize\MPSIndexTop};
  \draw [mid arrow] (A) -- (OpT);
  \draw (OpT) -- (Hp);
\fi
\end{tikzpicture}
 =
    
  \pgfkeys{
    label/.store in=\MPSTensorLabel,
    label=\(\), 
    label inner/.store in=\MPSTensorLabelInner, 
    label inner=\(\),
    rep top/.store in=\TopLegOperator,
    rep top = 1,
    vector top/.store in=\TopLegVector,
    vector top = 1,
    rep left/.store in=\LeftLegOperator,
    rep left = 1,
    rep right/.store in=\RightLegOperator,
    rep right = 1
  }
  \pgfkeys{label = \(B_e\), rep right = \(V_{e^{\mathrm{o}}}(g)\)}%
  \begin{tikzpicture}[baseline={([yshift=-\the\fontdimen22\textfont2]A.center)}] 
\node[tensor, label=below:{\scriptsize\MPSTensorLabel}] (A) at (0, 0) {\MPSTensorLabelInner};
\if 1\RightLegOperator
  \node[empty] (Hv2) at (1, 0) {\scriptsize\MPSIndexRight};
  \draw [mid arrow] (Hv2) -- (A);
\else
  \node[matrix, label=above:{\scriptsize\RightLegOperator}, right = 1 em of A] (OpR) {};
  \node[empty] (Hv2) at (1.5, 0) {\scriptsize\MPSIndexRight};
  \draw [mid arrow] (OpR) -- (A);
  \draw (Hv2) -- (OpR);
\fi
\if 1\LeftLegOperator
  \node[empty] (Hv1) at (-1, 0) {\scriptsize\MPSIndexLeft};
  \draw [mid arrow] (A) -- (Hv1);
\else
  \node[matrix, label=above:{\scriptsize\LeftLegOperator}, left = 1 em of A] (OpL) {};
  \node[empty] (Hv1) at (-1.5, 0) {\scriptsize\MPSIndexLeft};
  \draw [mid arrow] (A) -- (OpL);
  \draw (OpL) -- (Hv1);
\fi
\if 1\TopLegOperator
  \if 1\TopLegVector
    \node[empty] (Hp) at (0, 1) {\scriptsize\MPSIndexTop};
    \draw [mid arrow] (A) -- (Hp);
  \else
    \node[matrix, label=right:{\scriptsize\TopLegVector}, above = 1 em of A] (VT) {};
    \draw [mid arrow] (A) -- (VT);
  \fi
\else
  \node[matrix, label=right:{\scriptsize\TopLegOperator}, above = 1 em of A] (OpT) {};
  \node[empty] (Hp) at (0, 1.5) {\scriptsize\MPSIndexTop};
  \draw [mid arrow] (A) -- (OpT);
  \draw (OpT) -- (Hp);
\fi
\end{tikzpicture}
,
  \end{align}
  for all \(g \in G\).
\end{theorem}
\begin{proof}
  Without loss of generality we may assume that the subgraphs
  \(\Lambda^{\mathrm{t}}\) and \(\Lambda^{\mathrm{o}}\),
  defining the invertible neighborhoods,
  are full, connected and contain all vertices in \(\Gamma_{\mathrm{b}}\setminus \left\{ e \right\}\)
  adjacent to \(\operatorname{t}(e)\) and \(\operatorname{o}(e)\), respectively.
  We sketch an example of the general situation in Figure \ref{fig:injective-neighbourhood-edge}.
  Let \(J\) denote the set of all vertices in \(\Gamma_{\mathrm{b}}\setminus \left\{ e \right\}\)
  not included in \(\Lambda^{\mathrm{t}}\) or \(\Lambda^{\mathrm{o}}\)
  and let \(\Lambda^{\mathrm{j}} := \Gamma_{\mathrm{b}}[J] \subseteq \Gamma_{\mathrm{b}}\)
  be the full subgraph induced by the set \(J\).
  For brevity we drop the explicit dependence of the blocked tensors on the PEPS tensors from our notation.
  Let \(S\) be the set of edges in \(\Gamma_{\mathrm{b}}\) not included in
  any of \(\Lambda^{\mathrm{t}}\), \(\Lambda^{\mathrm{o}}\) or \(\Lambda^{\mathrm{j}}\),
  then we have
  \begin{align}
    \ket{\Psi_{\Gamma}(A,B)}
    &=\nonumber
    \operatorname{tr}_S\left(
      \mathbb{T}_{\Lambda^{\mathrm{t}}}\otimes
      B_e\otimes
      \mathbb{T}_{\Lambda^{\mathrm{o}}}\otimes
      \mathbb{T}_{\Lambda^{\mathrm{j}}}
    \right)\\
    &=
    
  \pgfkeys{
    label left/.store in=\MPSTensorLabelL,
    label left=\(\), 
    label center/.store in=\MPSTensorLabelC,
    label center=\(\), 
    label right/.store in=\MPSTensorLabelR,
    label right=\(\), 
    rep left/.store in=\RepL,
    rep left = 1,
    op left/.store in=\LegOperatorL,
    op left = 1,
    op center/.store in=\LegOperatorC,
    op center = 1,
    op right/.store in=\LegOperatorR,
    op right = 1,
    op bottom/.store in=\LegOperatorB,
    op bottom = 1
  }
  \pgfkeys{label left= \(\mathbb{T}_{\Lambda^{\mathrm{t}}}\),
          label center=\(B_e\),
          label right = \(\mathbb{T}_{\Lambda^{\mathrm{o}}}\),
          op bottom = \(\mathbb{T}_{\Lambda^{\mathrm{j}}}\)}%
  \begin{tikzpicture}[baseline={([yshift=-\the\fontdimen22\textfont2]A0.center)}] 
\node[tensor, label=below:{\scriptsize\MPSTensorLabelL}] (A0) at (0 cm, 0) {};
\node[tensor, label=below:{\scriptsize\MPSTensorLabelC}] (A1) at (1 cm, 0) {};
\node[tensor, label=below:{\scriptsize\MPSTensorLabelR}] (A2) at (2 cm, 0) {};
\foreach \i in {0, ..., 2}
{
  \node[empty] (H\i) at (\i cm, 1.4) {};
  \draw [mid arrow] (A\i) -- (H\i);
}
\if1\LegOperatorL
\else
  \node[matrix, label={[xshift=2pt]left:{\scriptsize\LegOperatorL}}] (O0) at (0 cm, .8) {};
\fi
\if1\LegOperatorC
\else
  \node[matrix, label={[xshift=2pt]left:{\scriptsize\LegOperatorC}}] (O1) at (1 cm, .8) {};
\fi
\if1\LegOperatorR
\else
  \node[matrix, label={[xshift=2pt]left:{\scriptsize\LegOperatorR}}] (O2) at (2 cm, .8) {};
\fi

\draw [mid arrow] (A2.west) to (A1.east);
\draw [mid arrow] (A1.west) to (A0.east);

\if1\LegOperatorB
  \draw [mid arrow] (A0.west) to[out=180,in=180] (-.3em,-2.2em) to (2cm + .3em, -2.2em) to[out=0,in=0] (A2.east);
\else
  \node[matrix, label=below:{\scriptsize\LegOperatorB}] (OB) at (1 cm, -2.2em) {};
  \draw (A0.west) to[out=180,in=180] (-.3em,-2.2em);
  \draw [mid arrow] (-.3em,-2.2em) to (OB.east);
  \draw [mid arrow] (OB.west) to (2cm + .3em, -2.2em);
  \draw (2cm + .3em, -2.2em) to[out=0,in=0] (A2.east);
\fi
\if1\RepL
\else
  \node[matrix, label=below:{\scriptsize\RepL}] (OB) at (0.5 cm, 0) {};
\fi
\end{tikzpicture}
,
  \end{align}
  where we do not draw the (generally numerous) physical legs of \(\mathbb{T}_{\Lambda^{\mathrm{j}}}\).
  The blocked tensors \(\mathbb{T}_{\Lambda^{\mathrm{o}}}\) and \(\mathbb{T}_{\Lambda^{\mathrm{t}}}\)
  associated with the invertible neighborhoods are injective by assumption
  and the symmetry properties
  \begin{equation}
    \widehat{U}_{\operatorname{t}(e)}\ket{\Psi_{\Gamma}(A,B)}=
    \widehat{U}_{\operatorname{o}(e)}\ket{\Psi_{\Gamma}(A,B)}=
    \ket{\Psi_{\Gamma}(A,B)}
  \end{equation}
  translate to 
  \begin{align}
    
  \pgfkeys{
    label left/.store in=\MPSTensorLabelL,
    label left=\(\), 
    label center/.store in=\MPSTensorLabelC,
    label center=\(\), 
    label right/.store in=\MPSTensorLabelR,
    label right=\(\), 
    rep left/.store in=\RepL,
    rep left = 1,
    op left/.store in=\LegOperatorL,
    op left = 1,
    op center/.store in=\LegOperatorC,
    op center = 1,
    op right/.store in=\LegOperatorR,
    op right = 1,
    op bottom/.store in=\LegOperatorB,
    op bottom = 1
  }
  \pgfkeys{label left= \(\mathbb{T}_{\Lambda^{\mathrm{t}}}\),
          label center=\(B_e\),
          label right = \(\mathbb{T}_{\Lambda^{\mathrm{o}}}\),
          op left = \(\widetilde{U}_{\mathrm{t}}(g)\),
          op center = \(L_e(g)\),
          op bottom = \(\mathbb{T}_{\Lambda^{\mathrm{j}}}\)}%
  \begin{tikzpicture}[baseline={([yshift=-\the\fontdimen22\textfont2]A0.center)}] 
\node[tensor, label=below:{\scriptsize\MPSTensorLabelL}] (A0) at (0 cm, 0) {};
\node[tensor, label=below:{\scriptsize\MPSTensorLabelC}] (A1) at (1 cm, 0) {};
\node[tensor, label=below:{\scriptsize\MPSTensorLabelR}] (A2) at (2 cm, 0) {};
\foreach \i in {0, ..., 2}
{
  \node[empty] (H\i) at (\i cm, 1.4) {};
  \draw [mid arrow] (A\i) -- (H\i);
}
\if1\LegOperatorL
\else
  \node[matrix, label={[xshift=2pt]left:{\scriptsize\LegOperatorL}}] (O0) at (0 cm, .8) {};
\fi
\if1\LegOperatorC
\else
  \node[matrix, label={[xshift=2pt]left:{\scriptsize\LegOperatorC}}] (O1) at (1 cm, .8) {};
\fi
\if1\LegOperatorR
\else
  \node[matrix, label={[xshift=2pt]left:{\scriptsize\LegOperatorR}}] (O2) at (2 cm, .8) {};
\fi

\draw [mid arrow] (A2.west) to (A1.east);
\draw [mid arrow] (A1.west) to (A0.east);

\if1\LegOperatorB
  \draw [mid arrow] (A0.west) to[out=180,in=180] (-.3em,-2.2em) to (2cm + .3em, -2.2em) to[out=0,in=0] (A2.east);
\else
  \node[matrix, label=below:{\scriptsize\LegOperatorB}] (OB) at (1 cm, -2.2em) {};
  \draw (A0.west) to[out=180,in=180] (-.3em,-2.2em);
  \draw [mid arrow] (-.3em,-2.2em) to (OB.east);
  \draw [mid arrow] (OB.west) to (2cm + .3em, -2.2em);
  \draw (2cm + .3em, -2.2em) to[out=0,in=0] (A2.east);
\fi
\if1\RepL
\else
  \node[matrix, label=below:{\scriptsize\RepL}] (OB) at (0.5 cm, 0) {};
\fi
\end{tikzpicture}
 =
    
  \pgfkeys{
    label left/.store in=\MPSTensorLabelL,
    label left=\(\), 
    label center/.store in=\MPSTensorLabelC,
    label center=\(\), 
    label right/.store in=\MPSTensorLabelR,
    label right=\(\), 
    rep left/.store in=\RepL,
    rep left = 1,
    op left/.store in=\LegOperatorL,
    op left = 1,
    op center/.store in=\LegOperatorC,
    op center = 1,
    op right/.store in=\LegOperatorR,
    op right = 1,
    op bottom/.store in=\LegOperatorB,
    op bottom = 1
  }
  \pgfkeys{label left= \(\mathbb{T}_{\Lambda^{\mathrm{t}}}\),
          label center=\(B_e\),
          label right = \(\mathbb{T}_{\Lambda^{\mathrm{o}}}\),
          op bottom = \(\mathbb{T}_{\Lambda^{\mathrm{j}}}\)}%
  \begin{tikzpicture}[baseline={([yshift=-\the\fontdimen22\textfont2]A0.center)}] 
\node[tensor, label=below:{\scriptsize\MPSTensorLabelL}] (A0) at (0 cm, 0) {};
\node[tensor, label=below:{\scriptsize\MPSTensorLabelC}] (A1) at (1 cm, 0) {};
\node[tensor, label=below:{\scriptsize\MPSTensorLabelR}] (A2) at (2 cm, 0) {};
\foreach \i in {0, ..., 2}
{
  \node[empty] (H\i) at (\i cm, 1.4) {};
  \draw [mid arrow] (A\i) -- (H\i);
}
\if1\LegOperatorL
\else
  \node[matrix, label={[xshift=2pt]left:{\scriptsize\LegOperatorL}}] (O0) at (0 cm, .8) {};
\fi
\if1\LegOperatorC
\else
  \node[matrix, label={[xshift=2pt]left:{\scriptsize\LegOperatorC}}] (O1) at (1 cm, .8) {};
\fi
\if1\LegOperatorR
\else
  \node[matrix, label={[xshift=2pt]left:{\scriptsize\LegOperatorR}}] (O2) at (2 cm, .8) {};
\fi

\draw [mid arrow] (A2.west) to (A1.east);
\draw [mid arrow] (A1.west) to (A0.east);

\if1\LegOperatorB
  \draw [mid arrow] (A0.west) to[out=180,in=180] (-.3em,-2.2em) to (2cm + .3em, -2.2em) to[out=0,in=0] (A2.east);
\else
  \node[matrix, label=below:{\scriptsize\LegOperatorB}] (OB) at (1 cm, -2.2em) {};
  \draw (A0.west) to[out=180,in=180] (-.3em,-2.2em);
  \draw [mid arrow] (-.3em,-2.2em) to (OB.east);
  \draw [mid arrow] (OB.west) to (2cm + .3em, -2.2em);
  \draw (2cm + .3em, -2.2em) to[out=0,in=0] (A2.east);
\fi
\if1\RepL
\else
  \node[matrix, label=below:{\scriptsize\RepL}] (OB) at (0.5 cm, 0) {};
\fi
\end{tikzpicture}

  \end{align}
  and
  \begin{align}
    
  \pgfkeys{
    label left/.store in=\MPSTensorLabelL,
    label left=\(\), 
    label center/.store in=\MPSTensorLabelC,
    label center=\(\), 
    label right/.store in=\MPSTensorLabelR,
    label right=\(\), 
    rep left/.store in=\RepL,
    rep left = 1,
    op left/.store in=\LegOperatorL,
    op left = 1,
    op center/.store in=\LegOperatorC,
    op center = 1,
    op right/.store in=\LegOperatorR,
    op right = 1,
    op bottom/.store in=\LegOperatorB,
    op bottom = 1
  }
  \pgfkeys{label left= \(\mathbb{T}_{\Lambda^{\mathrm{t}}}\),
          label center=\(B_e\),
          label right = \(\mathbb{T}_{\Lambda^{\mathrm{o}}}\),
          op right = \(\widetilde{U}_{\mathrm{o}}(g)\),
          op center = \(R_e(g)\),
          op bottom = \(\mathbb{T}_{\Lambda^{\mathrm{j}}}\)}%
  \begin{tikzpicture}[baseline={([yshift=-\the\fontdimen22\textfont2]A0.center)}] 
\node[tensor, label=below:{\scriptsize\MPSTensorLabelL}] (A0) at (0 cm, 0) {};
\node[tensor, label=below:{\scriptsize\MPSTensorLabelC}] (A1) at (1 cm, 0) {};
\node[tensor, label=below:{\scriptsize\MPSTensorLabelR}] (A2) at (2 cm, 0) {};
\foreach \i in {0, ..., 2}
{
  \node[empty] (H\i) at (\i cm, 1.4) {};
  \draw [mid arrow] (A\i) -- (H\i);
}
\if1\LegOperatorL
\else
  \node[matrix, label={[xshift=2pt]left:{\scriptsize\LegOperatorL}}] (O0) at (0 cm, .8) {};
\fi
\if1\LegOperatorC
\else
  \node[matrix, label={[xshift=2pt]left:{\scriptsize\LegOperatorC}}] (O1) at (1 cm, .8) {};
\fi
\if1\LegOperatorR
\else
  \node[matrix, label={[xshift=2pt]left:{\scriptsize\LegOperatorR}}] (O2) at (2 cm, .8) {};
\fi

\draw [mid arrow] (A2.west) to (A1.east);
\draw [mid arrow] (A1.west) to (A0.east);

\if1\LegOperatorB
  \draw [mid arrow] (A0.west) to[out=180,in=180] (-.3em,-2.2em) to (2cm + .3em, -2.2em) to[out=0,in=0] (A2.east);
\else
  \node[matrix, label=below:{\scriptsize\LegOperatorB}] (OB) at (1 cm, -2.2em) {};
  \draw (A0.west) to[out=180,in=180] (-.3em,-2.2em);
  \draw [mid arrow] (-.3em,-2.2em) to (OB.east);
  \draw [mid arrow] (OB.west) to (2cm + .3em, -2.2em);
  \draw (2cm + .3em, -2.2em) to[out=0,in=0] (A2.east);
\fi
\if1\RepL
\else
  \node[matrix, label=below:{\scriptsize\RepL}] (OB) at (0.5 cm, 0) {};
\fi
\end{tikzpicture}
 =
    
  \pgfkeys{
    label left/.store in=\MPSTensorLabelL,
    label left=\(\), 
    label center/.store in=\MPSTensorLabelC,
    label center=\(\), 
    label right/.store in=\MPSTensorLabelR,
    label right=\(\), 
    rep left/.store in=\RepL,
    rep left = 1,
    op left/.store in=\LegOperatorL,
    op left = 1,
    op center/.store in=\LegOperatorC,
    op center = 1,
    op right/.store in=\LegOperatorR,
    op right = 1,
    op bottom/.store in=\LegOperatorB,
    op bottom = 1
  }
  \pgfkeys{label left= \(\mathbb{T}_{\Lambda^{\mathrm{t}}}\),
          label center=\(B_e\),
          label right = \(\mathbb{T}_{\Lambda^{\mathrm{o}}}\),
          op bottom = \(\mathbb{T}_{\Lambda^{\mathrm{j}}}\)}%
  \begin{tikzpicture}[baseline={([yshift=-\the\fontdimen22\textfont2]A0.center)}] 
\node[tensor, label=below:{\scriptsize\MPSTensorLabelL}] (A0) at (0 cm, 0) {};
\node[tensor, label=below:{\scriptsize\MPSTensorLabelC}] (A1) at (1 cm, 0) {};
\node[tensor, label=below:{\scriptsize\MPSTensorLabelR}] (A2) at (2 cm, 0) {};
\foreach \i in {0, ..., 2}
{
  \node[empty] (H\i) at (\i cm, 1.4) {};
  \draw [mid arrow] (A\i) -- (H\i);
}
\if1\LegOperatorL
\else
  \node[matrix, label={[xshift=2pt]left:{\scriptsize\LegOperatorL}}] (O0) at (0 cm, .8) {};
\fi
\if1\LegOperatorC
\else
  \node[matrix, label={[xshift=2pt]left:{\scriptsize\LegOperatorC}}] (O1) at (1 cm, .8) {};
\fi
\if1\LegOperatorR
\else
  \node[matrix, label={[xshift=2pt]left:{\scriptsize\LegOperatorR}}] (O2) at (2 cm, .8) {};
\fi

\draw [mid arrow] (A2.west) to (A1.east);
\draw [mid arrow] (A1.west) to (A0.east);

\if1\LegOperatorB
  \draw [mid arrow] (A0.west) to[out=180,in=180] (-.3em,-2.2em) to (2cm + .3em, -2.2em) to[out=0,in=0] (A2.east);
\else
  \node[matrix, label=below:{\scriptsize\LegOperatorB}] (OB) at (1 cm, -2.2em) {};
  \draw (A0.west) to[out=180,in=180] (-.3em,-2.2em);
  \draw [mid arrow] (-.3em,-2.2em) to (OB.east);
  \draw [mid arrow] (OB.west) to (2cm + .3em, -2.2em);
  \draw (2cm + .3em, -2.2em) to[out=0,in=0] (A2.east);
\fi
\if1\RepL
\else
  \node[matrix, label=below:{\scriptsize\RepL}] (OB) at (0.5 cm, 0) {};
\fi
\end{tikzpicture}
,
  \end{align}
  where \(\widetilde{U}_{\mathrm{t}}(g) \otimes L_e(g) := \widehat{U}_{\operatorname{t}(e)}(g)\) and
  \(R_e(g)\otimes\widetilde{U}_{\mathrm{o}}(g)  := \widehat{U}_{\operatorname{o}(e)}(g)\),
  hence the Fundamental Lemma \ref{lem:fundamental} gives the claimed intertwiner properties. \qedhere
  \begin{figure}
    \centering
      \begin{tikzpicture}[rotate = 35]
\begin{scope}
  \clip plot [smooth cycle, tension=.6] coordinates {(-1,-1) (-5,4) (-5.5, 9.5) (-0, 11.7) (7,8.5) (9.5,4) (6, -1)};
  \draw [teal, fill=teal!30!white] plot [smooth cycle, tension=0.8] coordinates {(-4,5.5) (0.9, 6.4) (1,4) (.8,-.2) (-2, -2) (-5,2) (-5.5, 4)};
  \draw [orange, fill=orange!30!white] plot [smooth cycle, tension=0.8] coordinates {(9,7) (6, 9) (1.1, 7.9) (3,4) (3.2,-0.2) (5,-2) (9,0) (10,4)};
  \draw [green, fill=green!30!white] plot [smooth cycle, tension=1] coordinates {(1, -2) (2, 1) (3,-2)};
  \draw [green, fill=green!30!white] plot [smooth cycle, tension=1] coordinates {(-7, 8) (-4, 6) (2,9.5) (1,13) (-5,12)};
\end{scope}
\node [tensor, fill=white] (00) at (0,0) {};
\node [tensor, fill=white] (01) at (0,4) {};
\node [tensor, fill=white] (10) at (4,0) {};
\node [tensor, fill=white] (11) at (4,4) {};
\node [tensor, fill=white] (21) at (8, 4) {};
\node [tensor, fill=white] (t1) at (2, {sqrt(3)*2+4}) {};
\node [tensor, fill=white] (r1) at (6, {sqrt(3)*2+4}) {};
\node [tensor, fill=white] (p1) at ({-4*cos(12)},{4+4*sin(12)}) {};
\node [tensor, fill=white] (p2) at ($(p1)+({-4*sin(6)},{4*cos(6)})$) {};
\node [tensor, fill=white] (p3) at ($(t1)+({-4*sin(42)},{4*cos(42)})$) {};
\draw [mid bullet] (10)--(00);
\draw [mid bullet] (01)--(00);
\draw [mid bullet] (11)--(10);
\draw [mid bullet] (11)--(01);
\draw [mid bullet] (21)--(11);
\draw [mid bullet] (21)--(r1);
\draw [mid bullet] (r1)--(t1);
\draw [mid bullet] (11)--(t1);
\draw [mid bullet] (t1)--(01);
\draw [mid bullet] (t1)--(p3);
\draw [mid bullet] (p3)--(p2);
\draw [mid bullet] (p1)--(p2);
\draw [mid bullet] (01)--(p1);
\draw [dash pattern=on .82pt off .90pt, line width = .3pt] (00) --++ (0,-1) (00) --++ (-1,0);
\draw [dash pattern=on .82pt off .90pt, line width = .3pt] (10) --++ (0,-1) (10) --++ (1,0);
\draw [dash pattern=on .82pt off .90pt, line width = .3pt] (21) --++ (0,-1) (21) --++ (1,0);
\draw [dash pattern=on .82pt off .90pt, line width = .3pt] (p1) --++ ({-sin(47)},{-cos(47)});
\draw [dash pattern=on .82pt off .90pt, line width = .3pt] (p2) --++ ({-sin(64)},{cos(64)});
\draw [dash pattern=on .82pt off .90pt, line width = .3pt] (p3) --++ ({sin(17)},{cos(17)});
\draw [dash pattern=on .82pt off .90pt, line width = .3pt] (r1) --++ ({cos(60)},{sin(60)});
\node (e) at ([shift={(2,-2/3)}]01) {\(e_0\)};
\node (x) at ([shift={(1/2,2)}]01) {\(x\)};
\node (t) at ([shift={(-3/4,-1/2)}]01) {\(\operatorname{t}(e_0)\)};
\node (lo) at ([shift={(1/3,3/4)}]11) {\(\operatorname{o}(e_0)\)};
\node (lt) at (-2,2) {\(\color{teal}\Lambda^{\mathrm{t}}\)};
\node (l0) at (6,2) {\(\color{orange}\Lambda^{\mathrm{o}}\)};
\node (lj) at (-2,8) {\(\color{green!80!black}\Lambda^{\mathrm{j}}\)};
\end{tikzpicture}
      \caption{
        It is convenient to think of the graph \(\Gamma_{\mathrm{b}}\) as being embedded in Euclidean space
        and label subgraphs by the spatial regions they are contained in.
        Since the edge tensors are all assumed to be LRI,
        whether the vertex \(x\) is seen as belonging to the orange or the teal region
        does not influence the injectivity of the blocked tensors associated with either subgraph.
        Only because the gauge operator centered at \(\operatorname{t}(e_0)\)
        will also act on the degrees of freedom associated with \(x\)
        are we forced to include it in the teal region.
      }
      \label{fig:injective-neighbourhood-edge}
  \end{figure}
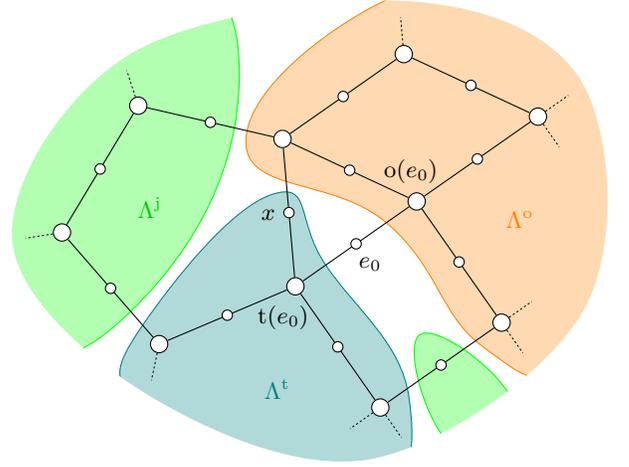
\end{proof}
\begin{theorem}\label{thm:2}
  Let \(\ket{\Psi_{\Gamma}(A,B)}\) be a gauge-invariant PEPS on \(\Gamma=(V,E)\).
  If all edges of \(\Gamma\), with vertex \(v\in V\) as an endpoint, have invertible neighborhoods,
  then the vertex tensor \(A_v\) will satisfy the intertwiner property
  \begin{align}
    U_v(g) \circ A_v = A_v \circ \bigotimes_{f\in E_v^{\mathrm{o}}} V_{f^{\mathrm{o}}}^{-1}(g) \otimes
    \bigotimes_{e\in E_v^{\mathrm{t}}} V_{e^{\mathrm{t}}}(g),
  \end{align}
  for all \(g \in G\),
  where \(\{(\mathcal{V}, V)\}\) are the representations of Theorem \ref{thm:1}
  and we view \(A_v\) as a linear map
  \begin{equation}
    A_v : \bigotimes_{f\in E_v^{\mathrm{o}}} \mathcal{V}_{f^{\mathrm{o}}}^{*}\otimes
    \bigotimes_{e\in E_v^{\mathrm{t}}} \mathcal{V}_{e^{\mathrm{t}}}
    \longrightarrow \mathcal{H}_v.
  \end{equation}
\end{theorem}
\begin{proof}
  By assumption, every \(e\in E\) has invertible neighborhoods,
  which we shall respectively label \(\Lambda^{\mathrm{t}}_e\) and \(\Lambda^{\mathrm{o}}_e\).
  Let \(v_0 \in V\) and use the invertible neighborhoods of the surrounding edges to construct the subgraph
  \begin{equation}
    \Lambda_{v_0} := \bigcup_{e\in E_{v_0}^{\mathrm{t}}} \Lambda^{\mathrm{o}}_e
    \cup \bigcup_{f\in E_{v_0}^{\mathrm{o}}} \Lambda^{\mathrm{t}}_f \subseteq \Gamma_{\mathrm{b}}.
  \end{equation}
  According to Lemma \ref{lem:union-injective},
  the blocked tensor \(\mathbb{T}_{\Lambda_{v_0}}(A,B)\) is injective 
  and since \(\Lambda_{v_0}\) contains all vertices adjacent to \({v_0}\) in \(\Gamma\),
  the condition
  \begin{equation}
    \widehat{U}_{v_0}(g) \ket{\Psi_{\Gamma}(A,B)} = \ket{\Psi_{\Gamma}(A,B)}
  \end{equation}
  is equivalent (by inverting \(\mathbb{T}_{\Lambda_{v_0}}\) and tracing out the remaining degrees of freedom)
  to the equality
  \begin{equation}\label{eq:8}
\begin{tikzpicture}[baseline={([yshift=-\the\fontdimen22\textfont2]0.center)}, y={(-1em,1.5em)},x={(1em,1.5em)}, z={(0em,.5em)}]
\node [tensor] (0) at (0, 0) {\(\)};
\node [stensor] (L1) at (0, -2) {};
\node [stensor] (L2) at ({2*sin(30)}, {-2*cos(30)}) {};
\node [stensor] (Ln) at (3/2, -1/2) {};
\node [stensor] (R1) at (-2, 0) {};
\node [stensor] (R2) at ({-2*cos(30)}, {2*sin(30)}) {};
\node [stensor] (Rm) at (-1/2, 3/2) {};

\node (VL1) at (0, -3) {};
\node (VL2) at ({3*sin(30)}, {-3*cos(30)}) {};
\node (VLn) at (9/4, -3/4) {};
\node (VR1) at (-3, 0) {};
\node (VR2) at ({-3*cos(30)}, {3*sin(30)}) {};
\node (VRm) at (-3/4, 9/4) {};

\foreach \x/\n in {L/1, L/2, L/n} 
{
  \draw [mid arrow] (\x\n) -- (0);
  \draw (V\x\n) -- (\x\n);
  \node [matrix, label = {[label distance = -3pt]right:{\contour{white}{\scriptsize\(L_{e_{\n}}(g)\)}}}] (m\x\n) at ($(\x\n)+(0,0,2)$) {};
  \draw (\x\n) -- (m\x\n);
  \draw (m\x\n) --++ (0,0,1.5);
}
\foreach \x/\n in {R/1, R/2, R/m} 
{
  \draw [mid arrow] (0) -- (\x\n);
  \draw (\x\n) -- (V\x\n);
  \node [matrix, label = {[label distance = -3pt]left:{\contour{white}{\scriptsize\(R_{f_{\n}}(g)\)}}}] (m\x\n) at ($(\x\n)+(0,0,2)$) {};
  \draw (\x\n) -- (m\x\n);
  \draw (m\x\n) --++ (0,0,1.5);
}
\draw [dash pattern=on .85pt off 1pt, line width = .3pt] (0) --++ ({-sin(40)}, {cos(40)});
\draw [dash pattern=on .85pt off 1pt, line width = .3pt] (0) --++ ({cos(40)}, {-sin(40)});

\node [matrix, label = {[label distance = -3pt]left:{\contour{white}{\scriptsize\(U_{v_0}(g)\)}}}] (m0) at (0,0,2.5) {};
\draw (m0) --++ (0, 0, 2);
\draw (0) -- (m0);

\node (v) at (-1/3,-1/3) {\scriptsize\(v_0\)};
\end{tikzpicture} = \hspace{-.5em}
\begin{tikzpicture}[baseline={([yshift=-\the\fontdimen22\textfont2]0.center)}, y={(-1em,1.5em)},x={(1em,1.5em)}, z={(0em,.5em)}]
\node [tensor] (0) at (0, 0) {\(\)};
\node [stensor] (L1) at (0, -2) {};
\node [stensor] (L2) at ({2*sin(30)}, {-2*cos(30)}) {};
\node [stensor] (Ln) at (3/2, -1/2) {};
\node [stensor] (R1) at (-2, 0) {};
\node [stensor] (R2) at ({-2*cos(30)}, {2*sin(30)}) {};
\node [stensor] (Rm) at (-1/2, 3/2) {};

\node (VL1) at (0, -3) {};
\node (VL2) at ({3*sin(30)}, {-3*cos(30)}) {};
\node (VLn) at (9/4, -3/4) {};
\node (VR1) at (-3, 0) {};
\node (VR2) at ({-3*cos(30)}, {3*sin(30)}) {};
\node (VRm) at (-3/4, 9/4) {};

\foreach \x/\n in {L/1, L/2, L/n} 
{
  \draw [mid arrow] (\x\n) -- (0);
  \draw (V\x\n) -- (\x\n);
  \draw (\x\n) --++ (0,0,1.5);
}
\foreach \x/\n in {R/1, R/2, R/m} 
{
  \draw [mid arrow] (0) -- (\x\n);
  \draw (\x\n) -- (V\x\n);
  \draw (\x\n) --++ (0,0,1.5);
}
\draw [dash pattern=on .85pt off 1pt, line width = .3pt] (0) --++ ({-sin(40)}, {cos(40)});
\draw [dash pattern=on .85pt off 1pt, line width = .3pt] (0) --++ ({cos(40)}, {-sin(40)});

\draw (0) --++ (0, 0, 2);

\node (v) at (-1/3,-1/3) {\scriptsize\(v_0\)};
\end{tikzpicture}\hspace{-.5em},
  \end{equation}
  where \(E_{v_0}^{\mathrm{t}} = \left\{ e_1, \ldots, e_n \right\}\) and \(E_{v_0}^{\mathrm{o}} = \left\{ f_1, \ldots, f_m \right\}\).
  After using Theorem \ref{thm:1} to move the representations \(R\) and \(L\) to the virtual legs and
  inverting the edge tensors, which are assumed to be LRI, we are left with
  \begin{equation}
\begin{tikzpicture}[baseline={([yshift=-\the\fontdimen22\textfont2]0.center)}, y={(-1em,1.5em)},x={(1em,1.5em)}, z={(0em,.5em)}]
\node [tensor] (0) at (0, 0) {\(\)};
\node (L1) at (0, -2) {};
\node (L2) at ({2*sin(30)}, {-2*cos(30)}) {};
\node (Ln) at (3/2, -1/2) {};
\node (R1) at (-2, 0) {};
\node (R2) at ({-2*cos(30)}, {2*sin(30)}) {};
\node (Rm) at ({.97*-1/2}, {.97*3/2}) {};

\foreach \x/\n in {L/1, L/2, L/n} 
{
  \draw [mid arrow] (\x\n) -- (0);
}
\foreach \x/\n in {R/1, R/2, R/m} 
{
  \draw [mid arrow] (0) -- (\x\n);
}
\draw [dash pattern=on .85pt off 1pt, line width = .3pt] (0) --++ ({-sin(40)}, {cos(40)});
\draw [dash pattern=on .85pt off 1pt, line width = .3pt] (0) --++ ({cos(40)}, {-sin(40)});

\node [matrix, label = {[label distance = -3pt]left:{\contour{white}{\scriptsize\(U_{v_0}(g)\)}}}] (m0) at (0,0,2.5) {};
\draw (m0) --++ (0, 0, 2);
\draw (0) -- (m0);

\node (v) at (-1/3,-1/3) {\scriptsize\(v_0\)};
\end{tikzpicture} = 
\begin{tikzpicture}[baseline={([yshift=-\the\fontdimen22\textfont2]0.center)}, y={(-1em,1.5em)},x={(1em,1.5em)}, z={(0em,.5em)}]
\node [tensor] (0) at (0, 0) {\(\)};
\node [matrix, label = {[label distance = -2pt]right:{\contour{white}{\scriptsize\(V_{e_1^{\mathrm{t}}}(g)\)}}}] (L1) at (0, -3/2) {};
\node [matrix, label = {[shift = {(+1/4,-2/3)}]above:{\contour{white}{\scriptsize\(V_{e_2^{\mathrm{t}}}(g)\)}}}] (L2) at ({3/2*sin(30)}, {-3/2*cos(30)}) {};
\node [matrix, label = {[label distance = -1pt]right:{\contour{white}{\scriptsize\(V_{e_n^{\mathrm{t}}}(g)\)}}}] (Ln) at (1, -1/3) {};
\node [matrix, label = {[label distance = -2pt]left:{\contour{white}{\scriptsize\(V_{f_1^{\mathrm{o}}}(g)^{-1}\)}}}] (R1) at (-3/2, 0) {};
\node [matrix, label = {[shift = {(-2/3,+1/4)}]above:{\contour{white}{\scriptsize\(V_{f_2^{\mathrm{o}}}(g)^{-1}\)}}}] (R2) at ({-3/2*cos(30)}, {3/2*sin(30)}) {};
\node [matrix, label = {[label distance = -1pt]left:{\contour{white}{\scriptsize\(V_{f_m^{\mathrm{o}}}(g)^{-1}\)}}}] (Rm) at (-1/3, 1) {};

\node (VL1) at (0, -5/2) {};
\node (VL2) at ({5/2*sin(30)}, {-5/2*cos(30)}) {};
\node (VLn) at (1.6, {-1.6/3}) {};
\node (VR1) at (-5/2, 0) {};
\node (VR2) at ({-5/2*cos(30)}, {5/2*sin(30)}) {};
\node (VRm) at ({-1.65/3}, 1.65) {};

\foreach \x/\n in {L/1, L/2, L/n} 
{
  \draw [mid arrow] (\x\n) -- (0);
  \draw (V\x\n) -- (\x\n);
}
\foreach \x/\n in {R/1, R/2, R/m} 
{
  \draw [mid arrow] (0) -- (\x\n);
  \draw (\x\n) -- (V\x\n);
}
\draw [dash pattern=on .85pt off 1pt, line width = .3pt] (0) --++ ({-sin(40)}, {cos(40)});
\draw [dash pattern=on .85pt off 1pt, line width = .3pt] (0) --++ ({cos(40)}, {-sin(40)});

\draw (0) --++ (0, 0, 3);

\node (v) at (-1/3,-1/3) {\scriptsize\(v_0\)};
\end{tikzpicture}.\qedhere
  \end{equation}
  \begin{figure}
    \centering
    \begin{tikzpicture}[rotate = 45, scale = 1.2]
\clip [rotate = -45] (0,2/3) ellipse (7 and 5);
\clip (0,0) ellipse (5 and 5);
\draw [purple, fill = purple!30!white] (5,0) ellipse (9/5 and 3/2);
\draw [purple, fill = purple!30!white] (0,5) ellipse (1 and 9/5);
\draw [purple, fill = purple!30!white, rotate = 60] (0,5) ellipse (1 and 9/5);
\draw [purple, fill = purple!30!white] plot [smooth cycle, tension=.8] coordinates {(-6,-1/2) (-3.7,1/2) (-2,-3) (2.6, -3.0) (1,-6)};
\node [tensor] (0) at (0, 0) {\(\)};
\node [stensor] (L1) at (0, -2) {};
\node [stensor] (L2) at ({2*sin(30)}, {-2*cos(30)}) {};
\node [stensor] (Ln) at (2, 0) {};
\node [stensor] (R1) at (-2, 0) {};
\node [stensor] (R2) at ({-2*cos(30)}, {2*sin(30)}) {};
\node [stensor] (Rm) at (0, 2) {};

\node [tensor, fill = white] (VL1) at (0, -4) {};
\node [tensor, fill = white] (VL2) at ({4*sin(30)}, {-4*cos(30)}) {};
\node [tensor, fill = white] (VLn) at (4, 0) {};
\node [tensor, fill = white] (VR1) at (-4, 0) {};
\node [tensor, fill = white] (VR2) at ({-4*cos(30)}, {4*sin(30)}) {};
\node [tensor, fill = white] (VRm) at (0, 4) {};

\foreach \x/\n in {L/1, L/2, L/n} 
{
  \draw [mid arrow] (\x\n) -- (0);
  \draw [mid arrow] (V\x\n) -- (\x\n);
}
\foreach \x/\n in {R/1, R/2, R/m} 
{
  \draw [mid arrow] (0) -- (\x\n);
  \draw [mid arrow] (\x\n) -- (V\x\n);
}
\draw [dash pattern=on .82pt off 1pt, line width = .3pt] (VL1) --++ ({1/sqrt(2)}, {-1/sqrt(2)});
\draw [dash pattern=on .82pt off 1pt, line width = .3pt] (VL1) --++ ({-1/sqrt(2)}, {-1/sqrt(2)});
\draw [dash pattern=on .82pt off 1pt, line width = .3pt] (VL1) --++ (0, -1);
\draw [dash pattern=on .82pt off 1pt, line width = .3pt] (VL2) --++ ({sin(10)}, {-cos(10)});
\draw [dash pattern=on .82pt off 1pt, line width = .3pt] (VL2) --++ ({sin(50)}, {-cos(50)});
\draw [dash pattern=on .82pt off 1pt, line width = .3pt] (VLn) --++ ({cos(45)}, {sin(45)}) {};
\draw [dash pattern=on .82pt off 1pt, line width = .3pt] (VLn) --++ ({cos(45)}, {-sin(45)}) {};
\draw [dash pattern=on .82pt off 1pt, line width = .3pt] (VLn) --++ (1, 0);
\draw [dash pattern=on .82pt off 1pt, line width = .3pt] (VR2) --++ ({-cos(30)}, {sin(30)}) {};
\draw [dash pattern=on .82pt off 1pt, line width = .3pt] (VR2) --++ ({-cos(75)}, {sin(75)}) {};
\draw [dash pattern=on .82pt off 1pt, line width = .3pt] (VR2) --++ ({-cos(-15)}, {sin(-15)}) {};
\draw [dash pattern=on .82pt off 1pt, line width = .3pt] (VR1) --++ (-1, 0);
\draw [dash pattern=on .82pt off 1pt, line width = .3pt] (VRm) --++ ({sin(20)}, {cos(20)});
\draw [dash pattern=on .82pt off 1pt, line width = .3pt] (VRm) --++ ({-sin(20)}, {cos(20)});

\draw [dash pattern=on .82pt off 1pt, line width = .3pt] (0) --++ ({-sin(30)}, {cos(30)});
\draw [dash pattern=on .82pt off 1pt, line width = .3pt] (0) --++ ({cos(30)}, {-sin(30)});
\draw [domain=-40:-20, dash pattern=on .4pt off 4pt] plot ({2*cos(\x)}, {2*sin(\x)});
\draw [domain=-40:-20, dash pattern=on .4pt off 8pt] plot ({4*cos(\x)}, {4*sin(\x)});
\draw [domain=110:130, dash pattern=on .4pt off 4pt] plot ({2*cos(\x)}, {2*sin(\x)});
\draw [domain=110:130, dash pattern=on .4pt off 8pt] plot ({4*cos(\x)}, {4*sin(\x)});

\node (v) at (-1/2,-1/2) {\(v_0\)};
\node (L) at (-3.6,-3/2) {\(\color{purple}\Lambda_{v_0}\)};
\end{tikzpicture}
      \caption{
        We sketch a general example of how \(\Lambda_{v_0}\) looks like.
        While the individual invertible neighborhoods of the edges can be chosen to be full and connected,
        there is no reason why their union should remain to be so.
        Crucially, \(\Lambda_{v_0}\) contains all vertices adjacent to \(v_0\) as a vertex of \(\Gamma\),
        but none as a vertex of \(\Gamma_{\mathrm{b}}\).
      }
      \label{fig:injective-neighbourhood-vertex}
  \end{figure}
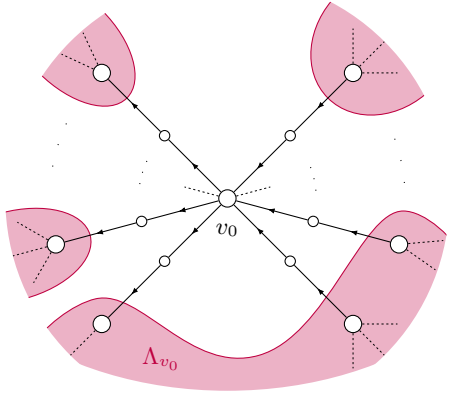
\end{proof}
\begin{corollary}
  Let \(\ket{\Psi_{\Gamma}(A)}\) be a normal PEPS
  on a \emph{big enough} graph \(\Gamma\)
  and \(B_e\) be an arbitrary collection of unital edge tensors,
  such that the resulting gPEPS \(\ket{\Psi_{\Gamma}(A, B)}\)
  is gauge-invariant.
  Then all vertex and edge tensors satisfy their appropriate intertwiner relations.
\end{corollary}
\begin{proof}
  \emph{Big enough} means that the injective regions,
  which can be constructed for each edge by the normality of the vertex tensors,
  do not overlap and form injective neighborhoods in the sense of \ref{def:inj-nbhd}.
  Then the conditions of Theorem \ref{thm:2} are satisfied for every vertex
  and the intertwiner properties hold everywhere.
\end{proof}
\begin{corollary}
  Let \(\ket{\psi_{\Gamma}(A)}\) be a normal PEPS with local onsite symmetry
  on a \emph{big enough} graph \(\Gamma\).
  Then the PEPS tensors \(A_v\) transform trivially under the symmetry, i.e.\
  \begin{equation}
    U_v (g) \circ A_v = A_v,
  \end{equation}
  for all \(g\in G\) and \(v \in \operatorname{vert}\Gamma\).
\end{corollary}
\begin{proof}
  The PEPS \(\ket{\psi_{\Gamma}(A)}\) can be viewed as a gauge-invariant
  gPEPS \(\ket{\psi_{\Gamma_{\mathrm{b}}}(A, B = \mathbbm{1})}\),
  with trivial edge tensors and Hilbert spaces.
  Since the edge tensors transform trivially,
  Theorem \ref{thm:2} implies the same for the vertex tensors.
\end{proof}

\section{Conclusions and outlook}
In this work we have analyzed the internal structure of tensor network states with a local (gauge) symmetry. The results derived here are valid for any dimension and geometry, and in particular the important case of three-space-dimensional cubic lattices. The main technical assumptions of this work are the LRI property of the edge tensors and the invertibility condition, cf.\ Definition \ref{def:inj-nbhd}, imposed on all edges.

How these conditions can be weakened, e.g., by using other notions of injectivity, like $G$-injectivity that in 2D results in topologically ordered PEPS~\cite{schuch_peps_2010}, we leave for future works.
Another interesting problem is the construction of parent Hamiltonians, possessing these PEPS as their unique ground states.

\ack{
We thank Alberto Ruiz de Alarc\'on for helpful discussions and input at the beginning of the project.
This research has been funded in part by 
the  European Union’s Horizon 2020 research and innovation programme through Grant No.\ 863476 (ERC-CoG SEQUAM).
D.B.\ acknowledges funding by the FWF Quantum Austria Funding Initiative project ``Entanglement Order Parameters'' (\href{https://doi.org/10.55776/P36305}{10.55776/P36305}), funded through the European Union (NextGenerationEU).
J.G.R.\ acknowledges funding by the FWF Erwin Schrödinger Program (\href{https://doi.org/10.55776/J4796}{10.55776/J4796}).
E.Z.\ acknowledges funding by the European Union (ERC-CoG OverSign, 101122583). Views and opinions expressed are, however, those of the author(s) only and do not necessarily reflect those of the European Union or the European Research Council. Neither the European Union nor the granting authority can be held responsible for them.}

\begin{appendices}
\section{Graphs}\label{sec:graphs}
\begin{definition}[Directed graph]
  A \emph{directed graph} is an ordered pair \(\Gamma = (V, E)\) of sets \(V\) and \(E\subseteq V\times V\).
  The elements of the set \(\operatorname{vert}\Gamma := V\) are called \emph{vertices} and
  the elements of the set \(\operatorname{edge}\Gamma := E\),
  which are ordered pairs of vertices, are called \emph{edges}.
  For any edge \(e\equiv (v_1,v_2) \in E\),
  the vertex \(\operatorname{o}(e) := v_1 \in V\) is called the \emph{origin} of \(e\) and
  the vertex \(\operatorname{t}(e) := v_2 \in V\) is called the \emph{terminus} of \(e\).
  For any vertex \(v\in V\) let
  \(E_v^{\mathrm{o}}\) denote the set of edges
  that have \(v\) as their origin
  (i.e.\ the outgoing edges),
  \(E_v^{\mathrm{t}}\) denote the set of edges
  that have \(v\) as their terminus
  (i.e.\ the incoming edges)
  and \(E_v\) denote the set of all adjacent edges.
  In formulae these sets are given as
  \begin{align*}
  E_v^{\mathrm{o}} &:= \left\{ e\in E \mid v = \operatorname{o}(e)\right\},\\
  E_v^{\mathrm{t}} &:= \left\{ e\in E \mid v = \operatorname{t}(e)\right\},\\
  E_v &:= E_v^{\mathrm{o}} \cup E_v^{\mathrm{t}}.
  \end{align*}
\end{definition}
\begin{definition}[Oriented graph]
  A directed graph \(\Gamma\) for which
  \((v_1, v_2) \in \operatorname{edge}\Gamma \) implies
  \((v_2, v_1) \notin \operatorname{edge}\Gamma\),
  is called an \emph{oriented graph}.
\end{definition}
\begin{definition}[Subgraph]
  We say that
  the graph \(\Gamma^{\prime}\) is a \emph{subgraph} of the graph \(\Gamma\),
  denoted by \(\Gamma^{\prime} \subseteq \Gamma\), if
  \(\operatorname{vert}\Gamma^{\prime}\subseteq \operatorname{vert}\Gamma\) and
  \(\operatorname{edge}\Gamma^{\prime}\subseteq \operatorname{edge}\Gamma\).
  Subgraphs are said to be \emph{disjoint} if their vertex sets are disjoint.
\end{definition}
\begin{remark}
  Clearly any subgraph of an oriented graph will also be oriented.
  Given a graph \(\Gamma\) one can construct all subgraphs by taking subsets
  \(V \subseteq \operatorname{vert}\Gamma\) and
  \(E \subseteq \operatorname{edge}\Gamma\) satisfying
  \(\operatorname{o}(e), \operatorname{t}(e) \in V\) for all \(e \in E\).
\end{remark}
\begin{definition}[Union of subgraphs]\label{def:subgraph-union}
  Let \(\Gamma_1 = (V_1, E_1)\) and \(\Gamma_2 = (V_2, E_2)\) be subgraphs of \(\Gamma\).
  We define the union of \(\Gamma_1\) with \(\Gamma_2\) to be the subgraph
  \begin{equation}
    \Gamma_1 \cup \Gamma_2 := (V_1 \cup V_2, E_1 \cup E_2) \subseteq \Gamma.
  \end{equation}
\end{definition}
\begin{definition}[Induced subgraph]\label{def:full-subgraph}
  A subgraph \(\Gamma^{\prime} \subseteq \Gamma\) is called \emph{induced} or \emph{full}
  if it contains all edges in \(\Gamma\) connecting vertices in \(\operatorname{vert}\Gamma^{\prime}\).
  It is often denoted \(\Gamma[S]\),
  where \(S := \operatorname{vert}\Gamma^{\prime}\subseteq \operatorname{vert}\Gamma\).
\end{definition}
\begin{definition}[Boundary of a subgraph]\label{def:bdry-subgraph}
  Given a subgraph \(\Gamma^{\prime} = (V^{\prime}, E^{\prime})\subseteq \Gamma= (V,E)\),
  we define the \emph{internal}, \emph{terminal} and \emph{original boundaries}
  of \(\Gamma^{\prime}\) (with respect to \(\Gamma\)) to be
  \begin{align*}
    \partial_{\mathrm{i}}\Gamma^{\prime} &:= \left\{e\in E\setminus E^{\prime} \mid
    \operatorname{o}(e)\in V^{\prime}\wedge\operatorname{t}(e)\in V^{\prime}\right\},\\
    \partial_{\mathrm{t}}\Gamma^{\prime} &:= \left\{e\in E\setminus E^{\prime} \mid
    \operatorname{o}(e)\notin V^{\prime}\wedge\operatorname{t}(e)\in V^{\prime}\right\},\\
    \partial_{\mathrm{o}}\Gamma^{\prime} &:= \left\{e\in E\setminus E^{\prime}
    \mid \operatorname{o}(e)\in V^{\prime}\wedge\operatorname{t}(e)\notin V^{\prime}\right\},
  \end{align*}
  respectively.  
\end{definition}
\begin{definition}[Bipartite graph]
  If the vertex set \(V\) of a graph \(\Gamma = (V,E)\)
  is the union of two disjoint subsets \(V = V_1 \cup V_2\) and 
  every edge in \(E\) connects a vertex in \(V_1\) with one in \(V_2\),
  then we say \(\Gamma\) is a \emph{bipartite} graph and we commonly write \(\Gamma = (V_1, V_2, E)\).
\end{definition}
\begin{definition}[Subdivision]
  Let \(\Gamma=(V,E)\) be an oriented graph.
  The \emph{subdivision} of \(\Gamma\) with respect to an edge \(e = (v_1,v_2) \in E\)
  is an oriented graph \(\Gamma_e\) defined by
  \begin{align}
    \operatorname{vert}\Gamma_e &:= V \cup \left\{ e \right\}\\
    \operatorname{edge}\Gamma_e &:= E \cup \left\{ (v_1,e), (e,v_2) \right\} \setminus \left\{ e \right\}.
  \end{align}
\end{definition}
\begin{remark}
  Intuitively, we cut an edge in half and introduce a new vertex at the cut.
  The orientation of the new edges is induced from the original edge.
  The new edges \(e^{\mathrm{o}} := (v_1, e)\) and \(e^{\mathrm{t}} := (e, v_2)\)
  can respectively be viewed as the first and second half of what used to be the edge \(e\).
  Pictorially:
  
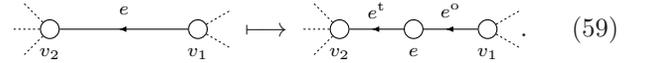
\begin{equation}
    \label{eq:14}
    \begin{tikzpicture}[baseline={([yshift=-\the\fontdimen22\textfont2]1.center)}]
\node [tensor, label = below:{\scriptsize \(v_1\)}] (1) at (4, 0) {\(\)};
\node [tensor, label = below:{\scriptsize \(v_2\)}] (2) at (0, 0) {\(\)};
\node (e) at (2,1/2) {\scriptsize\(e\)};
\draw [mid arrow] (1) -- (2);
\draw [dash pattern=on .82pt off 1pt, line width = .3pt] (2) --++ (-1, 0);
\draw [dash pattern=on .82pt off 1pt, line width = .3pt] (2) --++ ({-1/sqrt(2)}, {1/sqrt(2)});
\draw [dash pattern=on .82pt off 1pt, line width = .3pt] (2) --++ ({-1/sqrt(2)}, {-1/sqrt(2)});
\draw [dash pattern=on .82pt off 1pt, line width = .3pt] (1) --++ ({cos(30)}, {sin(30)}) {};
\draw [dash pattern=on .82pt off 1pt, line width = .3pt] (1) --++ ({cos(30)}, {-sin(30)}) {};
\end{tikzpicture} \;\longmapsto\; \begin{tikzpicture}[baseline={([yshift=-\the\fontdimen22\textfont2]1.center)}]
\node [tensor, label = below:{\scriptsize \(v_1\)}] (1) at (4, 0) {\(\)};
\node [tensor, label = below:{\scriptsize \(v_2\)}] (2) at (0, 0) {\(\)};
\node [tensor, label = below:{\scriptsize \(e\)}] (e) at (2, 0) {\(\)};
\node (eo) at (3,1/2) {\scriptsize\(e^{\mathrm{o}}\)};
\node (et) at (1,1/2) {\scriptsize\(e^{\mathrm{t}}\)};
\draw [mid arrow] (1) -- (e);
\draw [mid arrow] (e) -- (2);
\draw [dash pattern=on .82pt off 1pt, line width = .3pt] (2) --++ (-1, 0);
\draw [dash pattern=on .82pt off 1pt, line width = .3pt] (2) --++ ({-1/sqrt(2)}, {1/sqrt(2)});
\draw [dash pattern=on .82pt off 1pt, line width = .3pt] (2) --++ ({-1/sqrt(2)}, {-1/sqrt(2)});
\draw [dash pattern=on .82pt off 1pt, line width = .3pt] (1) --++ ({cos(30)}, {sin(30)}) {};
\draw [dash pattern=on .82pt off 1pt, line width = .3pt] (1) --++ ({cos(30)}, {-sin(30)}) {};
\end{tikzpicture}.
  \end{equation}
\end{remark}
\begin{definition}[Barycentric subdivision]\label{def:barycentric}
  Subdividing every edge of a graph \(\Gamma\) yields a new graph \(\Gamma_{\mathrm{b}}\),
  called the \emph{barycentric} subdivision of \(\Gamma\).
\end{definition}
\begin{remark}
  The barycentric subdivision of any graph \(\Gamma = (V,E)\) is bipartite and it is convenient to write
  \(\Gamma_{\mathrm{b}} = (V, E, \mathcal{E})\), where
\begin{align}\label{eq:7}
  \mathcal{E}
  &= \bigcup_{e\in E} \left\{ e^{\mathrm{o}}, e^{\mathrm{t}} \right\},&
  e^{\mathrm{t}} &:= (e, \operatorname{t}(e)),&
  e^{\mathrm{o}} &:= (\operatorname{o}(e), e).
\end{align}
\end{remark}
\end{appendices}
\bibliography{new-bibtex}
\end{document}